\documentclass[10pt]{article}
\usepackage{algorithm,algorithmicx,algpseudocode,amssymb,amsmath}
\usepackage{bm,color,comment,dsfont,epsfig,fullpage,graphicx,pifont}
\usepackage{url}
\usepackage{authblk}
\usepackage{makecell}
\usepackage{enumitem}
\usepackage{xcolor}
\usepackage{epstopdf}
\usepackage[caption=false]{subfig}
\usepackage{graphicx}
\graphicspath{{figures/}}
\newcommand{\RomanNumeralCaps}[1]{\MakeUppercase{\romannumeral #1}}
\usepackage{bm}
\usepackage{lipsum}
\usepackage{amsfonts}
\usepackage{graphicx}
\usepackage{epstopdf}
\usepackage{cleveref}
\allowdisplaybreaks
\ifpdf
  \DeclareGraphicsExtensions{.eps,.pdf,.png,.jpg}
\else
  \DeclareGraphicsExtensions{.eps}
\fi

\newcommand\numberthis{\addtocounter{equation}{1}\tag{\theequation}}
\newtheorem{assumption}{Assumption}

\newtheorem{lemma}{Lemma}
\newtheorem{theorem}{Theorem}
\newtheorem{remark}{Remark}
\def\lb{\left(}
\def\rb{\right)}
\def\lcb{\left\{}
\def\rcb{\right\}}
\def\ln{\left\|}
\def\rn{\right\|}
\def\lsb{\left[}
\def\rsb{\right]}
\def\P{\mathcal{P}}
\def\Po{\mathcal{P}_{\Omega}}
\def\Pt{\mathcal{P}_{T^t}}
\def\Pti{\mathcal{P}_{T^{t,i}}}

\def\Ho{\mathcal{H}_{\Omega}}
\def\I{\mathcal{I}}
\def\T{\mathcal{T}}
\def\De{\Delta}
\def\La{\Lambda}
\def\Si{\Sigma}

\DeclareMathOperator{\diag}{diag}

\begin{document}

\title{Fast and Provable Nonconvex Robust Matrix Completion}

\author[1]{Yichen Fu}
\author[2]{Tianming Wang}
\author[1]{Ke Wei}
\affil[1]{School of Data Science, Fudan University, Shanghai, China (22110980005@m.fudan.edu.cn,~kewei@fudan.edu.cn)}
\affil[2]{School of Mathematics, Southwestern University of Finance and Economics, Chengdu, Sichuan, China (wangtm@swufe.edu.cn)}

\maketitle

\begin{abstract}
We study the robust matrix completion (RMC)  problem subject to both sparse outliers and stochastic noise. A non-convex method termed Accelerated Robust Matrix Completion (ARMC) is proposed, which accelerates a prior non-convex approach by incorporating an explicit subspace projection step into the low-rank update, leading to significantly improved computational efficiency.  Through a delicate  analysis based on the leave-one-out technique,  the entrywise linear convergence guarantee of ARMC has been established. Notably,  the derived bounds for sample complexity and outlier sparsity improve upon existing guarantees of the convex relaxation approach that also accounts for both sparse outliers and stochastic noise. Moreover, numerical experiments on synthetic and real data show that ARMC is superior to existing non-convex RMC methods.
\end{abstract}

\section{Introduction}
This paper studies the robust matrix completion  (RMC) problem which is about reconstructing a low rank matrix from partially revealed entries that have been contaminated by sparse outliers as well as stochastic additive noise. Without loss of generality, assume $L^{\star}\in\mathbb{R}^{n\times n}$ in RMC is a square matrix of rank $r$. 
The goal is to recover $L^{\star}$ from the following measures:
\begin{equation}\label{eq:RMC-model}
    M_{ij} = L^{\star}_{ij} + S^{\star}_{ij} + N_{ij},\quad (i,j)\in \Omega,
\end{equation}
where  $\Omega\subseteq[n]\times[n]$ with $[n]:=[1,2,\dots,n]$ is an index subset, $S^{\star}_{ij}$ denotes the outlier, and $N_{ij}$ denotes the stochastic noise.
The above RMC problem covers a wide range of low rank matrix reconstruction problems  (e.g., low rank matrix completion and robust principal component analysis), and has many real applications, such as 
recommendation and rating systems with adversarial or abnormal users \cite{van2010manipulation}, video foreground/background separation from incompletely sampled frames \cite{de2003framework,bouwmans2018applications}, sensor and traffic data recovery with occasional spikes \cite{mardani2013recovery}.  


When there are only missing entries without corrupted outliers, RMC reduces to the classical low rank matrix completion problem, which has received a lot of investigations since the pioneering work in \cite{Candes2009,RFP:SIREV:10}. 
Convex relaxation via nuclear norm minimization is a natural approach for low rank matrix completion, whose theoretical recovery guarantee has been established in 
\cite{Candes2009,Rec:JMLR:11,Chen2015}  based on the construction of dual  certificate. In addition to the convex approach, there have also been many developments of the non-convex approaches for low rank matrix completion. 
A partial list includes gradient descent on Burer--Monteiro factorizations \cite{zheng-pg,procrustesflow2016}, SVP/iterative hard thresholding (IHT)  \cite{Jain2010,tanner2013normalized}, and Riemannian optimization on low-rank manifolds \cite{vandereycken2013low, wei2016guarantees,KMO:TIT:10}, which are overall gradient descent methods for certain non-convex loss functions associated with matrix completion. 
The analysis of the nonconvex methods  often relies on the incoherence of each iterate, which are earlier  imposed directly via trimming/projection steps such as in OptSpace \cite{Jain2015,KMO:TIT:10}, or ensured by sample splitting to maintain statistical independence across iterations \cite{Hardt2014,Jain2015,Wei2020}. 
Recent development of the leave-one-out  technique -{}- rooted in high-dimensional regression and random matrix eigenvector analysis  -{}- paves the way for the {incoherence-projection-free and sample-splitting-free} convergence guarantee analysis \cite{Abbe2020,Zhong2018,Ma2019,Ding2020,Ling2022}. 
For instance, Ma et al.\ \cite{Ma2019} have established the linear convergence of vanilla gradient descent, while Ding and Chen \cite{Ding2020} have presented an entrywise linear convergence of  IHT/SVP-type  without explicit incoherence projection. 

On the other hand, when the data matrix is fully observed but a small fraction of entries are contaminated by large-magnitude outliers, it becomes the robust PCA (RPCA) problem, which seeks to decompose a matrix into a low rank part and a sparse part.
It is shown in  \cite{Candes2011} that a convex approach based on the $\ell_1$-regularized nuclear norm minimization can achieve  exact recovery  under proper incoherence and sparsity assumptions. 
There are also non-convex methods developed for RPCA, which usually alternate between the updates of the low-rank part and the sparse part, achieving significant acceleration compared to convex approach while maintaining rigorous statistical guarantees \cite{Yi2016,accaltprj2019,Tanner2023}. Beyond computational efficiency, a recent work has conceptually bridged convex and nonconvex viewpoints: iterative nonconvex schemes are not only practical algorithms, but can also be the analytical tool for certifying the optimality of the convex approach, even when noise or missing data are present \cite{Chen2021}. 

RMC generalizes both matrix completion and RPCA, with the goal of recovering the underlying low-rank component \(L^\star\) from partially observed data subject to sparse outliers. 
Non-convex approaches for RMC usually conduct a low rank update based on the methods for low rank matrix completion, followed by  the sparse estimation via thresholding. 
For example, the gradient-descent–based robust PCA method has been adapted to the missing-data setting \cite{Yi2016}, and a Gauss-Newton method with support identification for the sparse part has been introduced in \cite{gaussnewton}. Note that the analysis in both \cite{Yi2016} and \cite{gaussnewton} require  explicit incoherence projection/constraint.
Another representative method, R-RMC, which combines the SVP update with hard thresholding for the sparse part is introduced in \cite{Cherapanamjeri2017}, whose theoretical guarantee requires sample-splitting. To overcome the  limitations of explicit incoherence projection and sample splitting, as inspired by \cite{Chen2021}, \cite{wang2024leave} studies a RMC algorithm with continuous thresholding strategies such as soft-thresholding or SCAD  and has established the incoherence-projection-free and sample-splitting-free guarantee based on the leave-one-out technique.

\subsection{Motivation and Main Contributions}
The convex approach analyzed in \cite{Chen2021} offers a theoretical guarantee for RMC in the presence of both outliers and stochastic noise.  However, no fast non-convex method has been shown to achieve a similar guarantee without requiring explicit incoherence projection or sample splitting, to the best of our knowledge. Our work fills this gap, and the main contributions of this paper are summarized as follows.

\begin{itemize}
  \item 
 An accelerated non-convex RMC method termed ARMC is proposed. It improves upon the method in \cite{wang2024leave} by introducing a subspace projection into the low-rank update, which confines the update to the tangent space of the low-rank manifold at the current estimate. This allows the best rank-\(r\) truncation to be computed more efficiently and thus reduces the per-iteration computational cost. Empirical studies on both synthetic and real data demonstrate the superiority of ARMC over prior non-convex methods, especially in terms of computational efficiency.
  \item 
  Building on the projection-free and sample-splitting-free leave-one-out 
framework of \cite{wang2024leave,Chen2021}, we establish the entrywise 
linear convergence of ARMC in the presence of both outliers and stochastic 
noise.  
Table~\ref{tab:comparison} summarizes the theoretical guarantees of 
representative convex and non-convex RMC methods. As shown therein, only 
the convex approach studied in \cite{Chen2021} and  ARMC can handle outliers and 
noise simultaneously, with ARMC being substantially faster as a non-convex method. 

In addition to its computational advantages, the subspace projection is also instrumental in obtaining our improved theoretical guarantees: it enables us to control the infinity norm of the estimation error more effectively. 
Consequently, our bounds on the sample complexity $p$ and outlier sparsity $\alpha$ 
improve upon those in \cite{Chen2021}, exhibiting reduced dependency on 
both $\kappa$ and $\log n$, while the noise tolerance $\sigma$ is 
comparable. 
Additionally, while our requirements on $p$ and $\alpha$ are more stringent 
than those  convex guarantees that only consider outliers 
\cite{Li2012,Chen2013}, the bound for $p$ is better  than that for the non-convex approaches in \cite{Yi2016, wang2024leave} in terms of $\kappa$.
\end{itemize}

\begin{table}[!htp]
\caption{Summary of Theoretical Guarantees for Related RMC Methods.}
\label{tab:comparison}
\begin{center}
\small
\begin{tabular}{| c | c | c | c |}
\hline 
& Sample Complexity & Outlier Sparsity & Noise Magnitude \\
\hline
\makecell[cc]{Convex approach \\ \cite{Li2012}} & $p\gtrsim\frac{\mu r\log^2 n}{n}$ & $\alpha\lesssim 1$ & \scalebox{1.5}{\ding{55}} \\
\hline
\makecell[cc]{Convex approach \\ \cite{Chen2013}}
& $p\gtrsim \frac{\mu r \log^6 n}{n}$ 
& $\alpha \lesssim 1$
& \scalebox{1.5}{\ding{55}} \\
\hline
\makecell[cc]{Convex approach \\ \cite{Chen2021}} & $p\gtrsim\frac{\kappa^4\mu^2r^2\log^6 n}{n}$ & $\alpha\lesssim\frac{1}{\kappa^3\mu r\log n}$ & $\sigma\lesssim\sqrt{\frac{np}{\kappa^4\mu r\log n}}\cdot\frac{\sigma_r^{\star}}{n}$ \\
\hline
RPCA–GD \cite{Yi2016} & $p\gtrsim\frac{\kappa^4\mu^2r^2\log n}{n}$ & $\alpha\lesssim\min\left\{\frac{1}{\kappa^2\mu r},\frac{1}{\kappa^{3/2}\mu r^{3/2}}\right\}$ & \scalebox{1.5}{\ding{55}} \\
\hline
RMC \cite{wang2024leave} & $p\gtrsim\frac{\kappa^4\mu^2r^2\log n}{n}$ & $\alpha\lesssim\frac{1}{\kappa^2\mu r}$ & \scalebox{1.5}{\ding{55}} \\
\hline
\textbf{ARMC} & $p\gtrsim\frac{\kappa^3\mu^2r^2\log^2 n}{n}$ & $\alpha\lesssim\frac{1}{\kappa^2\mu r}$ & $\sigma\lesssim\min\left\{\frac{1 / \alpha}{\kappa\sqrt{\log n}},\sqrt{\frac{n p}{\kappa^2\log n}}\right\} \cdot \frac{\sigma_{r}^{\star}}{n}$ \\
\hline
\end{tabular}
\end{center}
\end{table}

\subsection{Notation and Organization}
We adopt standard notation throughout the paper. Let \( e_i \in \mathbb{R}^n \) denote the \( i \)-th standard basis vector, and \( \bm{1} \in \mathbb{R}^n \) be the all-one vector. The identity matrix in \( \mathbb{R}^{n \times n} \) is denoted by \( I \), and for any matrix \( Z \in \mathbb{R}^{n \times n} \), we use \( Z_{i,:} \) and \( Z_{:,j} \) to denote its \( i \)-th row and \( j \)-th column, respectively. $\text{Supp}(\cdot)$ denotes the support of a matrix. The inner product between two matrices is defined as \( \langle A, B \rangle = \mathrm{trace}(A^\top B) \). For matrix norms, \( \| Z \|_{\infty} \) represents the maximum absolute value of entries, \( \| Z \|_2 \) the spectral norm (i.e., the largest singular value),  \( \| Z \|_{\mathrm{F}} \) the Frobenius norm, and $\|Z\|_{2,\infty}$ is defined as \(\max _i\left\|Z_{i,:}\right\|_2\). The notation $f\gtrsim g$ means there exists a universal constant $C>0$ such that $f\ge C\cdot g$, similarly for $\lesssim$. The statement ``with high probability'' indicates that the associated event holds with probability at least \( 1 - C_1 n^{-C_{2}} \) for some absolute constants \( C_1, C_{2} > 0 \) and $C_{2}$ being sufficiently large. In the analysis, we also use the symmetrization of a  matrix: for  $ Z\in\mathbb{R}^{n\times n}$, its symmetrization is defined as 
\begin{equation}\label{eq:symmetry}
\widehat{Z} = \left[\begin{array}{cc}
0   & Z \\
Z^T & 0
\end{array}\right].  
\end{equation}


The rest of this paper is organized as follows. Section~\ref{sec:alg} presents the proposed algorithm, its theoretical guarantee, and numerical experiments. Section~\ref{sec:proof_outline} contains the proof strategy, while Section~\ref{sec:proofs} provides the proof sketches of the main result and the new lemmas that enable our improved bounds. The detailed proofs of all supporting lemmas are deferred to the appendices. We conclude this paper with a discussion in Section~\ref{sec:conclusion}.

\section{ARMC and Its Theoretical Guarantee}\label{sec:alg}
\subsection{Accelerated Robust Matrix Completion Method}
When there are no outliers, it is natural to consider the following constrained least-squares problem
\begin{equation}\label{form:ls}
\min_{Z}~ \frac{1}{2} \left\| \Po(Z - M) \right\|_F^2 \quad \text{subject to} \quad \operatorname{rank}(Z) = r,
\end{equation}
where \( \mathcal{P}_\Omega \) is the projection operator onto the observed entries.
SVP  (also known as iterative hard thresholding \cite{Jain2010,tanner2013normalized})  is a standard algorithm for solving \eqref{form:ls}, whose update is given by 
\[
L^{t+1} = \P_r\left( L^t - p^{-1} \Po\left(L^t - M\right) \right),
\]
where \(\P_r\) computes the best rank-\(r\) approximation of a matrix via singular value decomposition (SVD).


In the presence of sparse corruptions (or outliers), the R-RMC algorithm introduced in \cite{Cherapanamjeri2017} augments SVP with a hard-thresholding step to identify the outliers. Thus the update is given by 
\begin{equation}\label{eq:R-RMC}
\begin{aligned}
S^t &= \mathcal{T}_{\eta^t} \left( \Po\left(M - L^t\right) \right), \\
L^{t+1} &= \P_r \left( L^t - p^{-1} \Po\left( L^t + S^t - M \right) \right),
\end{aligned}
\end{equation}
where \(\mathcal{T}_\eta(\cdot)\) in R-RMC refers to the hard-thresholding operator defined by
\[
\mathcal{T}_\eta(z) =
\begin{cases}
0, & |z| \leq \eta \\
z, & \text{otherwise}
\end{cases}.
\]

Theoretical recovery guarantee of \eqref{eq:R-RMC}  has also been established in \cite{Cherapanamjeri2017}, which requires {sample splitting} to decouple the statistical dependencies across different iterations. However, while theoretically convenient, sample splitting is often unnecessary in practice. In contrast, \cite{wang2024leave}  considers a broad family of $\mathcal{T}_{\eta}$ satisfying certain smooth properties (including soft-thresholding and SCAD \cite{Fan2001}, but not including hard thresholding) and then establishes the theoretical guarantee of \eqref{eq:R-RMC} without sample splitting by employing the leave–one–out technique.

Note that in order to perform $\mathcal{P}_r(\cdot)$ in \eqref{eq:R-RMC}, one needs to first compute the first $r$  principal singular values and singular vectors of an $n\times n$ matrix, and the computational cost is overall $O(n^2r)$ (typically with a large hidden constant). In order to improve the computational efficiency, as inspired by \cite{vandereycken2013low,wei2016guarantees,Wei2020}, we propose an accelerated (nonconvex) robust matrix completion (ARMC) method based on subspace projection, see Algorithm~\ref{Alg1} for a complete description. 

\begin{algorithm}[!ht]
\caption{Accelerated (Nonconvex) Robust Matrix Completion (ARMC)} \label{Alg1}
\begin{algorithmic}[1]
\State \textbf{Input:} Thresholding parameters \(\beta_1, \beta_2\) and geometric decay rate \(\gamma \in (0,1)\);
\State \textbf{Initialization:} Set \(\xi^0 = \beta_1 + \beta_2\), compute \(S^0 = \mathcal{T}_{\xi^0}\left( \Po(M) \right)\), and set \(L^1 = \P_r\left( p^{-1} \Po(M - S^0) \right)\);
\For{\(t = 1, 2, \dots\)}
    \State Update threshold: \(\xi^t = \beta_1 \cdot \gamma^t + \beta_2\);
    \State Update sparse component: \(S^t = \mathcal{T}_{\xi^t}\left( \Po(M - L^t) \right)\);
    \State Update low-rank component: \(L^{t+1} = \P_r\mathcal{P}_{T^t}\left( L^t - p^{-1} \Po(L^t + S^t - M) \right) \).
\EndFor
\end{algorithmic}
\end{algorithm}

Compared with \eqref{eq:R-RMC}, the key difference lies in the introduction of the subspace $T^t$ when updating the low rank part. In particular, we will choose $T^t$ as the tangent space of the rank-$r$ matrix manifold at the current estimate  $L^t$. Letting $L^t = U^t\Sigma^t (V^t)^T$ be the compact SVD of $L^t$, the tangent space $T^t$ is given by 
\[
 T^t = \left\{U^t A^T+B (V^t)^T:~A,B\in\mathbb{R}^{n\times r}\right\}.
\]
For any matrix $Z$, the projection onto $T^t$ can be computed  as follows:
\begin{equation} \label{eq:projection}
\mathcal{P}_{T^t}(Z) = U^t(U^t)^T Z + ZV^t(V^t)^T - U^t(U^t)^T  ZV^t(V^t)^T.
\end{equation}
After the projection, the truncated SVD can be carried out at low cost. Basically, noting that all the matrices in $T^t$ are of rank at most $2r$ and they also share partial column and row spaces, it is possible to exploit this structure to  compute the SVD more efficiently. We refer interested readers to    \cite{vandereycken2013low,wei2016guarantees,Wei2020} for details.

The main theoretical guarantee of ARMC is presented in the next section. We would like to emphasize that the introduction of the additional tangent-space projection not only yields a much more efficient algorithm, but also enables us to derive an improved sample complexity compared to existing ones.

\subsection{Theoretical Guarantee of ARMC} In this section, we will establish the theoretical guarantee of ARMC for the general setting in \eqref{eq:RMC-model}, where there exist both outliers and additive noise. To the best of our knowledge, this is the first recovery guarantee of a nonconvex method for RMC which considers both outliers and stability under stochastic noise.

We begin by introducing some standard assumptions  about the structural conditions of the ground truth matrix, the observation model, as well as the statistical properties of corruption and noise.
Recall that \( L^{\star} \in \mathbb{R}^{n \times n} \) is the unknown target matrix of rank $r$. Suppose the compact SVD of $L^{\star}$ is given by
$
L^{\star} = U^{\star} \Sigma^{\star} (V^{\star})^T,
$
where \( U^{\star}, V^{\star} \in \mathbb{R}^{n \times r} \) are orthonormal matrices, and \( \Sigma^{\star} = \mathrm{diag}(\sigma_1^{\star}, \ldots, \sigma_r^{\star}) \) is a diagonal matrix with decreasing singular values. The condition number of \( L^{\star} \) is denoted as  \( \kappa := \sigma_1^{\star} / \sigma_r^{\star} \).


\begin{assumption}\label{assump1}
The left and right singular vectors of \( L^{\star} \) are assumed to spread out across all coordinates, rather than concentrate on a few entries, which can be expressed as
\[
\| U^{\star} \|_{2,\infty} \leq \sqrt{\frac{\mu r}{n}}, \quad \| V^{\star} \|_{2,\infty} \leq \sqrt{\frac{\mu r}{n}},
\]
where we recall that \( \| Z \|_{2,\infty} := \max_i \| Z_{i,:} \|_2 \) denotes the largest row norm of a matrix \( Z \). 
As a result, it follows immediately that the maximum absolute entry of \( L^{\star} \), as well as the maximum row/column length of \( L^{\star} \), can be bounded as follows:
\begin{align*}
\| L^{\star} \|_{\infty} \leq \frac{\mu r}{n} \sigma_1^{\star}, \quad \| L^{\star} \|_{2,\infty}\leq\sqrt{\frac{\mu r}{n}} \sigma_1^{\star},\quad \text{and~}  \| (L^{\star})^T \|_{2,\infty} \leq \sqrt{\frac{\mu r}{n}} \sigma_1^{\star}.\numberthis\label{eq:tmp001}
\end{align*}
\end{assumption}

\begin{assumption}\label{assump2}
The entries of the matrix are observed independently with equal probability. Specifically, each entry \( (i,j) \in [n] \times [n] \) is revealed with probability \( p \), independent of the other entries.
\end{assumption}

\begin{assumption}\label{assump3}
Let \( \Omega_{S^{\star}} \subseteq \Omega \) be the set of corrupted indices within the observed entries. We assume that the number of outliers per row and per column does not exceed \( 2 \alpha p n \), where \( \alpha \in (0,1) \) is a small parameter. 
Meanwhile, we assume that the noise matrix \( N \in \mathbb{R}^{n \times n} \) has independent entries, and each  entry \( N_{ij} \)  is a symmetric mean-zero sub-Gaussian random variable with the sub-Gaussian norm being at most \( \sigma \), see for example \cite{vershynin2009high} for more about sub-Gaussian random variables. 
\end{assumption}

\begin{assumption}\label{assump4}
For the thresholding operator \(\mathcal{T}_\lambda\),  it should satisfy:
\begin{enumerate}[label=\textbf{(P.\arabic*)},left=0.6cm]
    \item \label{P1}for any input \(x\) with \(|x| \leq \lambda\), there holds \(\mathcal{T}_\lambda(x) = 0\);
    \item \label{P2}there exists a constant \(K \) such that \(|\mathcal{T}_\lambda(x) - \mathcal{T}_\lambda(y)| \leq K |x - y|\) for all \(x, y\);
    \item \label{P3}there exists a constant \(B\) such that \(|\mathcal{T}_\lambda(x) - x| \leq B\lambda\) for all \(x\).
\end{enumerate}
\end{assumption}

\begin{remark}
Both Assumptions~\ref{assump1}~and~\ref{assump2} are standard assumptions in low rank matrix completion. Moreover, Assumption~\ref{assump1} is also known as the incoherence condition \cite{Candes2009} that can prevent degenerate cases where the rank structure is  localized.  Assumption~\ref{assump3} guarantees that corruptions are sparse relative to the overall observation pattern. This is a more general assumption  than the uniform random corruption model and allows for mild dependencies in outlier patterns. The sub-Gaussian noise model in Assumption~\ref{assump3}  ensures that the additive noise behaves well in high probability, allowing us to use concentration inequalities in the analysis. 
Assumption~\ref{assump4} is firstly proposed in \cite{wang2024leave}.
Notably, both the soft-thresholding operator and the SCAD shrinkage function \cite{Fan2001} satisfy this assumption, as formally shown in \cite[Lemma~14]{wang2024leave}.
\end{remark}


Based on the above assumptions, we are now ready to present the main theoretical guarantee for ARMC.
\begin{theorem}\label{thm:main}
Suppose $\beta_1$ and $\beta_2$ in Algorithm~\ref{Alg1} satisfy
\begin{align*}
\frac{\mu r}{n}\sigma^{\star}_1&\leq\beta_1\leq C_{\emph{init}}\cdot\frac{\mu r}{n}\sigma^{\star}_1,\\
(1+\gamma)C_N^{(1)} \sigma\sqrt{\log n}&\leq \beta_2\leq C_{\emph{init}}\cdot (1+\gamma)C_N^{(1)} \sigma\sqrt{\log n},
\end{align*}
where $C_{\emph{init}}\geq 1$ and $C_N^{(1)}>0$ are two constants \textup{(}$C_N^{(1)}$ is specified in equation \eqref{eq:noise_infty}\textup{)}. 
Let $C_{\mathrm{thresh}}:=(K+B)\cdot C_{\emph{init}}$, where $K$ and $B$ are parameters in Assumption~\ref{assump4}. Assume
$$
p\geq\frac{C_{\emph{sample}}}{\gamma^2}\cdot\frac{\kappa^3\mu^2r^2\log^2 n}{n},\quad
\alpha\leq\frac{c_{\emph{outlier}}}{\kappa^2\mu r}\cdot\frac{\gamma}{C_{\mathrm{thresh}}}
$$
for some sufficiently large constant $C_{\emph{sample}}>0$ and some sufficiently small constant $c_{\emph{outlier}}>0$, and
$$
\sigma \lesssim \min\left\{\frac{\gamma/\alpha}{C_{\mathrm{thresh}}\kappa\sqrt{\log n}},\sqrt{\frac{\gamma^2 np}{\kappa^2\log n}}\right\}\cdot\frac{\sigma_r^{\star}}{n}.
$$
Then the iterates of Algorithm~\ref{Alg1} with $L^0:=0$ satisfy
$$
\ln L^{t}-L^{\star} \rn_{\infty}\leq \lb\frac{\mu r}{n}\sigma_{1}^{\star}\rb\gamma^t+\gamma C_N^{(1)}\sigma\sqrt{\log n},
$$
and
$$
\emph{Supp}\lb S^{t}\rb\subseteq\Omega_{S^{\star}},~\ln \Po\lb S^{t}-S^{\star}\rb\rn_{\infty}\leq C_{\mathrm{thresh}}\lsb\lb\frac{\mu r}{n}\sigma_{1}^{\star}\rb\gamma^t+\lb 1+\gamma\rb C_N^{(1)}\sigma\sqrt{\log n}\rsb
$$
with high probability for iteration $0\leq t\leq T$, where $T=n^{O(1)}$.
\end{theorem}

When $\sigma=0$, Theorem~\ref{thm:main} implies that ARMC converges to the ground truth matrix linearly.  It can also be seen that larger $\gamma$ allows us to solve the robust matrix completion problem with fewer samples and more outliers, but with a slower convergence rate. We will outline the proof of Theorem~\ref{thm:main} in Section~\ref{sec:proof_outline} and defer the proof details to Section~\ref{sec:proofs}. 



\subsection{Numerical Experiments}
Here we empirically evaluate the performance of ARMC against two nonconvex methods with theoretical recovery guarantees when there are no additive noise, namely nonconvex robust matrix completion without tangent-space projection \cite{wang2024leave} (RMC) and the fast gradient-based RPCA method \cite{Yi2016} (RPCA-GD). For RMC or ARMC, preliminary tests show that using the soft-thresholding operator and SCAD for the sparse part overall presents similar performance, thus we only report the results with the soft-thresholding operator.  Tests have been conducted on a laptop with Apple M2 Max processor and 32G memory, and run in MATLAB R2021b. We first investigate the phase transition, computational time and stability of ARMC on synthetic data, and then test its performance on  real data. 

\subsubsection{Synthetic Data} 
The synthetic data is generated in the following way. The ground truth rank $r$ matrix $L^{\star}$ is represented as  $L^{\star}=U^{\star} \Sigma^{\star} (V^{\star})^T$. The singular vector matrices  $U^{\star}$ and $V^{\star}$ are generated from $n\times r$ random Gaussian matrices followed by orthogonalization and projection  onto the set of matrices with row norms no more than $\sqrt{\frac{r}{n}}$ and re-orthogonalization in order to yield a small incoherence. The singular values are uniformly distributed between $1$ and $1/\kappa$ so that the condition number of $L^{\star}$ is $\kappa$. Each location is then included in $\Omega$ with probability $p$, and with probability $\alpha$ every entry in $\Omega$ is corrupted with an outlier $S^{\star}_{ij}$ chosen uniformly in $[-\|L^{\star}\|_\infty,\|L^{\star}\|_\infty]$. For the phase transition and computational time tests, we do not consider the additive sub-Gaussian noise. We set $\beta_1=1.1\cdot\frac{\mu r}{n}\sigma_1^*$ and $\gamma=0.9$ in ARMC (the same for RMC). The parameters of RPCA-GD are sufficiently fine-tuned.

\begin{figure}[ht!]
\centering
\includegraphics[width=0.45\textwidth]{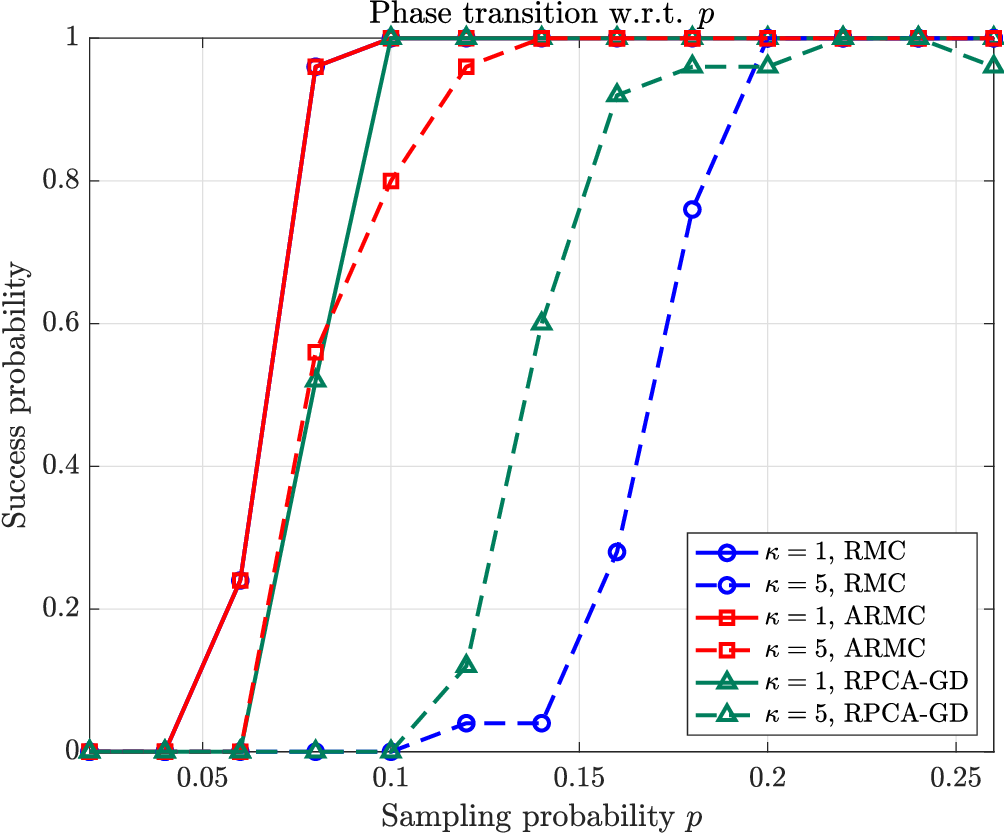}
\hspace{0.5cm}
\includegraphics[width=0.45\textwidth]{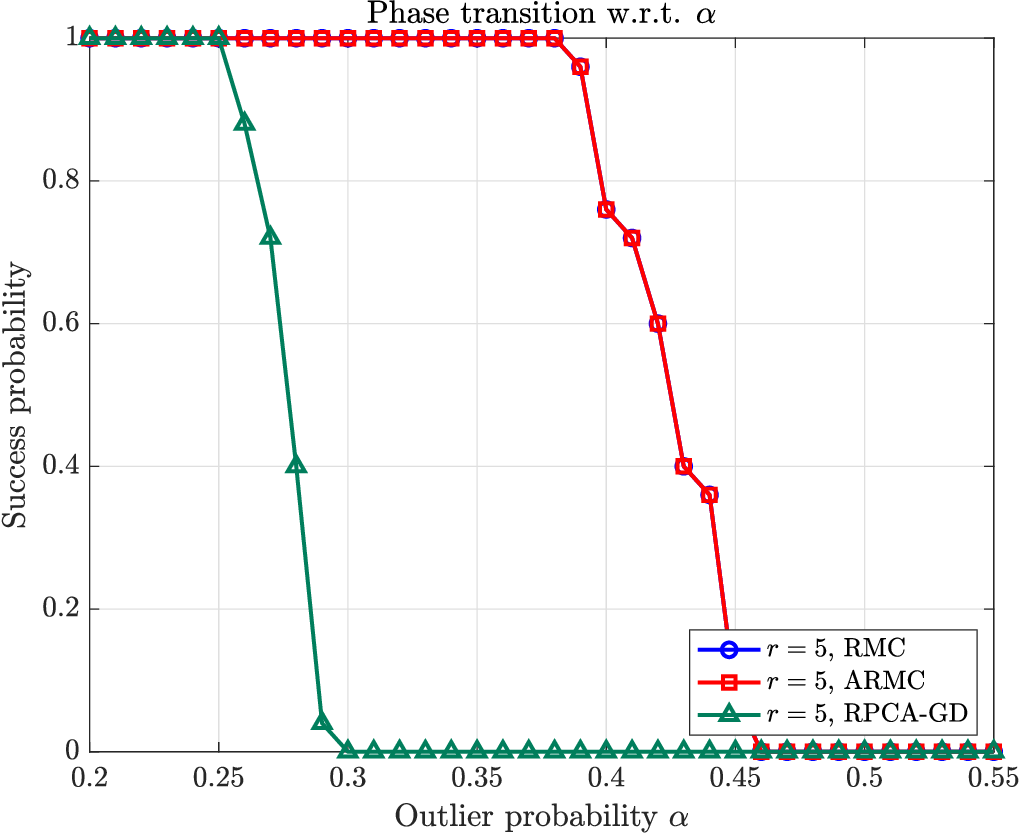}
\caption{Empirical phase transitions for the tested algorithms with $n=1000$ and $r=5$. Left: Success rates for $\alpha=0.15$ and $p\in\{0.02,0.04,\cdots,0.26\}$ with $\kappa\in\{1,5\}$; Right: Success rates for $p=0.2$ and $\alpha\in\{0.2,0.21,\cdots,0.55\}$ with $\kappa=2$.}\label{fig:phase}
\end{figure}

\paragraph{\textup{\textbf{Phase Transition}}}
We first fix $n=1000$, $r=5$, $\alpha=0.15$,   vary $p$ from $0.02$ to $0.26$, and test two condition numbers $\kappa\in\{1,5\}$. For each problem instance, $25$ random trials are repeated and a test is considered to be successful if the output low rank matrix $L_{\text{out}}$ satisfies $\|L_{\text{out}}-L^{\star}\|_\infty/\|L^{\star}\|_\infty\leq 10^{-3}$. The phase transitions of the tested algorithms are presented in the left subfigure of Fig.~\ref{fig:phase}. For $\kappa=1$, the three algorithms display similar phase transitions.  However, for $\kappa=5$, ARMC has better phase transition than the other two (note that the theoretical sample complexity of ARMC also has better dependency on $\kappa$ than that of RMC and RPCA-GD).

Next we fix $r=5$, $p=0.2$ and $\kappa=2$, and vary $\alpha$ from $0.2$ to $0.55$. The results are presented in the right subfigure of Fig.~\ref{fig:phase}. It can be seen that RMC and ARMC overall have the same phase transition and are able to tolerate more outliers than RPCA-GD. 

\begin{figure}[ht!]
\centering
\includegraphics[width=1\textwidth]{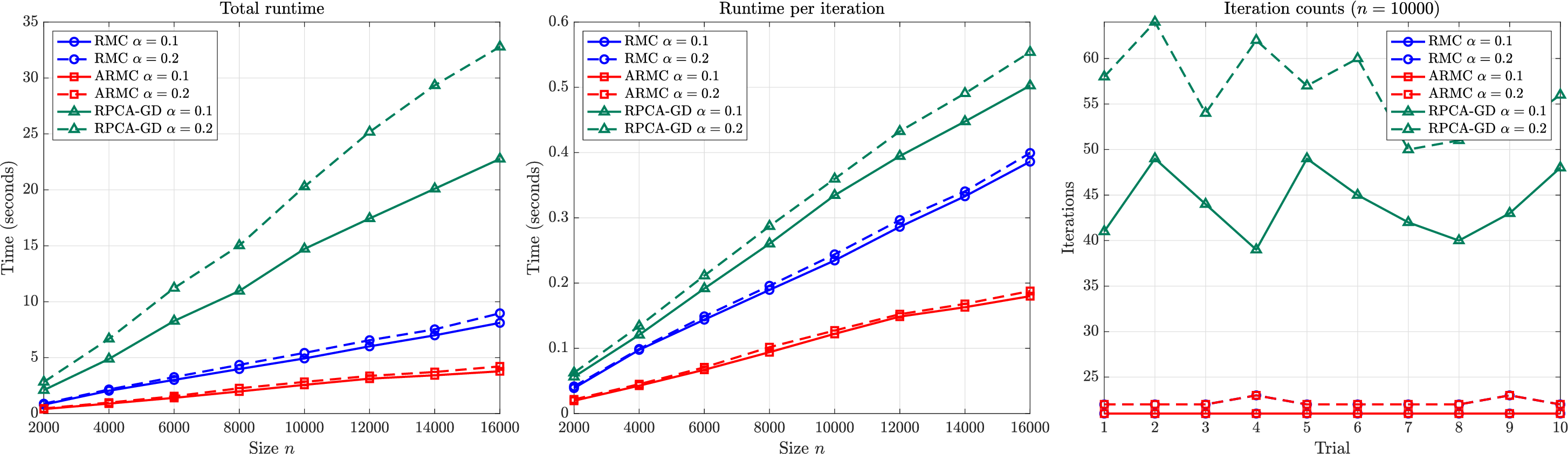}
\caption{Computational efficiency of the tested algorithms with $r=10$, $\kappa=2$ and $\alpha\in\{0.1,0.2\}$. Left: Total runtime comparisons; Middle: Runtime per iteration comparisons; Right: Iteration counts comparisons.}\label{fig:runtime}
\end{figure}

\paragraph{\textup{\textbf{Computational Time}}} To compare the computational efficiency, we consider the case $r=10$, $\kappa=2$, $\alpha\in\{0.1,0.2\}$, vary $n$ from $2000$ to $16000$ and compute $p=40\frac{r}{n}$ so that the oversampling ratio remains unchanged for different $n$. All the three algorithms are terminated when $\|L_{\text{out}}-L^{\star}\|_\infty/\|L^{\star}\|_\infty\leq 10^{-3}$.
The total computational time averaged out of $10$ random trials as well as the average runtime for each iteration are presented in Fig.~\ref{fig:runtime}. It is evident that ARMC has significantly lower per iteration computational cost and is also much faster than RMC and RPCA-GD. The per iteration computational cost of RMC and ARMC almost remain unchanged in contrast to RPCA-GD, as the latter relies on sorting to perform the thresholding operations. The numbers of iterations required to achieve the same accuracy for all the  $10$ random trials with $n=10000$ are also included in Fig.~\ref{fig:runtime}. It can be seen that RMC  and ARMC take the same number of iterations to converge which is desirable since this means that the  additional subspace projection in ARMC overall does not slow down the convergence  while can reduce the computational cost  substantially.

\begin{figure}[ht!]
\centering
\includegraphics[width=0.45\textwidth]{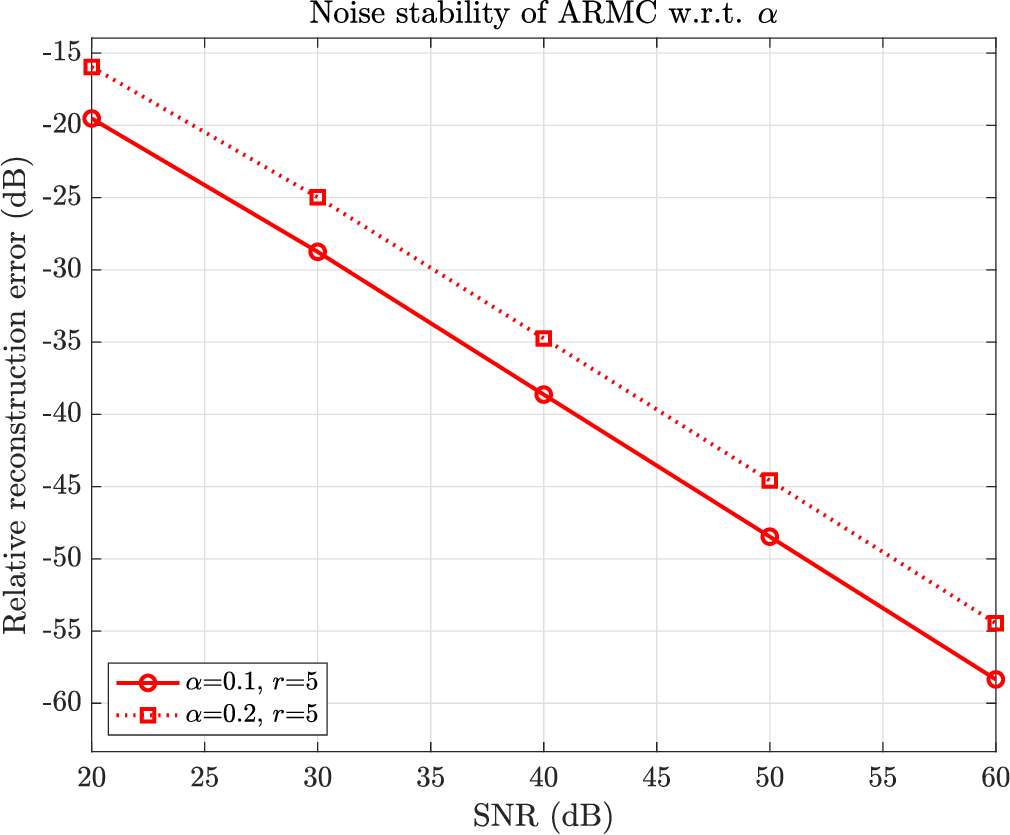}
\hspace{0.5cm}
\includegraphics[width=0.45\textwidth]{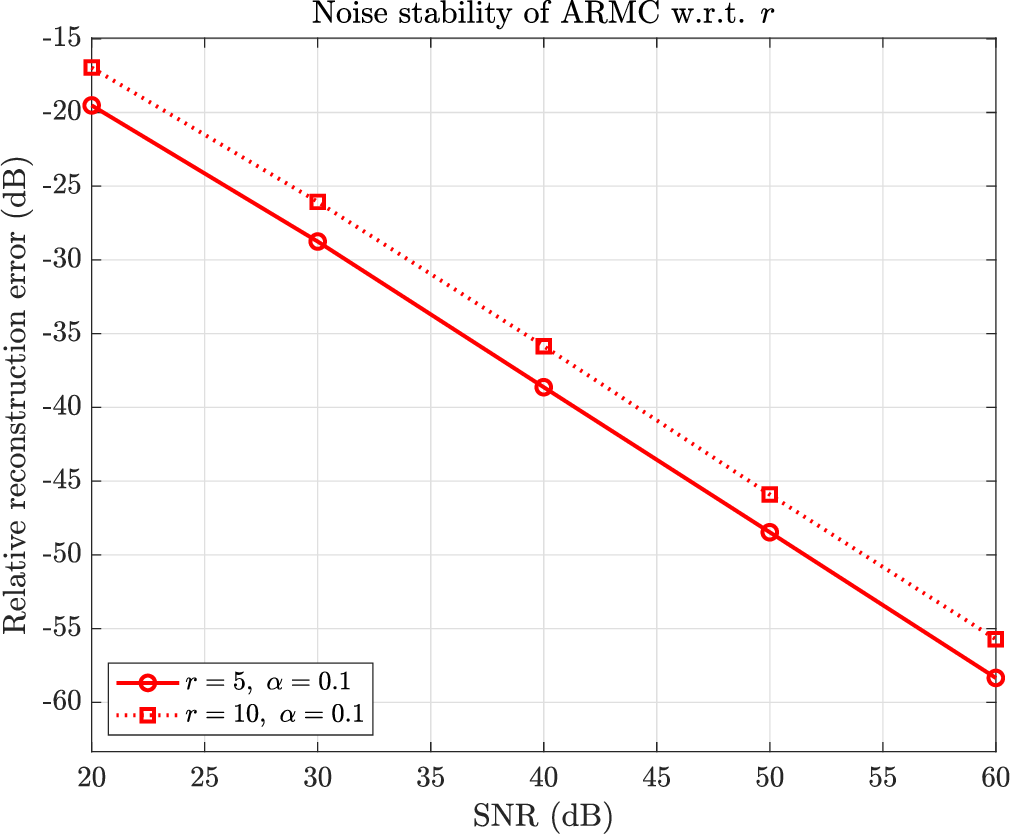}
\caption{Reconstruction stability of ARMC with respect to different noise levels in the observed samples. Left: Results with $r=5$ and $\alpha\in\{0.1,0.2\}$; Right: Results with $\alpha=0.1$ and $r\in\{5,10\}$. Here the relative reconstruction error is the noise-to-signal ratio calculated with $L_{\text{out}}-L^{\star}$ and $L^{\star}$.}\label{fig:snr}
\end{figure}

\paragraph{\textup{\textbf{Stability against Additive Noise}}} Here we only test the robustness of ARMC under the mean zero Gaussian noise with varying variance $\sigma^2$ so that the signal-to-noise ratio (SNR) ranges from $20$ to $60$. The other parameter $\beta_2$ in ARMC is set to be $\beta_2=1.1\cdot (1+\gamma)C_N^{(1)} \sigma\sqrt{\log n}$ following the condition in Theorem~\ref{thm:main}. Two scenarios with $n=1000$ and $p=0.3$ are tested: $r=5$, $\alpha\in\{0.1,0.2\}$ and $\alpha=0.1$, $r\in\{5,10\}$, and the results are presented in Fig.~\ref{fig:snr}, which shows the desirable linear dependency between the relative reconstruction error (i.e., noise-to-signal ratio, calculated with $L_{\text{out}}-L^{\star}$ and $L^{\star}$) and the SNR. The results also agree with the bound \eqref{eq:L_infty} we derive later in Theorem~\ref{thm:induction} for $\ln L_{\text{out}}-L^{\star}\rn_{\infty}$ since at convergence, the error is proportional to $C_{\mathrm{noise}}\sigma\frac{\sqrt{\kappa}\mu r}{n}$, and together with the definition of $C_{\mathrm{noise}}$ in \eqref{eq:C_noise} one can see that it increases linearly as $\alpha$ or $r$ increases.

\subsubsection{Real Data} 
\begin{figure}[!ht]
\subfloat{\includegraphics[width=.25\linewidth]{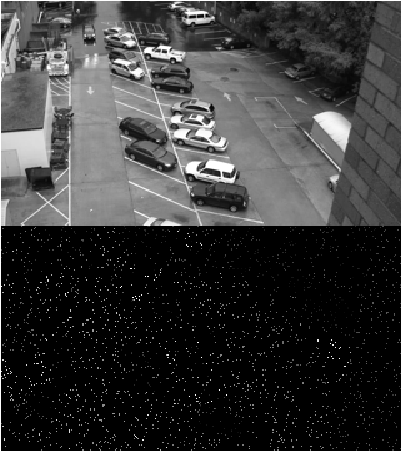}}\hfill
\subfloat{\includegraphics[width=.25\linewidth]{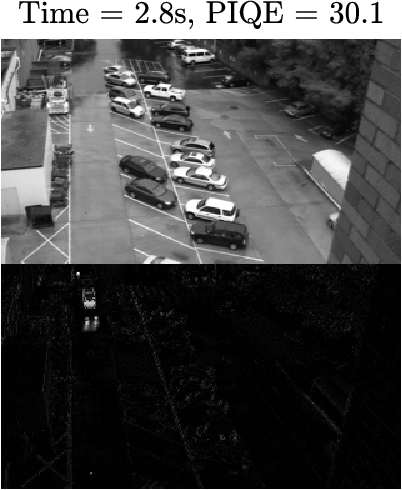}}\hfill
\subfloat{\includegraphics[width=.25\linewidth]{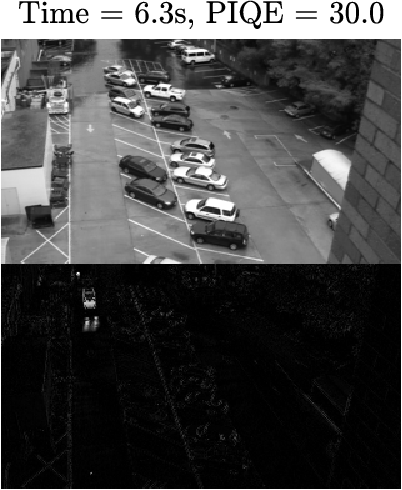}}%
\subfloat{\includegraphics[width=.25\linewidth]{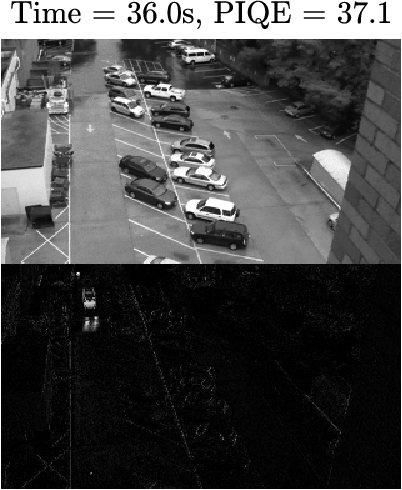}}%
\caption{Performances of the three algorithms on the test video with $p=0.04$. The first column shows one selected frame from the test video and the samples on that frame from a random trail. The second, third and fourth column show the separated background and foreground by ARMC, RMC and RPCA-GD, respectively, along with their averaged runtime and PIQE values over 10 trails.}\label{fig:real}
\end{figure}

Lastly, we show the performance of ARMC on the foreground/background separation task for a test video from the VIRAT dataset\footnote{Available at \url{https://viratdata.org/}, with the ID S$\_$050201$\_$10$\_$001992$\_$002056.}. The video has 1907 color frames, and each frame is converted to grayscale and downsampled to be of spatial resolution $180\times 320$. After vectorization, we get a data matrix $M\in\mathbb{R}^{57600\times 1907}$. We then run the three tested algorithms in recovering the static background and identifying  the moving sparse foreground from partial samples of $M$. After inspecting the singular values of $M$, we choose $r=2$ for the background. This choice allows us to account for relatively frequent camera movement due to the wind and produce sparser foreground estimates. For ARMC and RMC, we simply set $\beta_1$ to be  the maximum value of the data matrix (also see \eqref{eq:tmp001} for the validity) and observe that the output background estimates remain stable for a large range of values. The incoherence parameter used in RPCA-GD is computed from the output of ARMC. In Fig.~\ref{fig:real}, we show the performances of the three algorithms with a low sampling rate $p=0.04$. The evaluation metrics include the averaged runtime and averaged PIQE \cite{7084843} values computed on the estimated background over 10 random trials, as well as the visual result on a selected frame from one random trail. Note that PIQE index is a non-reference image quality evaluator that assigns lower values to higher quality images. One can see that ARMC is the fastest with the averaged PIQE value very similar to that of RMC. From the frame shown, one can also see that the separation results of the three algorithms are similar. The background estimates produced by ARMC and RMC are slightly cleaner than that of RPCA-GD, agreeing with the averaged PIQE values.

\section{Proof Strategy}\label{sec:proof_outline}
To prove Theorem~\ref{thm:main}, we need to control not only the entrywise error, but also the spectral norm of the perturbation and the incoherence of the iterates. In particular, establishing the incoherence property requires the leave-one-out technique \cite{Ma2019,Ding2020,wang2024leave}, which decouples the dependency between the iterates and the sampling pattern. This leads to a more detailed convergence result, stated as Theorem~\ref{thm:induction} at the end of this section, from which Theorem~\ref{thm:main} follows immediately.

The overall idea of leave-one-out is to construct an auxiliary sequence by running ARMC on observations based on $\Omega$ with one row or one  column of $L^\star$ fully given. 
Recall that the update of the low rank part in  Algorithm~\ref{Alg1} is given by 
\[
L^{t+1} = \P_r\mathcal{P}_{T^t}\big( L^t + \underbrace{p^{-1} \Po(M-(L^t + S^t))}_{\text{residual}} \big).
\]
When the entries in $i$-th (when $1\leq i\le n$) row or  the $(i-n)$-th (when $n+1\leq i\leq 2n$)  column are known, it is natural to replace the residual by $\mathcal{P}_i(L^\star-L^t)$ where $\mathcal{P}_i(\cdot)$ only sample the entries in the $i$-th row or $(i-n)$-th column. In this case, one also only needs to update the outliers out of the $i$-th  row or  the $(i-n)$-th   column. Define $\mathcal{P}_\Omega^{(-i)}(\cdot)$ as follows: 
\begin{align*}
\mbox{when $1\leq i\le n$}:\quad\lb\Po^{(-i)}(Z)\rb_{jk} &= \begin{cases}
\delta_{jk} Z_{jk} & j\neq i \\
0 & j = i
\end{cases},\\
\mbox{when $n+1\leq i\le 2n$}:\quad\lb\Po^{(-i)}(Z)\rb_{jk} &= \begin{cases}
\delta_{jk} Z_{jk} & k\neq i-n \\
0 & k = i-n
\end{cases}.
\end{align*}
Denote by $L^{t,i}$, $S^{t,i}$ the iterates obtained by the new update. One has 
\begin{align*}
 S^{t,i}=\T_{\xi^{t}}\lb\Po^{(-i)}\lb M-L^{t,i} \rb\rb
\end{align*}
and 
\begin{align*}
    L^{t+1,i} & =  \P_r\mathcal{P}_{T^{t,i}}\big( L^{t,i} + p^{-1} \Po^{(-i)}(M-(L^{t,i} + S^{t,i}))+\mathcal{P}_i(L^\star-L^{t,i}) \big)\\
    & =\P_r\big( L^{t,i} + p^{-1} \mathcal{P}_{T^{t,i}}\Po^{(-i)}(L^\star+S^\star+N-(L^{t,i} + S^{t,i}))+\mathcal{P}_{T^{t,i}}\mathcal{P}_i(L^\star-L^{t,i}) \big)\\
    & = \mathcal{P}_r\big(L^\star\underbrace{-E_1^{t,i}+E_2^{t,i}+E_3^{t,i}}_{:=E^{t,i}}\big),\numberthis\label{eq:defEti}
\end{align*}
where 
\begin{equation}\label{eq:tmp002}
E_1^{t,i}=\frac{1}{p}\Pti\Po^{(-i)}\lb S^{t,i}-S^{\star}\rb,~ 
E_2^{t,i}=\lb \I-\Pti\lb\I-\Ho^{(-i)}\rb \rb\lb L^{t,i}-L^{\star}\rb,~
E_3^{t,i}=\frac{1}{p}\Pti\Po^{(-i)}\lb N\rb.
\end{equation}
Note that in \eqref{eq:tmp002},  $\mathcal{H}_\Omega^{(-i)} = \mathcal{I}-p^{-1}\Po^{(-i)}-\mathcal{P}_i$, that is,
\begin{align*}
\mbox{when $1\leq i\le n$}:\quad\lb\Ho^{(-i)}(Z)\rb_{jk} &= \begin{cases}
(1-\frac1p\delta_{jk})Z_{jk} & j\neq i \\
0 & j = i
\end{cases},\\
\mbox{when $n+1\leq i\le 2n$}:\quad\lb\Ho^{(-i)}(Z)\rb_{jk} &= \begin{cases}
(1-\frac1p\delta_{jk})Z_{jk} & k\neq i-n \\
0 & k = i-n
\end{cases}.
\end{align*}

When the $i$-th row or $(i-n)$-th column of $L^\star$ is known, it is also natural to consider the following initialization:
\[
S^{0,i} =\T_{\xi^{0}}\lb\Po^{(-i)}\lb M \rb\rb
\]
and 
\begin{align*}
L^{1,i}&=\mathcal{P}_r\big(p^{-1}\mathcal{P}_\Omega^{(-i)}(M-S^{0,i})+\mathcal{P}_i(L^{\star})\big)=\mathcal{P}_r\big(L^{\star}\underbrace{-E_1^{0,i}+E_2^{0,i}+E_3^{0,i}}_{:=E^{0,i}}\big),\numberthis\label{eq:defE0i}
\end{align*}
where
\begin{align*}
  E_1^{0,i}=p^{-1}\Po^{(-i)}\lb S^{0,i}-S^{\star}\rb,~  E_2^{0,i}=\Ho^{(-i)}\lb -L^{\star}\rb,~E_3^{0,i}=p^{-1}\Po^{(-i)}\lb N \rb.
\end{align*}

It is not hard to see that the original sequence of the proposed algorithm can also be written in a unified way, with the definition $\Po^{(-0)}:=\Po$, and $\Ho^{(-0)}:=\Ho$. 

By the construction, $\{S^{t,i}\}_{t=0}^{\infty}$ and $\{L^{t,i}\}_{t=0}^{\infty}$ are independent with respect to the random variables $\left\{\delta_{ij}N_{ij}\right\}_{j=1}^n$ on the $i$-th row if $1\leq i\leq n$, and $\left\{\delta_{j(i-n)}N_{j(i-n)}\right\}_{j=1}^n$ on the $(i-n)$-th column if $n+1\leq i\leq 2n$. On the other hand, since only a small fraction is different when constructing the auxiliary sequence, it is anticipated the original and auxiliary sequences are close to each other. Moreover, the independence and proximity are key to establishing the incoherence property of the iterates.

Overall, the introduction of the auxiliary sequence enables us to prove Theorem~\ref{thm:main} in an inductive way. To this end, we need to introduce more notation.
Denote by  $U^{t+1,i}\Si^{t+1,i}\lb V^{t+1,i}\rb^T$ the compact SVD of $L^{t+1,i}$. The deviation between 
$$
F^{t+1,i} := \left[\begin{array}{c}
U^{t+1,i} \\
V^{t+1,i}
\end{array}\right]~\mathrm{and}~
F^{\star} := \left[\begin{array}{c}
U^{\star} \\
V^{\star}
\end{array}\right]
$$
can be measured by
$
\min_{R\in\mathcal{O}(r)}\ln F^{t+1,i}-F^{\star}R\rn_{\mathrm{F}},
$
where $\mathcal{O}(r):=\{R\in\mathbb{R}^{r\times r}~|~R^TR=I_r\}$. This is known as the orthogonal Procrustes problem \cite{procrustesflow2016,Ma2019}, and an optimal rotation matrix, denoted $G^{t+1,i}$, can be obtained by computing the SVD of $H^{t+1,i}:=\frac12\lb F^{\star}\rb^TF^{t+1,i}=A^{t+1,i}\widetilde{\Si}^{t+1,i}\lb B^{t+1,i}\rb^T$, and setting $G^{t+1,i}:=A^{t+1,i}\lb B^{t+1,i}\rb^T$. Further define
$\De^{t+1,i} := F^{t+1,i}-F^{\star}G^{t+1,i}$. 
For $i,m\in\{0,1,\cdots,2n\}$, the deviation between $F^{t+1,i}$ and $F^{t+1,m}$ can be measured by
$
\min_{R\in\mathcal{O}(r)}\ln F^{t+1,i}-F^{t+1,m}R\rn_{\mathrm{F}}.
$
Let $G^{t+1,i,m}$ be the optimal rotation matrix and $D^{t+1,i,m} := F^{t+1,i}-F^{t+1,m}G^{t+1,i,m}$. 
Lastly, define
$\ln E^{t,\infty}\rn_2=\max_{0\leq i\leq 2n}\ln E^{t,i} \rn_2$, $\ln \De^{t+1,\infty}\rn_{2,\infty}=\max_{0\leq i\leq 2n}\ln\De^{t+1,i}\rn_{2,\infty}$ and $\ln D^{t+1,\infty}\rn_{\mathrm{F}}=\max_{0\leq i,m\leq 2n}\ln D^{t+1,i,m}\rn_{\mathrm{F}}$.
We will show that the following induction hypotheses hold for all the sequences.

\begin{theorem}\label{thm:induction}
 Let  $C_1$ and $C_0$ be the constants satisfying \eqref{eq:constants} and  define $C_{\emph{noise}}$ as in \eqref{eq:C_noise}. Then, under the assumptions of Theorem~\ref{thm:main}, with high probability
\begin{subequations}
\begin{align}
    \ln E^{0,\infty}\rn_2 &\leq \frac{1}{C_0}\frac{\sigma_{r}^{\star}}{\sqrt{\kappa}}\gamma+4C_{\emph{noise}}\sigma\label{eq:op_t=0}\\
    \ln \De^{1,\infty}\rn_{2,\infty}&\leq\frac{5C_1}{C_0}\sqrt{\frac{\kappa\mu r}{n}}\gamma+10C_1C_{\emph{noise}}\lb\frac{\sigma}{\sigma_r^{\star}}\rb\sqrt{\frac{\kappa^2\mu r}{n}}\label{eq:l_2_infty_init}\\
    \ln D^{1,\infty}\rn_{\mathrm{F}}&\leq\frac{8C_1}{C_0}\sqrt{\frac{\mu r}{n}}\gamma+12C_1C_{\emph{noise}}\lb\frac{\sigma}{\sigma_r^{\star}}\rb\sqrt{\frac{\kappa\mu r}{n}}\label{eq:proximity_init} \\
    \max _{0 \leq i \leq 2 n} \left\|L^{1, i}-L^{\star}\right\|_{\infty} & \leq\left(\frac{\mu r}{n} \sigma_{r}^{\star}\right) \gamma+100C_1C_{\emph {noise }} \sigma\frac{\sqrt{\kappa} \mu r}{n} \label{eq:L_infty_init}
\end{align}
\end{subequations}
hold for $t=0$, and
\begin{subequations}
\begin{align}
    \ln E^{t,\infty}\rn_2 &\leq \frac{1}{C_0}\frac{\sigma_{r}^{\star}}{\kappa}\gamma^{t+1}+6C_{\emph{noise}}\sigma\label{eq:op}\\
    \ln \De^{t+1,\infty}\rn_{2,\infty}&\leq\frac{5C_1}{C_0}\sqrt{\frac{\mu r}{n}}\gamma^{t+1}+10C_1 C_{\emph{noise}}\lb\frac{\sigma}{\sigma_r^{\star}}\rb\sqrt{\frac{\kappa^2\mu r}{n}}\label{eq:l_2_infty}\\
    \ln D^{t+1,\infty}\rn_{\mathrm{F}}&\leq\frac{8C_1}{C_0}\sqrt{\frac{\kappa\mu r}{n}}\gamma^{t+1}+24C_1 C_{\emph{noise}}\lb\frac{\sigma}{\sigma_r^{\star}}\rb\sqrt{\frac{\kappa^2\mu r}{n}}\label{eq:proximity} \\
    \max _{0 \leq i \leq 2 n} \left\|L^{t+1, i}-L^{\star}\right\|_{\infty} & \leq\left(\frac{\mu r}{n} \sigma_{r}^{\star}\right) \gamma^{t+1}+100C_1 C_{\emph {noise }}\sigma \frac{\sqrt{\kappa} \mu r}{n}\label{eq:L_infty}
\end{align}
\end{subequations}
hold for $1\leq t\leq T$, where $T=n^{O(1)}$.
\end{theorem}
\begin{remark}\label{remark32}
    The conclusion of Theorem~\ref{thm:main} for $t=0$ follows easily as we show in the the proof of Lemma~\ref{lem:init_bounds}. For $t\geq 1$, in order to obtain Theorem~\ref{thm:main} from Theorem~\ref{thm:induction}, it requires
    \[
    100C_1C_{\emph{noise}}\sigma\frac{\sqrt{\kappa}\mu r}{n}\leq \gamma C_N^{(1)}\sigma\sqrt{\log n}.
    \]
    Noting the definition of $C_{\mathrm{noise}}$ in \eqref{eq:C_noise}, this is true if 
    \begin{equation}\label{eq:conditions}
      p \geq \lb\frac{200C_1C_N/C_N^{(1)}}{\gamma}\rb^2\cdot\frac{\kappa\mu^2r^2}{n}\quad\mathrm{and}\quad
    \alpha\leq \frac{C_N^{(1)}/C_N}{200C_1}\frac{1}{\sqrt{\kappa}\mu r}\cdot\frac{\gamma}{C_{\mathrm{thresh}}},  
    \end{equation}
    which can be satisfied by the assumptions on $p$ and $\alpha$ given in Theorem~\ref{thm:main}. It follows from Lemma~\ref{lem:thresh} that $\emph{Supp}\lb S^{t}\rb\subseteq\Omega_{S^{\star}}$ and
\begin{align*}
\ln\Po\lb S^{t}-S^{\star}\rb\rn_{\infty}\leq &  C_{\mathrm{thresh}}\lsb\lb\frac{\mu r}{n}\sigma_{1}^{\star}\rb\gamma^{t}+\lb 1+\gamma\rb C_N^{(1)}\sigma\sqrt{\log n}\rsb.
\end{align*}
\end{remark}

\section{Proof of Theorem~\ref{thm:induction}}\label{sec:proofs}

In this section, we outline the overall arguments for the proof of Theorem~2 and  highlight the key technical innovations that lead to the improved theoretical guarantees. We begin with several new lemmas that play a critical role in handling the noise and tightening our bounds. For conciseness, the details of the proofs are deferred to the appendices.

\subsection{Useful Lemmas}\label{sec:useful} 

The following lemma will be used to bound the noise terms. Note that, to ease the presentation, we often use this lemma with 
\begin{equation}\label{eq:C_N}
C_N :=\max\left\{\sqrt{2}+C_N^{(1)},C_N^{(2)}\right\}.
\end{equation}

\begin{lemma}\label{lem:noise}
Under Assumption~\ref{assump3},
\begin{equation}\label{eq:noise_infty}
|N_{ij}|\leq C_N^{(1)}\cdot\sigma\sqrt{\log n},\quad\forall i,j\in [n]
\end{equation}
holds with high probability for some  constant $C_N^{(1)}>0$. Moreover, conditioned on the event that \eqref{eq:noise_infty} holds, the following two claims further hold provided $p\geq \frac{\log^2n}{n}$: 
\begin{itemize}
    \item There exists a  constant $C_N^{(2)}>0$ such that
    $$
    \|\mathcal{P}_{\Omega}(N)\|_2\leq C_N^{(2)}\cdot\sigma\sqrt{np}
    $$
    holds with high probability.
\item Suppose $V\in\mathbb{R}^{n\times r}$ is independent with respect to $\{\delta_{mj}N_{mj}\}_{j=1}^n$ for some $1\leq m\leq n$. Then there exists a constant $C_1>0$ such that 
$$
\ln e_m^T\Po\lb N\rb V\rn \leq C_1\lb C_N\sqrt{np\log n}\rb\sigma\ln V\rn_{2,\infty}
$$
holds with high probability.
\end{itemize}
\end{lemma}

The next lemma is an extension of \cite[Lemma~2]{wang2024leave}, which is also used when handling the additive noise.
\begin{lemma}\label{lem:thresh}
Assume 
$\frac{\mu r}{n}\sigma^{\star}_1\leq\beta_1\leq C_{\emph{init}}\cdot\frac{\mu r}{n}\sigma^{\star}_1$ for some constant $C_{\emph{init}}\geq 1$, and 
$$
\lb 1+\gamma \rb C_N^{(1)}\sigma\sqrt{\log n}\leq \beta_2\leq C_{\emph{init}}\cdot\lb 1+\gamma \rb C_N^{(1)}\sigma\sqrt{\log n}
$$ 
for some $C_N^{(1)}>0$ such that \eqref{eq:noise_infty} holds. Then under Assumption~\ref{assump4}, if
$$
\ln L^{t,i}-L^{\star} \rn_{\infty}\leq \lb\frac{\mu r}{n}\sigma_{1}^{\star}\rb\gamma^t+\gamma C_N^{(1)}\sigma\sqrt{\log n},
$$
there hold $\emph{Supp}\lb S^{t,i}\rb\subseteq\Omega^{(-i)}\cap\Omega_{S^{\star}}$ and
$$
\ln\Po^{(-i)}\lb S^{t,i}-S^{\star}\rb\rn_{\infty}\leq C_{\mathrm{thresh}}\lsb\lb\frac{\mu r}{n}\sigma_{1}^{\star}\rb\gamma^t+\lb 1+\gamma \rb C_N^{(1)}\sigma\sqrt{\log n}\rsb.
$$
Here, $\Omega^{(-i)}$ is defined as follows: $\Omega^{(-i)}=\Omega$ 
if $i=0$; $\Omega^{(-i)}$ is $\Omega$ without the indices from the 
$i$-th row if $1\leq i\leq n$; and $\Omega^{(-i)}$ is $\Omega$ without 
the indices from the $(i-n)$-th column if $n+1\leq i\leq 2n$. 
We also define $C_{\mathrm{thresh}}:=(K+B)\cdot C_{\mathrm{init}}$.
\end{lemma}

To establish the incoherence property (i.e., \eqref{eq:l_2_infty_init} and \eqref{eq:l_2_infty}) of the iterates, we will use the following deterministic result together with the leave-one-out analysis to bound $\ln \cdot\rn_{2,\infty}$.
\begin{lemma}\label{lem:incoherence}
Suppose $L=\mathcal{P}_r\left(L^{\star}+E\right)$ for some perturbation matrix $E$ and let $L = U\Sigma V^T$ be its compact SVD. Define $F^{\star}=[(U^{\star})^T~(V^{\star})^T]^T$ and $F=[U^T~V^T]^T$. Denote the SVD of $H=\frac12\left(F^{\star}\right)^TF$ as $A\widetilde{\Si} B^T$. Set $G=AB^T$ and $\Delta=F-F^{\star}G$. If $\|E\|_{2} \leq\frac12\sigma_r^{\star}$. Then
\begin{align*}
\left\|\Delta\right\|_{2,\infty} \leq & 18\kappa\frac{\ln E\rn_2}{\sigma_r^{\star}}\sqrt{\frac{\mu r}{n}}+\max\lcb\ln EV\rn_{2,\infty},\ln E^TU\rn_{2,\infty}\rcb\cdot\ln \Si^{-1}\rn_2.
\end{align*}
\end{lemma}


Lemma~\ref{lem:L_infinity} is the main tool we use to sharpen the bound of the $\ln\cdot\rn_{\infty}$ norm, which in turn leads to the improved sample complexity dependence on $\kappa$. This lemma is similar 
in spirit to \cite[Lemma~9]{Cherapanamjeri2017}, except that we 
consider all $a\geq 0$ instead of restricting $a$ to be bounded 
by $\log n$. The proof follows the same argument and is provided 
in Appendix~\ref{subsec:Linfinity} for completeness.
\begin{lemma}\label{lem:L_infinity}
Let $L=\mathcal{P}_r(L^{\star}+E)$, where $E\in\mathbb{R}^{n\times n}$ is any perturbation matrix that satisfies:
\begin{enumerate}
    \item $\|E\|_2\leq \frac{\sigma_r^{\star}}{4}$;
    \item there exists some $v \leq \frac{\sigma_r^{\star}}{4}$ such that $\forall i\in[n]$, $\forall a\geq 0$,
    \begin{equation}\label{eq:induction}
    \begin{aligned}
\|e_i^T(E^TE)^aV^{\star}\|_2,~\|e_i^T(EE^T)^aU^{\star}\|_2\leq& v^{2a}\sqrt{\frac{\mu r}{n}}, \\
\|e_i^TE(E^TE)^aV^{\star}\|_2,~\|e_i^TE^T(EE^T)^aU^{\star}\|_2\leq&v^{2a+1}\sqrt{\frac{\mu r}{n}}.
    \end{aligned}
    \end{equation}
\end{enumerate}
Then one has
$$
\|L-L^{\star}\|_{\infty}\leq \frac{\mu r}{n}\lb 5\|E\|_2+24v\rb.
$$
\end{lemma}

Moreover, we can derive the following tight bound on the spectral norm estimation, which is used in obtaining \eqref{eq:op}.
\begin{lemma}\label{lem:Ho}
Under the same conditions of Lemma~\ref{lem:L_infinity}, there holds
$$
\begin{aligned}
\left\|\Ho\lb L-L^{\star}\rb\right\|_2 \leq & \ln\Ho\lb \bm{1}\bm{1}^T\rb\rn_2\cdot\frac{\mu r}{n}\lb 5\|E\|_2+24v\rb.
\end{aligned}
$$
\end{lemma}

\subsection{Proof Outline of Theorem~\ref{thm:induction} -- Base Case}\label{sec:base}
Since there is no tangent-space projection in the initialization, the establishment of the bounds for $\ln E^{0,\infty} \rn_2$, $\ln \De^{1,\infty} \rn_{2,\infty}$ and $\ln D^{1,\infty} \rn_{\mathrm{F}}$ is overall similar to that in \cite{wang2024leave}, except that we need to consider the noise terms which can be bounded with Lemma~\ref{lem:noise} and leave-one-out analysis. Our analysis for the initialization differs most significantly from previous works in the estimation of $\ln L^{1,i}-L^{\star} \rn_{\infty}$~$(0\leq i\leq 2n)$. By using  Lemma~\ref{lem:L_infinity},  we obtain a bound that is tighter by a factor of $\kappa^2$ compared to the argument in \cite{wang2024leave}. This sharper estimate is crucial, as it allows us to show that a less stringent initialization is sufficient for the algorithm to enter the contraction region in the induction steps.

\subsubsection{Bounds for $\ln E^{0,\infty} \rn_2$, $\ln \De^{1,\infty} \rn_{2,\infty}$ and $\ln D^{1,\infty} \rn_{\mathrm{F}}$}\label{sec:init_bounds} 
Despite the additional noise term, we can follow the argument in \cite{wang2024leave}, exploiting the independence induced by the auxiliary leave-one-out sequences, to establish the bounds in the following lemma. Note that \eqref{eq:2_infty} and \eqref{eq:Frobenius} jointly bound $\|\Delta^{1,\infty}\|_{2,\infty}$ and 
$\|D^{1,\infty}\|_{\mathrm{F}}$. 
The proof of Lemma~\ref{lem:init_bounds} is deferred to 
Appendix~\ref{appen:init_bounds}.
\begin{lemma}\label{lem:init_bounds}
Suppose Assumptions~\ref{assump1}~-~\ref{assump4} hold with $
p\geq\frac{16c_{12}^2C_0^2}{\gamma^2}\cdot\frac{\kappa^{3}\mu r\log^2 n}{n}$, $
\alpha\leq\frac{1}{4C_0}\frac{1}{\kappa^{1.5} \mu r}\cdot\frac{\gamma}{C_{\mathrm{thresh}}}$, and $C_{\mathrm{noise}} \sigma\leq \frac{\sigma^{\star}_r}{2C_0}$, where $c_{12}>0$ is a universal constant, $C_0\geq 42$, and
\begin{equation}\label{eq:C_noise}
C_{\emph{noise}} := C_{N}\lb C_{\mathrm{thresh}}\alpha n\sqrt{\log n}+\sqrt{\frac{n\log n}{p}}\rb.
\end{equation}
For $0\leq i\leq 2n$, $1\leq m\leq 2n$, and $C_1\geq 4$, it holds with high probability that
\begin{equation}\label{eq:base_op}
\begin{aligned}
\ln E^{0,i} \rn_2
\leq & \frac{1}{C_0}\frac{\sigma_{r}^{\star}}{\sqrt{\kappa}}\gamma+4C_{\emph{noise}}\sigma,
\end{aligned}
\end{equation}
\begin{equation}\label{eq:2_infty}
\begin{aligned}
    \ln \De^{1,i} \rn_{2,\infty}
    \leq & \lsb\frac{19+C_1}{C_0}\gamma+\lb 72+2C_1\rb\sqrt{\kappa} C_{\emph{noise}}\lb\frac{\sigma}{\sigma^{\star}_r}\rb\rsb\sqrt{\frac{\kappa\mu r}{n}}\\
    &+\lsb\frac{1+C_1}{C_0}+2C_1C_{\emph{noise}}\lb\frac{\sigma}{\sigma^{\star}_r}\rb\rsb\ln\De^{1,\infty} \rn_{2,\infty}+\frac{2}{C_0}\ln D^{1,\infty} \rn_{\mathrm{F}}, 
\end{aligned} 
\end{equation}
\begin{equation}\label{eq:Frobenius}
\begin{aligned}
\ln D^{1,i,m} \rn_{\mathrm{F}}
\leq&\lsb\frac{4+4C_1}{C_0}\frac{\gamma}{\sqrt{\kappa}}+8C_1C_{\emph{noise}}\lb\frac{\sigma}{\sigma^{\star}_r}\rb\rsb\lb\sqrt{\frac{\mu r}{n}} + \ln \De^{1,\infty} \rn_{2,\infty}\rb.
\end{aligned}
\end{equation}
\end{lemma}

Combining \eqref{eq:2_infty} with \eqref{eq:Frobenius},  we get (assuming $C_{\mathrm{noise}} \sigma\leq \frac{\sigma^{\star}_r}{2C_0}$)
\begin{align*}
    \ln \De^{1,\infty} \rn_{2,\infty}
    \leq &\lsb\frac{19+C_1}{C_0}\gamma+\lb 72+2C_1\rb\sqrt{\kappa} C_{\mathrm{noise}}\lb\frac{\sigma}{\sigma_r^{\star}}\rb\rsb\sqrt{\frac{\kappa\mu r}{n}}+\frac{1+2C_1}{C_0}\ln\De^{1,\infty}\rn_{2,\infty}\\
    &+\lsb\frac{1+C_1}{C_0}\frac{\gamma}{\sqrt{\kappa}}+2C_1C_{\mathrm{noise}}\lb\frac{\sigma}{\sigma_r^{\star}}\rb\rsb\lb\sqrt{\frac{\mu r}{n}} + \ln \De^{1,\infty} \rn_{2,\infty}\rb,
\end{align*}
where the bound for $\frac{2}{C_0}\ln D^{1,\infty} \rn_{\mathrm{F}}$ holds when $C_0\geq 8$. After further simplification, one gets
\begin{align*}
    \ln \De^{1,\infty} \rn_{2,\infty} \leq & \frac{20+2C_1}{C_0-2-4C_1}\sqrt{\frac{\kappa\mu r}{n}}\gamma+\frac{C_0\lb 72+4C_1\rb}{C_0-2-4C_1}C_{\mathrm{noise}}\lb\frac{\sigma}{\sigma_r^{\star}}\rb\sqrt{\frac{\kappa^2\mu r}{n}}.
\end{align*} 
When $C_1\geq 18$ and $C_0\geq 5(2+4C_1)$, we obtain the bound
\begin{align*}
\ln \De^{1,\infty} \rn_{2,\infty} \leq &\frac{5C_1}{C_0}\sqrt{\frac{\kappa\mu r}{n}}\gamma+10C_1 C_{\mathrm{noise}}\lb\frac{\sigma}{\sigma_r^{\star}}\rb\sqrt{\frac{\kappa^2\mu r}{n}},
\end{align*}
which is no more than $\frac12\sqrt{\frac{\kappa\mu r}{n}}$ if $C_{\mathrm{noise}} \sigma \leq \frac{1}{2C_0}\frac{\sigma^{\star}_r}{\sqrt{\kappa}}$. Plugging back into \eqref{eq:Frobenius} yields 
\begin{align*} 
\ln D^{1,\infty} \rn_{\mathrm{F}}
\leq & \frac{8C_1}{C_0}\sqrt{\frac{\mu r}{n}}\gamma+12C_1C_{\mathrm{noise}}\lb\frac{\sigma}{\sigma_r^{\star}}\rb\sqrt{\frac{\kappa\mu r}{n}}.
\end{align*}

\subsubsection{Bound for $\ln L^{1,i}-L^{\star} \rn_{\infty}$~$(0\leq i\leq 2n)$}\label{sec:init-analysis4}
 
With the bounds on $\|E^{0,i}\|_2$ and $\|\Delta^{1,i}\|_{2,\infty}$ 
already established, one could in principle bound 
$\|L^{1,i}-L^{\star}\|_{\infty}$ by applying \cite[Lemma~1]{wang2024leave}. 
For $\sigma=0$, however, this yields a bound of order 
$\bigl(\frac{\kappa\mu r}{n}\sigma_1^{\star}\bigr)\gamma$, 
which is looser by a factor of $\kappa^2$ than the target bound 
\eqref{eq:L_infty_init}.

An innovation of the analysis here is the use of Lemma~\ref{lem:L_infinity} to bound
$\ln L^{1,i}-L^{\star}\rn_{\infty}$. However, 
Lemma~\ref{lem:L_infinity} cannot be directly applied to 
$L^{1,i}=\mathcal{P}_r(L^{\star}+E^{0,i})$, since the complex 
statistical dependency makes it difficult to verify that $E^{0,i}$ 
satisfies \eqref{eq:induction}.  To circumvent this, we consider the following alternative representation:
\begin{align*}
L^{1,i} = & \mathcal{P}_{T^{1,i}}\lb L^{\star}+E^{0,i}\rb=\mathcal{P}_r\mathcal{P}_{T^{1,i}}\lb L^{\star}+E^{0,i}\rb\\
=& \mathcal{P}_r \lb L^{\star} + \lb\mathcal{P}_{T^{1,i}}-\I\rb\lb L^{\star}\rb+\mathcal{P}_{T^{1,i}}E^{0,i}\rb, 
\end{align*}
where the first two equalities hold because ${T^{1,i}}$ is the tangent space containing $L^{1,i}$. The key insight of this reformulation is that introducing $\mathcal{P}_{T^{1,i}}$  transfers the incoherence property of $L^{1,i}$ to $E^{0,i}$, which,  combined with the bound \eqref{eq:2_infty_init} already derived in proving Lemma~\ref{lem:init_bounds}, makes \eqref{eq:induction} verifiable. 
\begin{lemma}\label{lem:L_infty_init}
Conditioned on the event that Lemma~\ref{lem:init_bounds} holds, and further assuming $C_1\geq 18$, $C_0\geq 49C_1$, and $C_{\mathrm{noise}} \sigma \leq \frac{1}{2C_0}\frac{\sigma^{\star}_r}{\sqrt{\kappa}}$, it follows that
\begin{align*}
    \ln L^{1,i}-L^{\star}\rn_{\infty}
    \leq & \lb\frac{\mu r}{n}\sigma_r^{\star}\rb\gamma+100C_1C_{\emph{noise}}\sigma\frac{\sqrt{\kappa}\mu r}{n}.
\end{align*}
\end{lemma}

The proof of Lemma~\ref{lem:L_infty_init} is deferred to Appendix~\ref{appen:L_infty_init}.
As discussed in Remark~\ref{remark32}, 
\begin{align*}
   \ln L^{1,i}-L^{\star}\rn_{\infty} \leq \lb\frac{\mu r}{n}\sigma_r^{\star}\rb\gamma+\gamma C_N^{(1)}\sigma\sqrt{\log n}
\end{align*}
then holds under \eqref{eq:conditions}. Thus it follows from Lemma~\ref{lem:thresh} that $\text{Supp}\lb S^{1,i}\rb\subseteq\Omega^{(-i)}\cap\Omega_{S^{\star}}$,
\begin{align*}
\ln\Po^{(-i)} \lb S^{1,i}-S^{\star}\rb\rn_{\infty}\leq &  C_{\mathrm{thresh}}\lsb\lb\frac{\mu r}{n}\sigma_{1}^{\star}\rb\gamma+\lb 1+\gamma\rb C_N^{(1)}\sigma\sqrt{\log n}\rsb,
\end{align*}
which will be used in the proof for the next step.

\subsection{Proof Outline of Theorem~\ref{thm:induction} -- Induction Steps}\label{sec:induction}

Compared with the analysis in \cite{wang2024leave}, the tangent-space projection in ARMC introduces additional terms that require 
careful treatment. We show that these terms can be controlled using the incoherence 
$\|\Delta^{t,\infty}\|_{2,\infty}$, and the proximity 
$\|D^{t,\infty}\|_{\mathrm{F}}$ bounds already established for the 
$t$-th iterate, together with the bound on $\|E^{t,\infty}\|_2$. Notably, some parts of the analysis become simpler. 
For instance, we can bound $\|\Delta^{t+1,\infty}\|_{2,\infty}$ 
directly by leveraging the tangent-space projection inherited from the 
$t$-th iterate. This avoids the simultaneous analysis of 
$\|\Delta^{t+1,\infty}\|_{2,\infty}$ and 
$\|D^{t+1,\infty}\|_{\mathrm{F}}$ required in \cite{wang2024leave} 
to obtain their separate bounds, which would be particularly 
cumbersome in the presence of noise. 

Owing to the less stringent initialization, a desired bound on 
$\|\Delta^{t+1,\infty}\|_{2,\infty}$ requires a more refined 
exploitation of the independence induced by the leave-one-out 
sequences. In the course of this refined analysis, the direct 
control of $\|E^{t,i}\|_{2,\infty}$ also facilitates the 
verification of the assumptions of Lemma~\ref{lem:L_infinity} 
and thus the subsequent bound on $\|L^{t+1,i}-L^{\star}\|_\infty$.

To provide a unified treatment, since \cref{eq:op_t=0,eq:l_2_infty_init,eq:proximity_init,eq:L_infty_init} 
are used for the case $t=1$ while \cref{eq:op,eq:l_2_infty,eq:proximity,eq:L_infty} 
are used for $t\geq 2$, we will assume the following noise bound 
(the condition under which it holds will be given at the end of the proof):
\begin{equation}\label{eq:noise_bound2}
C_{\mathrm{noise}}\sigma \leq \frac{1}{24C_0} \frac{\sigma_r^{\star}}{\kappa}\gamma,
\end{equation}
and wherever possible, employ the following coarse bounds, which are derived by taking the maximum of the induction hypotheses 
for the two cases:
\begin{align*}
    \ln E^{t-1,\infty}\rn_2 &\leq \frac{1}{C_0}\frac{\sigma_{r}^{\star}}{\sqrt{\kappa}}\gamma^t+6C_{\mathrm{noise}}\sigma\leq \frac{2}{C_0}\frac{\sigma_r^{\star}}{\sqrt{\kappa}}\gamma,
    \\
    \ln \De^{t,\infty}\rn_{2,\infty}&\leq\frac{5C_1}{C_0}\sqrt{\frac{\kappa\mu r}{n}}\gamma^t+10C_1C_{\mathrm{noise}}\lb\frac{\sigma}{\sigma_r^{\star}}\rb\sqrt{\frac{\kappa^2\mu r}{n}}\leq \frac{6C_1}{C_0}\sqrt{\frac{\kappa\mu r}{n}}\gamma,\\
    \ln D^{t,\infty}\rn_{\mathrm{F}}&\leq\frac{8C_1}{C_0}\sqrt{\frac{\kappa\mu r}{n}}\gamma^t+24C_1 C_{\mathrm{noise}}\lb\frac{\sigma}{\sigma_r^{\star}}\rb\sqrt{\frac{\kappa^2\mu r}{n}}\leq\frac{9C_1}{C_0}\sqrt{\frac{\kappa\mu r}{n}}\gamma.
\end{align*}
We will explicitly indicate where the proof differs for the cases 
$t=1$ and $t\geq 2$, and where finer bounds are required.

\subsubsection{Bound for $\ln E^{t,\infty} \rn_2$} 
Recalling the definition of $E^{t,i}$ in \eqref{eq:defEti}, we can bound 
$E_1^{t,i}$, $E_2^{t,i}$ and $E_3^{t,i}$ separately, following the same 
strategy as in the base case. The additional tangent-space projection 
in the spectral norm estimates is handled by Lemmas~\ref{lem:T_diff} 
and~\ref{lem:T_op}.

The main novelty here is the treatment of 
$\|\mathcal{H}_{\Omega}(L^{t,i}-L^{\star})\|_2$ appearing in 
$\|E_2^{t,i}\|_2$. We bound this term using a new lemma, 
Lemma~\ref{lem:Ho}, whose proof builds on the same representation of 
$L^{t,i}-L^{\star}$ as in Lemma~\ref{lem:L_infinity}. This lemma is a 
key ingredient for establishing the quadratic dependence of $p$ on $r$. 
If one were to bound the same term using \cite[Lemma~22]{Ding2020}, the 
result would be $\sqrt{\frac{rn\log n}{p}}\cdot\|L^{t,i}-L^{\star}\|_{\infty}$, 
which is looser by a factor of $\sqrt{r}$ compared to the bound obtained 
from Lemma~\ref{lem:Ho}.

The final bound of $\ln E^{t,\infty} \rn_2$ in summarized in Lemma~\ref{lem:op_induct}, whose proof is deferred to Appendix~\ref{proof:lem:op_induct}.

\begin{lemma}\label{lem:op_induct} 
Suppose \cref{eq:op_t=0,eq:l_2_infty_init,eq:L_infty_init} hold if $t=1$ and \cref{eq:op,eq:l_2_infty,eq:L_infty} hold if $t\geq 2$, and Assumptions \ref{assump1}~-~\ref{assump4} are satisfied with
$p\geq\frac{86C_1^2C_{0}^2}{\gamma^2}\cdot\frac{\kappa^2\mu^2r^2\log^2 n}{n}$, $\alpha\leq\frac{1}{5 C_{0}} \frac{1}{\kappa^{2} \mu r}\cdot \frac{\gamma}{C_{\mathrm{thresh}}}$,
and \eqref{eq:noise_bound2} for some $C_1\geq 125$ and $C_0\geq 50C_1$. Then
\begin{align*}
\ln E^{t,i} \rn_2 \leq \frac{1}{C_0}\frac{\sigma_{r}^{\star}}{\kappa}\gamma^{t+1}+6C_{\emph{noise}}\sigma
\end{align*}
holds with high probability for $\forall i\in\{0,1,\cdots,2n\}$.
\end{lemma}

\subsubsection{Bound for $\ln \De^{t+1,\infty} \rn_{2,\infty}$}

Applying the deterministic result in Lemma~\ref{lem:incoherence}, one can get
\begin{equation}\label{eq:bounds}
\begin{aligned}
\ln \De^{t+1,i}\rn_{2,\infty} 
\leq & 18\kappa\frac{\ln E^{t,i}\rn_2}{\sigma_r^{\star}}\sqrt{\frac{\mu r}{n}}\\
&+
\max\lcb\ln E^{t,i}V^{t+1,i}\rn_{2,\infty},\ln \lb E^{t,i}\rb^T U^{t+1,i}\rn_{2,\infty}\rcb\cdot\ln \lb \Si^{t+1,i}\rb^{-1}\rn_2.
\end{aligned}
\end{equation}
Given the  bound of $\ln E^{t,i}\rn_2$, the remaining task is to bound the $\ln\cdot \rn_{2,\infty}$ norms. Thanks to the 
tangent-space projection built into $E^{t,i}$, unlike existing analysis, we can directly bound $\ln E^{t,i}\rn_{2,\infty}$. The same bound for $\|(E^{t,i})^{T}\|_{2,\infty}$ follows 
similarly. The key advantage  is that such bounds immediately verify the conditions in Lemma~\ref{lem:L_infinity} and make the estimation of $\ln L^{t+1,i}-L^{\star}\rn_{\infty}$ straightforward, as can be seen in Section~\ref{sec:L_infty_induction}.

The bound on $\|E^{t,i}\|_{2,\infty}$ is summarized in the following 
lemma. In its proof, additional terms arise compared to the base case 
due to the tangent-space projection. These terms can be controlled 
using the incoherence and spectral norm bounds already established 
for the $t$-th iterate. 

The main difficulty stems from the fact that, at $t=1$, the iterate 
lies in a larger neighborhood of the ground truth in terms of 
\eqref{eq:l_2_infty_init} compared to previous 
analyses such as \cite{wang2024leave}. Consequently, we must 
exploit the independence inherent in the leave-one-out construction 
more carefully and derive tighter estimates. For instance, 
consider the term 
$\|e_m^{T}\mathcal{H}_{\Omega}^{(-i)}(U^{t,i}\Sigma^{t,i}
(D_V^{t,i,m})^{T})V^{t,m}\|_2$ that arises in bounding 
$\|E_2^{t,i}\|_{2,\infty}$. If one were to bound this term 
by following \cite{wang2024leave} and applying 
Lemma~\ref{lem:bound1}, achieving the target order 
$\bigl(\sigma_r^{\star}\sqrt{\frac{\mu r}{n}}\bigr)\gamma^{2}$ 
(in the noiseless case $\sigma=0$) would require $p$ to scale at 
least as $\kappa^4$.

\begin{lemma}\label{lem:2_infty_induction}
Conditioned on the event that Lemma~\ref{lem:op_induct} holds, and further assuming \eqref{eq:proximity_init} if $t=1$ and \eqref{eq:proximity} if $t\geq 2$, it holds with high probability for $\forall i\in\{0,1,\cdots,2n\}$ that
\begin{equation}\label{eq:2_infinity_induction}
\begin{aligned}
    \ln E^{t,i} \rn_{2,\infty}
    \leq & \frac{5+2C_1}{C_0}\lb \sigma_r^{\star}\sqrt{\frac{\mu r}{n}}\rb\gamma^{t+1}+\lb 5+4C_1\rb C_{\mathrm{noise}}\sigma\sqrt{\frac{\kappa\mu r}{n}},
\end{aligned}
\end{equation}
provided $
p\geq \frac{9C_{0}^2}{\gamma^2}\cdot\frac{\kappa^3\mu^2r^2\log^2 n}{n}$, and $\alpha\leq\frac{1}{5 C_{0}} \frac{1}{\kappa^{1.5} \mu r}\cdot \frac{\gamma}{C_{\mathrm{thresh}}}$ for some $C_0\geq 100C_1$.
\end{lemma}

Together with \eqref{eq:bounds}, provided that $C_0\geq 12$ (therefore $\ln E^{t,\infty}\rn_2\leq \frac{2}{C_0}\sigma_r^{\star}\leq\frac16 \sigma_r^{\star}$),
\begin{equation*}
\begin{aligned}
    \ln \De^{t+1,\infty} \rn_{2,\infty} 
    \leq & 18\kappa\frac{\ln E^{t,\infty}\rn_2}{\sigma_r^{\star}}\sqrt{\frac{\mu r}{n}}+\frac{3C_1}{C_0}\sqrt{\frac{\mu r}{n}}\gamma^{t+1}+6C_1C_{\mathrm{noise}}\lb\frac{\sigma}{\sigma_r^{\star}}\rb\sqrt{\frac{\kappa\mu r}{n}} \\
    \leq &\frac{18+3C_1}{C_0}\sqrt{\frac{\mu r}{n}}\gamma^{t+1}+\lb 108+6C_1\rb C_{\mathrm{noise}}\lb\frac{\sigma}{\sigma_r^{\star}}\rb\sqrt{\frac{\kappa^2\mu r}{n}}\\
    \leq &\frac{5C_1}{C_0}\sqrt{\frac{\mu r}{n}}\gamma^{t+1}+10C_1C_{\mathrm{noise}}\lb\frac{\sigma}{\sigma_r^{\star}}\rb\sqrt{\frac{\kappa^2\mu r}{n}},
\end{aligned}
\end{equation*}
where the last inequality holds if $C_1\geq 27$.

\subsubsection{Bound for $\ln D^{t+1,\infty} \rn_{\mathrm{F}}$}
Following a similar argument as in the base case (Part~\RomanNumeralCaps{3} 
in the proof of Lemma~\ref{lem:init_bounds}), we obtain
\begin{equation}\label{eq:WF_induction}
\begin{aligned}
\ln D^{t+1,\infty}\rn_{\mathrm{F}}\leq & 2\cdot\max_{1\leq m\leq 2n} \ln D^{t+1,0,m}\rn_{\mathrm{F}} \\
\leq &\frac{4}{\sigma_r^{\star}}\lb \ln\lb E^{t}-E^{t,m}\rb V^{t+1,m}\rn_{\mathrm{F}}+\ln\lb E^{t}-E^{t,m}\rb^T U^{t+1,m}\rn_{\mathrm{F}}\rb.
\end{aligned}
\end{equation}
Here, unlike the base case, we proceed by deriving the bound for $\ln E^{t}-E^{t,m}\rn_{\mathrm{F}}$. Recalling the definitions of $E^{t}$ and $E^{t,m}$, the difference $E^t-E^{t,m}$ can be decomposed as
\begin{equation*}
\begin{aligned}
    E^t-E^{t,m}
    = & \underbrace{p^{-1}\mathcal{P}_{T^{t,m}}\mathcal{P}_\Omega^{(-m)}\lb S^{t,m}-S^\star\rb -p^{-1}\mathcal{P}_{T^t}\mathcal{P}_\Omega(S^t-S^\star)}_{\zeta}\\
    & + \underbrace{\left( \mathcal{I}-p^{-1}\mathcal{P}_{T^t}\mathcal{P}_\Omega\right)\lb L^t-L^\star\rb 
    -\left( \mathcal{I}-\mathcal{P}_{T^{t,m}}\left(\I-\Ho^{(-m)}\right)\right)\lb L^{t,m}-L^\star\rb}_{\xi} \\
    & +\underbrace{p^{-1}\Pt\Po\lb N\rb
    -p^{-1}\mathcal{P}_{T^{t,m}}\Po^{(-m)}\lb N \rb}_{\tau},
\end{aligned}
\end{equation*}
where the notation is slightly abused: we suppress the explicit 
dependence on $m$ inside $\zeta$, $\xi$, and $\tau$ for brevity. We then bound their Frobenius norms separately and combine them to get the following lemma. The additional terms arising from the tangent-space projection, once suitably split and rearranged, can 
be controlled using the proximity and spectral norm bounds already 
established for the $t$-th iterate. Note that \eqref{eq:proximity} is larger than \eqref{eq:proximity_init} 
by a factor of $\sqrt{\kappa}$. This is mainly because, at $t=1$, 
we can only work with the loose bound \eqref{eq:l_2_infty_init} 
for the term 
$\|\mathcal{H}_{\Omega}(U^{t}\Sigma^{t}(D_V^{t,0,m})^{T})V^{t,m}\|_{\mathrm{F}}$ 
that appears within $\xi$. The increase in \eqref{eq:proximity} 
is offset by the decrease in \eqref{eq:l_2_infty}, so that our 
argument carries through for $t\geq 2$.

\begin{lemma}\label{lem:D_bounds}
Conditioned on the event that Lemma~\ref{lem:op_induct} holds, and further assuming \eqref{eq:proximity_init} if $t=1$ and \eqref{eq:proximity} if $t\geq 2$, it holds with high probability for $1\leq m\leq 2n$ that
\begin{equation}\label{eq:WF_induct_bound}
\begin{aligned}
\ln E^t-E^{t,m}\rn_{\mathrm{F}} 
\leq & \frac{C_1}{C_0}\lb\sigma_{r}^{\star}\sqrt{\frac{\kappa\mu r}{n}}\rb\gamma^{t+1}+3C_1C_{\mathrm{noise}}\sigma\sqrt{\frac{\kappa^2\mu r}{n}},
\end{aligned}
\end{equation}
provided $p\geq \frac{\max\{900C_0^2,3600C_1^4\}}{\gamma^2}\cdot\frac{\kappa^3 \mu^2r^2\log^2 n}{n}$, and $\alpha \leq \frac{1}{3C_0}\frac{1}{\kappa^2\mu r}\cdot\frac{\gamma}{C_{\mathrm{thresh}}}$ for some
$C_1\geq 128$ and $C_0\geq 384C_1$. 

\end{lemma}

Plug this result into \eqref{eq:WF_induction} and one gets
$$
\ln D^{t+1,\infty} \rn_{\mathrm{F}} \leq \frac{8}{\sigma_r^{\star}}\cdot\ln E^t-E^m\rn_{\mathrm{F}} \leq \frac{8C_1}{C_0}\sqrt{\frac{\kappa\mu r}{n}}\gamma^{t+1}+24C_1C_{\mathrm{noise}}\lb\frac{\sigma}{\sigma_r^{\star}}\rb\sqrt{\frac{\kappa^2\mu r}{n}}.
$$

\subsubsection{Bound for $\ln L^{t+1,i}-L^{\star} \rn_{\infty}$~$(0\leq i\leq 2n)$}\label{sec:L_infty_induction}
As mentioned, the second condition of Lemma~\ref{lem:L_infinity} are readily satisfied with the bound of $\ln E^{t,i}\rn_{2,\infty}$. Take $\ln E^{t,i}\lsb\lb E^{t,i}\rb^TE^{t,i}\rsb^a V^{\star}\rn_{2,\infty}$ as an example. For any $1\leq m\leq n$ and $\forall a\geq 0$,
\begin{align*}
    & \ln e_m^TE^{t,i}\lsb\lb E^{t,i}\rb^TE^{t,i}\rsb^a V^{\star}\rn_2 \\
    \leq & \|E^{t,i}\|_{2,\infty}\cdot\ln E^{t,i}\rn_2^{2a} \\
    \leq &\lsb\frac{5+2C_1}{C_{0}}\lb \sigma_r^{\star}\sqrt{\frac{\mu r}{n}}\rb\gamma^{t+1}+\lb 5+4C_1\rb C_{\mathrm{noise}}\sigma\sqrt{\frac{\kappa\mu r}{n}}\rsb\cdot\lb \frac{1}{C_0}\frac{\sigma_r^{\star}}{\kappa}\gamma^{t+1}+6C_{\mathrm{noise}}\sigma\rb^{2a}\\
    \leq &\lsb \frac{5+2C_1}{C_0}\sigma_r^{\star}\gamma^{t+1}+\lb 5+4C_1\rb\sqrt{\kappa} C_{\mathrm{noise}}\sigma\rsb^{2a+1}\cdot\sqrt{\frac{\mu r}{n}},
\end{align*}
where the bound for $\|E^{t,i}\|_{2,\infty}$ in the second inequality follows from \eqref{eq:2_infinity_induction}. Thus, by Lemma~\ref{lem:L_infinity},
\begin{align*}
    \ln L^{t+1,i}-L^{\star}\rn_{\infty}
    \leq &\frac{\mu r}{n}\lsb \frac{5}{C_0}\frac{\sigma_r^{\star}} {\kappa}\gamma^{t+1}+30C_{\mathrm{noise}}\sigma+\frac{120+48C_1}{C_0}\sigma_r^{\star}\gamma^{t+1}+\lb 120+96C_1\rb\sqrt{\kappa} C_{\mathrm{noise}}\sigma\rsb \\
    \leq &\lb\frac{\mu r}{n}\sigma_r^{\star}\rb\gamma^{t+1}+100C_1C_{\mathrm{noise}}\sigma\frac{\sqrt{\kappa}\mu r}{n},
\end{align*}
where the last inequality holds if $C_1\geq 125$ and $C_0\geq 49C_1$. In addition, as in the base case (also see Remark~\ref{remark32}),
\begin{align*}
    \ln L^{t+1,i}-L^{\star}\rn_{\infty} \leq \lb\frac{\mu r}{n}\sigma_r^{\star}\rb\gamma^{t+1}+\gamma C_N^{(1)}\sigma\sqrt{\log n},
\end{align*}
and $\text{Supp}\lb S^{t+1,i}\rb\subseteq\Omega^{(-i)}\cap\Omega_{S^{\star}}$,
\begin{align*}
\ln\Po^{(-i)} \lb S^{t+1,i}-S^{\star}\rb\rn_{\infty}\leq &  C_{\mathrm{thresh}}\lsb\lb\frac{\mu r}{n}\sigma_{1}^{\star}\rb\gamma^{t+1}+\lb 1+\gamma\rb C_N^{(1)}\sigma\sqrt{\log n}\rsb,
\end{align*}
which will be used in the next induction step.

Going through the proofs for the base case and induction steps, one can see that provided
\begin{equation}\label{eq:constants}
C_1\geq \max\{128,C_{\text{prob}}\},\quad C_0\geq 384 C_1,
\end{equation}
where $C_{\text{prob}}$ is a constant that solely dependent on the maximum iteration number $T$, the universal constant $c_{15}>0$ from Lemma~\ref{lem:bernstein} and the failure probability of all the terms bounded using Lemma~\ref{lem:bernstein}\footnote{We use Lemma~\ref{lem:bernstein} with $C_1=a\cdot(2c_{15})$ and each time the failure probability is bounded by $n^{-a}$. If $T\leq n$ and choose $a=4$, the overall failure probability is no more than $8n^2\times n^{-4}=8n^{-2}$.},
in the noiseless case ($\sigma=0$) the induction hypotheses hold if $p$ and $\alpha$ satisfy
$$
p\gtrsim \frac{1}{\gamma^2}\cdot\frac{\kappa^3\mu^2r^2\log n}{n},\quad\alpha \lesssim \frac{1}{\kappa^2\mu r}\cdot\frac{\gamma}{C_{\mathrm{thresh}}}.
$$
If there is noise, we need to further assume $p\geq \frac{\log^2n}{n}$. Therefore, the final bounds are
$$
p\gtrsim\frac{1}{\gamma^2}\cdot\frac{\kappa^3\mu^2r^2\log^2 n}{n},\quad
\alpha\leq\frac{1}{\kappa^2\mu r}\cdot\frac{\gamma}{C_{\mathrm{thresh}}}.
$$
Furthermore, the sub-Gaussian norm $\sigma$ needs to satisfy \eqref{eq:noise_bound2}, which requires
$$
\sigma \lesssim \min\left\{\frac{\gamma/\alpha}{C_{\mathrm{thresh}}\kappa\sqrt{\log n}},\sqrt{\frac{\gamma^2 np}{\kappa^2\log n}}\right\}\cdot\frac{\sigma_r^{\star}}{n}.
$$

\section{Conclusion} \label{sec:conclusion}
In this paper, we have developed a fast non-convex method called ARMC for the robust matrix completion problem by introducing an additional subspace to an existing singular value thresholding based method for the update of the low rank part. A theoretical recovery guarantee of ARMC has been established for the scenario when there exist sparse outliers as well as stochastic noise, which improves that for a convex approach which considers the same setting. Numerical experiments on synthetic and real data have demonstrated the superiority of ARMC over other non-convex methods, especially in terms of computational efficiency.  

\bibliographystyle{siam}
\bibliography{ref}

@Misc{siam,
  key = {zzz},
  title =	 {{SIAM} Style Manual: For journals and books},
  year =	 2013,
  url = {https://epubs.siam.org/pb-assets/files/SIAM_STYLE_GUIDE_2019-1635349464967.pdf}
}

@Misc{amsmath,
  author =	 {{American Mathematical Society}},
  title =	 {User's Guide for the \texttt{amsmath} Package
                  (Version 2.0)},
  url =		 {ftp://ftp.ams.org/pub/tex/doc/amsmath/amsldoc.pdf},
  urldate =	 {2015-07-30},
  year =	 2002
}

@article{Abbe2020,
  title={Entrywise eigenvector analysis of random matrices with low expected rank},
  author={Abbe, Emmanuel and Fan, Jianqing and Wang, Kaizheng and Zhong, Yiqiao},
  journal={Ann. Stat.},
  volume={48},
  number={3},
  pages="1452--1474",
  year={2020},
  publisher={NIH Public Access}
}

@article{Cai2021,
  author={Cai, HanQin and Cai, Jian-Feng and Wang, Tianming and Yin, Guojian},
  journal={IEEE Trans. Signal Process.}, 
  title={Accelerated Structured Alternating Projections for Robust Spectrally Sparse Signal Recovery}, 
  year={2021},
  volume={69},
  number={},
  pages="809--821",
}

@article{Candes2011,
  title={Robust principal component analysis?},
  author={Candès,  Emmanuel J. and Li,  Xiaodong and Ma,  Yi and Wright,  John},
  journal={J. ACM},
  volume={58},
  number={3},
  pages = "1--37",
  year={2011},
  publisher={Association for Computing Machinery (ACM)}
}

@article{Chen2015,
  title = {Incoherence-Optimal Matrix Completion},
  volume = {61},
  number = {5},
  journal = {IEEE Trans. Inf. Theory},
  author = {Chen,  Yudong},
  year = {2015},
  pages = "2909--2923"
}

@article{Chen2014,
  title = {Robust Spectral Compressed Sensing via Structured Matrix Completion},
  volume = {60},
  number = {10},
  journal = {IEEE Trans. Inf. Theory},
  author = {Chen,  Yuxin and Chi,  Yuejie},
  year = {2014},
  pages = "6576--6601"
}

@article{Chen2021,
  year = {2021},
  volume = {49},
  number = {5},
  author = {Yuxin Chen and Jianqing Fan and Cong Ma and Yuling Yan},
  title = {Bridging convex and nonconvex optimization in robust {PCA}: Noise,  outliers and missing data},
  journal = {Ann. Stat.},
  pages = "2948--2971"
}

@inproceedings{Cherapanamjeri2017,
  title={Nearly optimal robust matrix completion},
  author={Cherapanamjeri, Yeshwanth and Gupta, Kartik and Jain, Prateek},
  booktitle={International Conference on Machine Learning},
  pages={797--805},
  year={2017}
}

@article{Ding2020,
  year = {2020},
  volume = {66},
  number = {11},
  pages = {7274--7301},
  author = {Lijun Ding and Yudong Chen},
  title = {Leave-One-Out Approach for Matrix Completion: Primal and Dual Analysis},
  journal = {IEEE Trans. Inf. Theory}
}

@article{Fan2001,
  title = {Variable Selection via Nonconcave Penalized Likelihood and its Oracle Properties},
  volume = {96},
  number = {456},
  journal = {J. Am. Stat. Assoc.},
  author = {Fan,  Jianqing and Li,  Runze},
  year = {2001},
  pages = "1348--1360"
}

@InProceedings{Hardt2014,
  title = 	 {Fast matrix completion without the condition number},
  author = 	 {Hardt, Moritz and Wootters, Mary},
  booktitle = {Conference on Learning Theory},
  pages = 	 {638--678},
  year = 	 {2014}
}

@inproceedings{Jain2010,
 author = {Jain, Prateek and Meka, Raghu and Dhillon, Inderjit},
 booktitle = {Advances in Neural Information Processing Systems},
 pages = "937--945",
 title = {Guaranteed Rank Minimization via Singular Value Projection},
 year = {2010}
}

@InProceedings{Jain2015,
  title = 	 {Fast Exact Matrix Completion with Finite Samples},
  author = 	 {Jain, Prateek and Netrapalli, Praneeth},
  booktitle = 	 {Conference on Learning Theory},
  pages = 	 {1007--1034},
  year = 	 {2015}
}

@article{Ling2022,
  title = {Near-optimal performance bounds for orthogonal and permutation group synchronization via spectral methods},
  volume = {60},
  ISSN = {1063-5203},
  url = {http://dx.doi.org/10.1016/j.acha.2022.02.003},
  DOI = {10.1016/j.acha.2022.02.003},
  journal = {Appl. Comput. Harmon. Anal.},
  publisher = {Elsevier BV},
  author = {Ling,  Shuyang},
  year = {2022},
  month = sep,
  pages = "20--52"
}

@article{Ma2019,
  title = {Implicit Regularization in Nonconvex Statistical Estimation: Gradient Descent Converges Linearly for Phase Retrieval,  Matrix Completion,  and Blind Deconvolution},
  volume = {20},
  number = {3},
  journal = {Found. Comput. Math.},
  author = {Ma,  Cong and Wang,  Kaizheng and Chi,  Yuejie and Chen,  Yuxin},
  year = {2019},
  pages = "451--632"
}

@article{Tropp2011,
  year = {2011},
  volume = {12},
  number = {4},
  pages = {389--434},
  author = {Joel A. Tropp},
  title = {User-Friendly Tail Bounds for Sums of Random Matrices},
  journal = {Found. Comput. Math.}
}

@article{Wei2020,
  title = {Guarantees of {R}iemannian optimization for low rank matrix completion},
  volume = {14},
  number = {2},
  journal = {Inverse Probl. Imaging},
  publisher = {American Institute of Mathematical Sciences (AIMS)},
  author = {Wei,  Ke and Cai,  Jian-Feng and F. Chan,  Tony and Leung,  Shingyu},
  year = {2020},
  pages = "233--265"
}

@article{wei2016guarantees,
  title={Guarantees of {R}iemannian optimization for low rank matrix recovery},
  author={Wei, Ke and Cai, Jian-Feng and Chan, Tony F and Leung, Shingyu},
  journal={SIAM J. Matrix Anal. Appl.},
  volume={37},
  number={3},
  pages={1198--1222},
  year={2016},
  publisher={SIAM}
}

@inproceedings{Yi2016,
author = {Yi, Xinyang and Park, Dohyung and Chen, Yudong and Caramanis, Constantine},
title = {Fast algorithms for robust {PCA} via gradient descent},
year = {2016},
booktitle = {Advances in Neural Information Processing Systems},
pages = "4159--4167"
}

@incollection{Ver12,
author = {Vershynin, Roman},
title = {Introduction to the non-asymptotic analysis of random matrices},
year = {2012},
booktitle = {Compressed Sensing, Theory and Applications},
pages = "210--268",
publisher = {Cambridge University Press}
}

@article{Chen2019,
  title={Model-free nonconvex matrix completion: Local minima analysis and applications in memory-efficient kernel {PCA}},
  author={Chen, Ji and Li, Xiaodong},
  journal={J. Mach. Learn. Res.},
  volume={20},
  number={142},
  pages="1--39",
  year={2019}
}

@article{Candes2009,
  title={Exact matrix completion via convex optimization},
  author={Cand{\`e}s, Emmanuel J and Recht, Benjamin},
  journal={Found. Comput. Math.},
  volume={9},
  number={6},
  pages="717--772",
  year={2009}
}

@article{KMO:TIT:10,
  title={Matrix completion from a few entries},
  author={Keshavan, Raghunandan H and Montanari, Andrea and Oh, Sewoong},
  journal={IEEE Trans. Inf. Theory},
  volume={56},
  number={6},
  pages={2980--2998},
  year={2010},
  publisher={IEEE}
}

@article{Rec:JMLR:11,
  title={A simpler approach to matrix completion},
  author={Recht, B.},
  journal={J. Mach. Learn. Res.},
  volume={12},
  pages={3413--3430},
  year={2011},
  publisher={JMLR.org}
}

@article{RFP:SIREV:10,
  title={Guaranteed minimum-rank solutions of linear matrix equations via nuclear norm minimization},
  author={Recht, B. and Fazel, M. and Parrilo, P.A.},
  journal={SIAM Rev.},
  pages={471--501},
  volume={52},
  number={3},
  year={2010}
}

@article{Zhong2018,
  title={Near-optimal bounds for phase synchronization},
  author={Zhong, Yiqiao and Boumal, Nicolas},
  journal={SIAM J. Optim.},
  volume={28},
  number={2},
  pages="989--1016",
  year={2018},
  publisher={SIAM}
}

@article{zheng-pg,
    author = {Q. Zheng and J. Lafferty},
    title = {Convergence analysis for rectangular matrix completion using {B}urer-{M}onteiro factorization and gradient descent},
    journal = {arXiv preprint arXiv:1605.07051},
    year = {2016}
}

@article{vandereycken2013low,
  title={Low-rank matrix completion by {R}iemannian optimization},
  author={Vandereycken, Bart},
  journal={SIAM J. Optim.},
  volume={23},
  number={2},
  pages={1214--1236},
  year={2013},
  publisher={SIAM}
}

@article{tanner2013normalized,
  title={Normalized iterative hard thresholding for matrix completion},
  author={Tanner, Jared and Wei, Ke},
  journal={SIAM J. Sci. Comput.},
  volume={35},
  number={5},
  pages={S104--S125},
  year={2013},
  publisher={SIAM}
}

@inproceedings{procrustesflow2016,
  title={Low-rank solutions of linear matrix equations via procrustes flow},
  author={Tu, Stephen and Boczar, Ross and Simchowitz, Max and Soltanolkotabi, Mahdi and Recht, Ben},
  booktitle={International Conference on Machine Learning},
  pages={964--973},
  year={2016}
}

@article{accaltprj2019,
    author = {HanQin Cai and Jian-Feng Cai and Ke Wei},
    title = {Accelerated alternating projections for robust principal component analysis},
    journal = {J. Mach. Learn. Res.},
    year = {2019},
    volume = {20},
    number = {20},
    pages = {1-33}
}

@article{Li2012,
  title = {Compressed Sensing and Matrix Completion with Constant Proportion of Corruptions},
  volume = {37},
  ISSN = {1432-0940},
  number = {1},
  journal = {Constructive Approximation},
  publisher = {Springer Science and Business Media LLC},
  author = {Li,  Xiaodong},
  year = {2012},
  month = dec,
  pages = "73--99"
}

@article{Chen2013,
  title = {Low-Rank Matrix Recovery From Errors and Erasures},
  volume = {59},
  ISSN = {1557-9654},
  number = {7},
  journal = {IEEE Trans. Inf. Theory},
  publisher = {Institute of Electrical and Electronics Engineers (IEEE)},
  author = {Chen,  Yudong and Jalali,  Ali and Sanghavi,  Sujay and Caramanis,  Constantine},
  year = {2013},
  month = jul,
  pages = "4324--4337"
}

@article{wang2024leave,
  title={Leave-One-Out Analysis for Nonconvex Robust Matrix Completion with General Thresholding Functions},
  author={Wang, Tianming and Wei, Ke},
  journal={arXiv preprint arXiv:2407.19446},
  year={2024}
}

@article{Bandeira2016,
author = {Afonso S. Bandeira and Ramon van Handel},
title = {{Sharp nonasymptotic bounds on the norm of random matrices with independent entries}},
volume = {44},
journal = {Ann. Probab.},
number = {4},
publisher = {Institute of Mathematical Statistics},
pages = {2479 -- 2506},
year = {2016}
}

@article{Klopp2015,
  title = {Matrix completion by singular value thresholding: Sharp bounds},
  volume = {9},
  ISSN = {1935-7524},
  number = {2},
  pages = "2348--2369",
  journal = {Electron. J. Stat.},
  publisher = {Institute of Mathematical Statistics},
  author = {Klopp,  Olga},
  year = {2015},
  month = jan 
}

@inproceedings{7084843,
  author={Venkatanath N and Praneeth D and Maruthi Chandrasekhar Bh and Channappayya, Sumohana S. and Medasani, Swarup S.},
  booktitle={National Conference on Communications}, 
  title={Blind image quality evaluation using perception based features}, 
  year={2015},
  pages={1-6}
}

@article{gaussnewton,
  title={{RGNMR}: A {Gauss-Newton} method for robust matrix completion with theoretical guarantees}, 
  author={Eilon Vaknin Laufer and Boaz Nadler},
  journal={arXiv preprint arXiv:2505.12919},
  year={2025}
}

@BOOK{vershynin2009high,
  TITLE = {High-dimensional probability},
  AUTHOR = {Vershynin, Roman},
  YEAR = {2009},
  PUBLISHER = {Cambridge University Press, UK}
}

@article{van2010manipulation,
  title={Manipulation robustness of collaborative filtering},
  author={Van Roy, Benjamin and Yan, Xiang},
  journal={Manage. Sci.},
  volume={56},
  number={11},
  pages={1911--1929},
  year={2010},
  publisher={INFORMS}
}

@article{de2003framework,
  title={A framework for robust subspace learning},
  author={De La Torre, Fernando and Black, Michael J},
  journal={Int. J. Comput. Vision},
  volume={54},
  number={1},
  pages={117--142},
  year={2003},
  publisher={Springer}
}

@article{bouwmans2018applications,
  title={On the applications of robust {PCA} in image and video processing},
  author={Bouwmans, Thierry and Javed, Sajid and Zhang, Hongyang and Lin, Zhouchen and Otazo, Ricardo},
  journal={Proc. IEEE},
  volume={106},
  number={8},
  pages={1427--1457},
  year={2018},
  publisher={IEEE}
}

@article{mardani2013recovery,
  title={Recovery of low-rank plus compressed sparse matrices with application to unveiling traffic anomalies},
  author={Mardani, Morteza and Mateos, Gonzalo and Giannakis, Georgios B},
  journal={IEEE Trans. Inf. Theory},
  volume={59},
  number={8},
  pages="5186--5205",
  year={2013},
  publisher={IEEE}
}

@article{Tanner2023,
  title = {Compressed sensing of low-rank plus sparse matrices},
  volume = {64},
  ISSN = {1063-5203},
  url = {http://dx.doi.org/10.1016/j.acha.2023.01.008},
  DOI = {10.1016/j.acha.2023.01.008},
  journal = {Appl. Comput. Harmon. Anal.},
  publisher = {Elsevier BV},
  author = {Tanner,  Jared and Vary,  Simon},
  year = {2023},
  month = May,
  pages = "254--293"
}

\section*{\appendixname}

The appendices are organized as follows. For the reader's convenience, 
we collect in Appendix~\ref{sec:more_lemmas} several auxiliary lemmas 
from the literature that are used in the proofs. Appendix~\ref{sec:appen_supp} 
contains the proofs of the new lemmas introduced in Section~\ref{sec:useful}. 
The detailed proofs of the lemmas in Section~\ref{sec:base} and 
Section~\ref{sec:induction} are given in Appendix~\ref{sec:proofs_base} 
and Appendix~\ref{sec:proofs_induction}, respectively.

\begin{appendix}\label{sec:appendix}

\section{Additional Lemmas}\label{sec:more_lemmas}

\begin{lemma}[{\cite[Lemma~3]{Cherapanamjeri2017}}]\label{lem:S_op}
If $S\in\mathbb{R}^{n\times n}$ is $\alpha$-sparse, i.e., $S$ has no more than $\alpha n$ nonzero entries per row and column, then $\ln S\rn_2\leq \alpha n\cdot\ln S\rn_{\infty}$.
\end{lemma}

\begin{lemma}[{\cite[Lemma~2]{Chen2015}}]\label{lem:init}
Suppose $Z\in\mathbb{R}^{n\times n}$ is a fixed matrix. There exists a universal constant $c_{12}>1$ such that  
$$
\ln \Ho(Z)\rn_2\leq c_{12}\lb \frac{\log n}{p}\ln Z\rn_{\infty}+\sqrt{\frac{\log n}{p}}\cdot\max\left\{\ln Z\rn_{2,\infty},\ln Z^T\rn_{2,\infty}\right\}\rb
$$
holds with high probability.
\end{lemma}

\begin{lemma}[{\cite[Lemma~7]{wang2024leave},~\cite[Lemma~1]{Ding2020}}]\label{lem:perturb_gt} 
Suppose that $Z=L^{\star}+E\in\mathbb{R}^{n\times n}$. Recall the definition of matrix symmetrization in \eqref{eq:symmetry}. Denote by $\frac{1}{\sqrt{2}}F^{\star}$ the top-$r$ ($r<n$) orthonormal eigenvectors of $\widehat{L^{\star}}$ and denote by $\frac{1}{\sqrt{2}}F$ the top-$r$ orthonormal eigenvectors of $\widehat{Z}$. Let the SVD of the matrix $H=\frac12(F^{\star})^T F$ be $A\widetilde{\Sigma} B^T$ and define $G=A B^T$. If $\|E\|_{2}\leq\frac{1}{2}\sigma_r^{\star}$,
$$
\begin{aligned}
\left\|\Si^{\star} H-G \Si^{\star}\right\|_{2} & \leq\left(2+\frac{\sigma_1^{\star}}{\sigma_r^{\star}-\|E\|_{2}}\right)\|E\|_{2},\\
\left\|\Si^{\star} G-G \Si^{\star}\right\|_{2} & \leq\left(2+\frac{2 \sigma_1^{\star}}{\sigma_r^{\star}-\|E\|_{2}}\right)\|E\|_{2}.
\end{aligned}
$$
\end{lemma}

\begin{lemma}[{\cite[Lemma~8]{wang2024leave},~\cite[Lemma~45]{Ma2019}}]\label{lem:op} 
Under the same conditions of Lemma~\ref{lem:perturb_gt},  
$$
\left\|F -F^{\star}G\right\|_2 \leq \frac{4\sqrt{2}}{\sigma_{r }^{\star}}\left\|E\right\|_2.
$$
\end{lemma}

\begin{lemma}[{\cite[Lemma~11]{Chen2014},~\cite[Theorem~1.6]{Tropp2011}}]\label{lem:bernstein}
Consider $m$ independent random matrices $M_l$ $(1 \leq l \leq m)$ of dimension $d_1 \times d_2$ that satisfy $\mathbb{E}\left[M_l\right]=0$ and $\left\|M_l\right\|_2 \leq B$. Define
$$
\sigma^2:=\max \left\{\left\|\sum_{l=1}^m \mathbb{E}\left[M_l M_l^T\right]\right\|_2,\left\|\sum_{l=1}^m \mathbb{E}\left[M_l^TM_l\right]\right\|_2\right\} .
$$
Then there exists a universal constant $c_{15}>0$ such that for any integer $a \geq 2$,
$$
\left\|\sum_{l=1}^m M_l\right\|_2 \leq c_{15}\left(\sqrt{a \sigma^2 \log \left(d_1+d_2\right)}+a B \log \left(d_1+d_2\right)\right)
$$
with probability at least $1-(d_1+d_2)^{-a}$.
\end{lemma}

\begin{lemma}[{\cite[Lemma~14]{Ding2020}}]\label{lem:perturb_F}
Suppose that $A$ and $\widetilde{A}=A+W$ are symmetric matrices, and $\lambda_1(A)\geq \cdots\geq \lambda_r(A)>0$. Denote by  $F\La F^T$ and $\widetilde{F}\widetilde{\La}(\widetilde{F})^T$ the top-$r$ eigen-decompositions of $A$ and $\widetilde{A}$ , respectively. Let the SVD of the matrix $H=F^T\widetilde{F}$ be $A\widetilde{\Sigma} B^T$, and define $G=A B^T$. If $\|W\|_2<\delta:=\lambda_r(A)-\lambda_{r+1}(A)$, then
$$
\ln \widetilde{F}-FG\rn_{\mathrm{F}} \leq \frac{\sqrt{2}\|W F\|_{\mathrm{F}}}{\delta-\|W\|_{2}}.
$$
\end{lemma}

\begin{lemma}[{\cite[Lemma~9]{wang2024leave},~\cite[Lemma~2]{Ding2020}}]\label{lem:perturb_S}
Under the same conditions of Lemma~\ref{lem:perturb_F},
$$
\left\|\La G-G\widetilde{\La}\right\| \leq\left(\frac{2 \lambda_1(A)+\|W\|_2}{\delta-\|W\|_2}+1\right)\ln W F\rn,
$$
where the norm can be either the Frobenius norm or the 2-norm.
\end{lemma}

\begin{lemma}[{\cite[Lemma~8]{Cai2021}}]\label{lem:T_diff}  
Suppose $L$ is a rank-$r$ matrix whose compact SVD is $U\Si V^T$. Let $\mathcal{P}_T$ be the projection onto the tangent space of the manifold of rank-$r$ matrices at $L$, as defined in \eqref{eq:projection}. Then,
$$
\|(\mathcal{I}-\mathcal{P}_T)\lb L^{\star}\rb \|_2\leq \frac{\|L-L^{\star}\|_2^2}{\sigma_r^{\star}}.
$$
\end{lemma}

\begin{lemma}[{\cite[Lemma~6]{accaltprj2019}}]\label{lem:T_op}
Under the same conditions of Lemma~\ref{lem:T_op},
$$
\|\mathcal{P}_T(Z)\|_2\leq \sqrt{\frac43}\|Z\|_2,~\forall Z\in\mathbb{R}^{n\times n}.
$$
\end{lemma}

\begin{lemma}[{\cite[Lemma~8]{Chen2019}}]\label{lem:bound2} 
Suppose $A,C\in\mathbb{R}^{n\times r_1}$ and $B,D\in\mathbb{R}^{n\times r_2}$. Then we have the deterministic bound
$$
\left|\left\langle \Ho\lb AC^T\rb,BD^T\right\rangle\right| \leq \ln \Ho\lb \bm{1}\bm{1}^T\rb\rn_2\cdot\ln A\rn_{2,\infty}\ln B\rn_{\mathrm{F}}\ln C\rn_{\mathrm{F}}\ln D\rn_{2,\infty}.
$$
\end{lemma}

\begin{lemma}[{\cite[Lemma~11]{wang2024leave}}]\label{lem:bound1}
Suppose $A,B\in\mathbb{R}^{n\times r}$. For $1\leq m\leq n$,  define
$$
H^{(m)} = \diag\lb 1-\delta_{m1}/p,\cdots,1-\delta_{mn}/p\rb\in\mathbb{R}^{n\times n},
$$
and $R^{(m)} = B^TH^{(m)}C$. Then we have the following deterministic bound
$$
\ln e_m^T\Ho\lb AB^T\rb C\rn_2 \leq \ln e_m^T A\rn_{2}\cdot \ln R^{(m)}\rn_2.
$$
Furthermore,
$$
\ln \Ho\lb AB^T\rb C\rn_{\mathrm{F}} \leq \ln A\rn_{\mathrm{F}}\cdot \max_m\ln R^{(m)}\rn_{2}.
$$
\end{lemma}

\begin{lemma}[{\cite[Lemma~13]{wang2024leave}}]\label{lem:P_Omega_AB}
Recall that $\Omega_{S^{\star}}$ is the support of $\Po(S^{\star})$. If $\Po(S^{\star})$ is $2\alpha p$-sparse, then
$$
\ln \P_{\Omega_{S^{\star}}}(AB^T)\rn_{\mathrm{F}}^2\leq 2\alpha pn\cdot\min\left\{\ln A\rn_{\mathrm{F}}^2\ln B\rn_{2,\infty}^2,\ln A\rn_{2,\infty}^2\ln B\rn_{\mathrm{F}}^2\right\}
$$
holds uniformly for all $A,B\in\mathbb{R}^{n\times r}$.
\end{lemma}

\section{Proofs of Lemmas in Section~\ref{sec:useful}}\label{sec:appen_supp}

\subsection{Proof of Lemma~\ref{lem:noise}}

Due to the equivalent characterization of sub-Gaussian random variables in \cite[Lemma~5.5]{Ver12}, one can get $|N_{ij}| \leq C_N^{(1)}\cdot\sigma\sqrt{\log n}$ with high probability. 

Conditioned on the event that \eqref{eq:noise_infty} holds, and apply \cite[Proposition~13]{Klopp2015}\footnote{This is an extension of \cite[Corollary~3.12]{Bandeira2016} to the rectangular case.} with $t=O(\sigma\log n)$, one has
$$
\|\mathcal{P}_{\Omega}(N)\|_2\leq C_N^{(2)}\cdot\sigma\sqrt{np} 
$$
with high probability given the assumption $\sqrt{np}\geq\log n$. 

Using a similar argument as in \eqref{eq:vector_bern}, one can get with high probability,
\begin{align*}
    \ln e_m^T\Po\lb N\rb V\rn_2 
    \leq & C_1\cdot\lb\sqrt{n\cdot 2p\sigma^2\ln V\rn_{2,\infty}^2\cdot\log n} + C_N^{(1)}\sigma\sqrt{\log n}\ln V\rn_{2,\infty}\cdot\log n\rb \\
    \leq & C_1\cdot\lb \sqrt{2}+C_N^{(1)}\rb\cdot\sigma\sqrt{\frac{n\log n}{p}}\ln V\rn_{2,\infty}\\
    \leq & C_1\lb C_N\sqrt{\frac{n\log n}{p}}\rb\sigma\ln V\rn_{2,\infty},
\end{align*}
where the second inequality holds if $\sqrt{np}\geq\log n$.
    
\subsection{Proof of Lemma~\ref{lem:thresh}}

For $(k,l)\in \Omega^{(-i)} \setminus \Omega_{S^{\star}}$,
\begin{align*}
|(M-L^{t,i})_{kl}|=&|(L^{\star}-L^{t,i})_{kl}|+|N_{kl}|\\
\leq &\lsb\lb\frac{\mu r}{n}\sigma_{1}^{\star}\rb\gamma^t+\gamma C_{N}^{(1)}\sigma\sqrt{\log n}\rsb+C_N^{(1)}\sigma\sqrt{\log n} \\
= & \lb\frac{\mu r}{n}\sigma_{1}^{\star}\rb\gamma^t+\lb 1+\gamma \rb C_N^{(1)}\sigma\sqrt{\log n} \leq \xi^t.
\end{align*}
Due to \labelcref{P1} of the thresholding function, $\text{Supp}\lb S^{t,i}\rb\subseteq\Omega^{(-i)}\cap\Omega_{S^{\star}}$. 

For $(k,l)\in \Omega^{(-i)}\cap\Omega_{S^{\star}}$,
\begin{align*}
    |(S^{t,i}-S^{\star})_{kl}|=&|\T_{\xi^t}((L^{\star}+S^{\star}+N-L^{t,i})_{kl})-S^{\star}_{kl}| \\
    \leq&|\T_{\xi^t}((L^{\star}+S^{\star}+N-L^{t,i})_{kl})-\T_{\xi^t}(S^{\star}_{kl})|+|\T_{\xi^t}(S^{\star}_{kl})-S^{\star}_{kl}|\\
    \leq&K\lb|(L^{\star}-L^{t,i})_{kl}|+|N_{kl}|\rb + B\xi^t
    \leq (K+B)\xi^t,
\end{align*}
where in the second inequality, we use \labelcref{P2,P3} of the thresholding function. The conclusion follows by noting that $\xi^t\leq C_{\text{init}}\lsb\lb\frac{\mu r}{n}\sigma_{1}^{\star}\rb\gamma^t+\lb 1+\gamma\rb C_N^{(1)}\sigma\sqrt{\log n}\rsb$.

\subsection{Proof of Lemma~\ref{lem:incoherence}} 
For any matrix $Z$, recall that $\widehat{Z}$ is its symmetrization as defined in \eqref{eq:symmetry}. It is easy to see the eigen-decomposition of $\widehat{L^{\star}}$ is
$$
\widehat{L^{\star}}=\lb\frac{1}{\sqrt{2}}F^{\star}\rb\Si^{\star}\lb\frac{1}{\sqrt{2}}F^{\star}\rb^T+\lb\frac{1}{\sqrt{2}}\widetilde{F}^{\star}\rb\lb-\Si^{\star}\rb\lb\frac{1}{\sqrt{2}}\widetilde{F}^{\star}\rb^T,
$$
where $\widetilde{F}^{\star}:=[(-U^{\star})^T~(V^{\star})^T]^T$. The top-$r$ eigen-decomposition of $\widehat{L^{\star}}+\widehat{E}$ is $\frac12F\Si F^T$, 
and $\Si$ is invertible since $\ln E \rn_2$ is bounded by $\frac12\sigma_r^{\star}$. Thus for $1\leq m\leq 2n$,
\begin{align*}
& \De_{m,:} = e_m^T\lb F-F^{\star}G\rb \\
= & e_m^T\lb\lb\widehat{L^{\star}}+\widehat{E}\rb F \Si^{-1}-F^{\star}G\rb \\
= & e_m^TF^{\star}\Si^{\star}\lsb\frac12\lb F^{\star}\rb^T F\Si^{-1}-\lb\Si^{\star}\rb^{-1}G\rsb\\
& +\frac12e_m^T\widetilde{F}^{\star}\lb-\Si^{\star}\rb\lb \widetilde{F}^{\star}\rb^T F \Si^{-1} +e_m^T\widehat{E} F\Si^{-1} \\
= &\underbrace{e_m^TF^{\star}\Si^{\star}\lsb H\lb\Si^{\star}\rb^{-1}-\lb\Si^{\star}\rb^{-1}G\rsb}_{T_1}+\underbrace{e_m^TF^{\star}\Si^{\star}H\lsb \Si^{-1}-\lb\Si^{\star}\rb^{-1}\rsb}_{T_2}\\
&-\frac12\underbrace{e_m^T\widetilde{F}^{\star}\Si^{\star}\lb \widetilde{F}^{\star}\rb^T F\Si^{-1}}_{T_3}+\underbrace{e_m^T\widehat{E} F\Si^{-1}}_{T_4}.
\end{align*}

\noindent\textbf{$\bullet$ Bounding $T_1$.} Consider
\begin{align*}
R := & \Si^{\star}\lsb H\lb\Si^{\star}\rb^{-1}-\lb\Si^{\star}\rb^{-1}G\rsb = \lb\Si^{\star}H-G\Si^{\star}\rb\lb\Si^{\star}\rb^{-1}.
\end{align*}
Applying Lemma~\ref{lem:perturb_gt} to $\widehat{L}$ and $\widehat{L^{\star}}$ yields
\begin{align*}
    \ln R \rn_2\leq\ln \Si^{\star}H-G\Si^{\star}\rn_2\cdot\ln\lb\Si^{\star}\rb^{-1}\rn_2\leq 4\kappa\frac{\ln E\rn_2}{\sigma_r^{\star}},
\end{align*}
where $\ln E \rn_2\leq \frac12\sigma_r^{\star}$ is used in the last inequality. Therefore,
\begin{align*}
    \ln T_1 \rn_2\leq 4\kappa\frac{\ln E \rn_2}{\sigma^{\star}_r}\sqrt{\frac{\mu r}{n}}.
\end{align*}

\noindent\textbf{$\bullet$ Bounding $T_2$.} Note that
\begin{align*}
    \ln \Si^{-1}-\lb\Si^{\star}\rb^{-1}\rn_2=\max_{1\leq k\leq r} \left|\frac{1}{\sigma_k}-\frac{1}{\sigma_k^{\star}}\right|= \max_{1\leq k\leq r} \frac{\left|\sigma_k-\sigma_k^{\star}\right|}{\sigma_k\sigma_k^{\star}} \leq 2\frac{\ln E\rn_2}{\lb\sigma^{\star}_r\rb^2}.
\end{align*}
Therefore,
\begin{align*}
    \ln T_2 \rn_2 \leq 2\kappa\frac{\ln E\rn_2}{\sigma^{\star}_r}\sqrt{\frac{\mu r}{n}}.
\end{align*}

\noindent\textbf{$\bullet$ Bounding $T_3$.} Note that
\begin{align*}
    \ln \lb\widetilde{F}^{\star}\rb^T F \rn_2 
    = & \ln \lb U^{\star}\rb^T U-\lb V^{\star}\rb^T V \rn_2 \\
    \leq &\ln\lb U^{\star}\rb^T \lb U-U^{\star}G \rb \rn_2+\ln\lb V^{\star}\rb^T \lb V-V^{\star}G\rb \rn_2\\
    \leq & \ln U-U^{\star}G\rn_2+\ln V-V^{\star}G \rn_2 \\
    \leq & 2\ln F-F^{\star}G\rn_2
    \leq \frac{8\sqrt{2}}{\sigma_r^{\star}} \ln E\rn_2,
\end{align*}
where last inequality follows from Lemma~\ref{lem:op}. Therefore,
\begin{align*}
    \ln T_3 \rn_2 \leq 16\sqrt{2}\kappa\frac{\ln E \rn_2}{\sigma^{\star}_r}\sqrt{\frac{\mu r}{n}}.
\end{align*} 

The proof is complete, noting that $\ln T_4 \rn_2 = \ln e_m^TEV\Si^{-1}\rn_2$ if $1\leq m\leq n$ and $\ln T_4 \rn_2 = \ln e_{(m-n)}^TE^TU\Si^{-1}\rn_2$ if $n+1\leq m\leq 2n$.

\subsection{Proof of Lemma~\ref{lem:L_infinity}}\label{subsec:Linfinity}

It can be verified that $\widehat{L}=\mathcal{P}_{2r}\lb\widehat{L^{\star}}+\widehat{E}\rb$, and we proceed to bound $\ln\widehat{L}-\widehat{L^{\star}}\rn_{\infty}$. 

Denote the compact SVD of $L$ as $U\Sigma V^T$. In the proof of this lemma and Lemma~\ref{lem:Ho}, denote the eigen-decomposition of $\widehat{L}$ and $\widehat{L^{\star}}$ as 
$\widehat{F}\widehat{\Lambda}\widehat{F}^T$ and $\widehat{F^{\star}}\widehat{\Lambda^{\star}}\lb\widehat{F^{\star}}\rb^T$ respectively,
where
$$
\widehat{F} = \frac{1}{\sqrt{2}}\left[\begin{array}{cc}
U & -U \\
V & V 
\end{array}\right]:=[w_1,\cdots,w_{2r}],\quad\widehat{F^{\star}} = \frac{1}{\sqrt{2}}\left[\begin{array}{cc}
U^{\star} & -U^{\star} \\
V^{\star} & V^{\star} 
\end{array}\right].
$$
For $1\leq k\leq 2r$, we have $\lb\widehat{L^{\star}}+\widehat{E}\rb w_k = \lambda_k w_k$, therefore
$$
w_k = \left(I-\frac{\widehat{E}}{\lambda_k}\right)^{-1}\frac{\widehat{L^{\star}} w_k}{\lambda_k}=\left[\sum_{p=0}^{\infty}\left(\frac{\widehat{E}}{\lambda_k}\right)^p\right]\frac{\widehat{L^{\star}} w_k}{\lambda_k}.
$$
With this expression, we can get
\begin{align*}
    \widehat{L}-\widehat{L^{\star}} = &\sum_{k=1}^{2r}\lambda_k w_kw_k^T - \widehat{L^{\star}}\\
    = & \sum_{k=1}^{2r} \lambda_k\left[\sum_{p=0}^{\infty} \left(\frac{\widehat{E}}{\lambda_k}\right)^p\right] \frac{\widehat{L^{\star}} w_k}{\lambda_k}\frac{w_k^T\widehat{L^{\star}}}{\lambda_k}\left[\sum_{q=0}^{\infty} \left(\frac{\widehat{E}}{\lambda_k}\right)^q\right]- \widehat{L^{\star}}\\
    = & \sum_{p=0}^{\infty}\sum_{q=0}^{\infty} \widehat{E}^p \widehat{L^{\star}}\left(\sum_{k=1}^{2r}\lambda_k^{-(p+q+1)}w_kw_k^T\right)\widehat{L^{\star}}\widehat{E}^q - \widehat{L^{\star}}\\
    = & \left(\widehat{L^{\star}}\widehat{F}\widehat{\Lambda}^{-1}\widehat{F}^T\widehat{L^{\star}}-\widehat{L^{\star}}\right)+\sum_{p+q\geq 1}\widehat{E}^p\widehat{L^{\star}}\widehat{F}\widehat{\Lambda}^{-(p+q+1)}\widehat{F}^T\widehat{L^{\star}}\widehat{E}^q\\
    = & \widehat{F^{\star}}(\widehat{F^{\star}})^T(\widehat{L^{\star}}\widehat{F}\widehat{\Lambda}^{-1}\widehat{F}^T\widehat{L^{\star}}-\widehat{L^{\star}})\widehat{F^{\star}}(\widehat{F^{\star}})^T+\sum_{p+q\geq 1}\widehat{E}^p\widehat{F^{\star}}(\widehat{F^{\star}})^T\widehat{L^{\star}}\widehat{F}\widehat{\Lambda}^{-(p+q+1)}\widehat{F}^T\widehat{L^{\star}}\widehat{F^{\star}}(\widehat{F^{\star}})^T\widehat{E}^q \\
    := & \sum_{p,q\geq 0} \widehat{E}^p\widehat{F^{\star}}(\widehat{F^{\star}})^TR_{p,q}\widehat{F^{\star}}(\widehat{F^{\star}})^T\widehat{E}^q.
\end{align*}

When $p=q=0$,
\begin{align*}
\left\|\widehat{F^{\star}}(\widehat{F^{\star}})^TR_{0,0}\widehat{F^{\star}}(\widehat{F^{\star}})^T\right\|_{\infty}
\leq & \ln\widehat{F^{\star}}\rn_{2,\infty}\cdot\ln R_{0,0}\rn_2\cdot\ln\widehat{F^{\star}}\rn_{2,\infty}\\
\leq & \frac{\mu r}{n}\cdot\left\|\widehat{L^{\star}}\widehat{F}\widehat{\Lambda}^{-1}\widehat{F}^T\widehat{L^{\star}}-\widehat{L^{\star}}\right\|_{2} \\
\leq & \frac{\mu r}{n}\cdot 5\|\widehat{E}\|_2 = \frac{5\mu r}{n}\|E\|_2,
\end{align*}
where the third inequality follows from \cite[Lemma~13]{Jain2015}. 

When $p+q\geq 1$,
\begin{align*}
\left\|\widehat{E}^p\widehat{F^{\star}}(\widehat{F^{\star}})^TR_{p,q}\widehat{F^{\star}}(\widehat{F^{\star}})^T\widehat{E}^q\right\|_{\infty}
\leq & \|\widehat{E}^p \widehat{F^{\star}}\|_{2,\infty}\cdot\left\|R_{p,q}\right\|_{2}\cdot\|\widehat{E}^q \widehat{F^{\star}}\|_{2,\infty}\\
\leq & (v)^{p+q}\frac{\mu r}{n}\cdot \left\|R_{p,q}\right\|_{2}\\
\leq & (v)^{p+q}\frac{\mu r}{n}\cdot 4\left(\frac{\sigma_r^{\star}}{2}\right)^{-(p+q+1)+2}\\
= & \frac{4\mu r}{n}v\left(v\frac{2}{\sigma_r^{\star}}\right)^{p+q-1}
\leq \frac{4\mu r}{n}v\left(\frac12\right)^{p+q-1},
\end{align*}
where the second inequality follows from assumption \eqref{eq:induction} since when $s$ is odd,
$$
\widehat{E}^s \widehat{F^{\star}} = \frac{1}{\sqrt{2}}\left[\begin{array}{cc}
E\lb E^TE\rb^{\lfloor\frac{s}{2}\rfloor} V^{\star} & E\lb E^TE\rb^{\lfloor\frac{s}{2}\rfloor} V^{\star} \\
E^T\lb EE^T\rb^{\lfloor\frac{s}{2}\rfloor} U^{\star} & - E^T\lb EE^T\rb^{\lfloor\frac{s}{2}\rfloor} U^{\star}
\end{array}\right]
$$
and when $s$ is even,
$$
\widehat{E}^s \widehat{F^{\star}} = \frac{1}{\sqrt{2}}\left[\begin{array}{cc}
\lb EE^T\rb^{\frac{s}{2}} U^{\star} & -\lb EE^T\rb^{\frac{s}{2}} U^{\star} \\
\lb E^TE\rb^{\frac{s}{2}} V^{\star} & \lb E^TE\rb^{\frac{s}{2}} V^{\star}
\end{array}\right];
$$
and the third inequality is due to \cite[Lemma~13]{Jain2015} again. Since
\begin{align*}
\sum_{p+q\geq 1} \left(\frac12\right)^{p+q-1} = \sum_{p\geq 1} \left(\frac12\right)^{p-1}+\sum_{q\geq 1} \left(\frac12\right)^{q-1}+\sum_{p,q\geq 1} \left(\frac12\right)^{p+q-1} = 6,
\end{align*}
\begin{align*}
\left\|\sum_{p+q\geq 1}\widehat{E}^p\widehat{F^{\star}}(\widehat{F^{\star}})^TR_{p,q}\widehat{F^{\star}}(\widehat{F^{\star}})^T\widehat{E}^q\right\|_{\infty}
\leq \sum_{p+q\geq 1} \frac{4\mu r}{n}v\left(\frac12\right)^{p+q-1}
\leq \frac{24\mu r}{n}v. 
\end{align*}

\subsection{Proof of Lemma~\ref{lem:Ho}}

Recall that the operator $\Ho(\cdot)$ acts on matrices of size $n$. Define $\widehat{\mathcal{H}}_{\widehat{\Omega}}(\cdot)$ that acts on symmetric matrices of size $2n$ as follows:
$$
\widehat{\mathcal{H}}_{\widehat{\Omega}}(Z) = \left[\begin{array}{cc}
0 & \Ho(Z_2) \\
\lsb\Ho(Z_2)\rsb^T & 0
\end{array}\right],~\text{where}~Z = \left[\begin{array}{cc}
Z_1 & Z_2 \\
Z_2^T & Z_3
\end{array}\right].
$$

The decomposition of $\widehat{L}-\widehat{L^{\star}}$ in Lemma~\ref{lem:L_infinity} yields that
$$
\widehat{L}-\widehat{L^{\star}} =\sum_{p,q\geq 0} \widehat{E}^p\widehat{F^{\star}}(\widehat{F^{\star}})^TR_{p,q}\widehat{F^{\star}}(\widehat{F^{\star}})^T\widehat{E}^q,
$$
we then get
\begin{align*}
\left\|\Ho\lb L-L^{\star}\rb\right\|_2 = & \left\|\widehat{\mathcal{H}}_{\widehat{\Omega}}\lb \widehat{L}-\widehat{L^{\star}}\rb\right\|_2
\leq \sum_{p,q\geq 0} \ln \widehat{\mathcal{H}}_{\widehat{\Omega}}\lb \widehat{E}^p\widehat{F^{\star}}(\widehat{F^{\star}})^TR_{p,q}\widehat{F^{\star}}(\widehat{F^{\star}})^T\widehat{E}^q\rb\rn_2.
\end{align*}

For $p,q\geq 0$,
\begin{align*}
    &\left\|\widehat{\mathcal{H}}_{\widehat{\Omega}}\lb \widehat{E}^p\widehat{F^{\star}}(\widehat{F^{\star}})^TR_{p,q}\widehat{F^{\star}}(\widehat{F^{\star}})^T\widehat{E}^q\rb\right\|_2 \\
    = &\max_{\ln z\rn_2=1} \left\langle \widehat{\mathcal{H}}_{\widehat{\Omega}}\lb \widehat{E}^p\widehat{F^{\star}}(\widehat{F^{\star}})^TR_{p,q}\widehat{F^{\star}}(\widehat{F^{\star}})^T\widehat{E}^q\rb,zz^T\right\rangle \\
    =&\max_{\ln z\rn_2=1} \left\langle\widehat{\mathcal{H}}_{\widehat{\Omega}}\lb \bm{1}\bm{1}^T\circ\widehat{E}^p\widehat{F^{\star}}(\widehat{F^{\star}})^TR_{p,q}\widehat{F^{\star}}(\widehat{F^{\star}})^T\widehat{E}^q\rb,zz^T\right\rangle\\
    =&\max_{\ln z\rn_2=1} \left\langle\widehat{\mathcal{H}}_{\widehat{\Omega}}\lb \bm{1}\bm{1}^T\rb,\widehat{E}^p\widehat{F^{\star}}(\widehat{F^{\star}})^TR_{p,q}\widehat{F^{\star}}(\widehat{F^{\star}})^T\widehat{E}^q\circ zz^T\right\rangle\\
    \leq & \ln \widehat{\mathcal{H}}_{\widehat{\Omega}}\lb \bm{1}\bm{1}^T\rb\rn_2\cdot\ln\lsb \widehat{E}^p\widehat{F^{\star}}\cdot\lb \widehat{E}^q\widehat{F^{\star}}(\widehat{F^{\star}})^TR_{p,q}\widehat{F^{\star}}\rb^T \rsb\circ zz^T\rn_{*} \\
    \leq &\ln \Ho\lb \bm{1}\bm{1}^T\rb\rn_2\cdot\lb\ln\widehat{E}^p \widehat{F^{\star}}\rn_{2,\infty}\ln z\rn_2\rb\cdot\lb\ln \widehat{E}^q\widehat{F^{\star}}(\widehat{F^{\star}})^TR_{p,q}\widehat{F^{\star}}\rn_{2,\infty}\ln z\rn_2\rb, \\
    \leq &\ln \Ho\lb \bm{1}\bm{1}^T\rb\rn_2\cdot\ln\widehat{E}^p \widehat{F^{\star}}\rn_{2,\infty}\cdot\ln \widehat{E}^q\widehat{F^{\star}}\rn_{2,\infty}\ln R_{p,q}\rn_2,
\end{align*} 
where the second inequality follows from the same argument as \cite[Lemma~8]{Chen2019} in bounding $\ln \lb AC^T\circ BD^T\rb\rn_{*}$, by writing the Hadamard product as the sum of rank-one matrices. The upper bound can then be derived based on the bounds of $\ln\widehat{E}^p \widehat{F^{\star}}\rn_{2,\infty}$ and $\ln\widehat{E}^q \widehat{F^{\star}}\rn_{2,\infty}$, and the bound of $\ln R_{p,q}\rn_2$ as in the proof of Lemma~\ref{lem:L_infinity}.

\section{Proofs of Lemmas in Section \ref{sec:base}}\label{sec:proofs_base}

\subsection{Proof of Lemma~\ref{lem:init_bounds}}\label{appen:init_bounds}
With $L^{0,i}:=0$,
\(
\ln L^{0,i}-L^{\star} \rn_{\infty} = \ln L^{\star} \rn_{\infty} \leq \frac{\mu r}{n} \sigma_1^{\star}
\)
under Assumption~\ref{assump1}, see \eqref{eq:tmp001}. Conditioned on the event \eqref{eq:noise_infty} holds and by Lemma~\ref{lem:thresh}, one gets $
\text{Supp}\lb S^{0,i}\rb\subseteq\Omega^{(-i)}\cap\Omega_{S^{\star}}$ and
$$
\ln\Po^{(-i)}\lb S^{0,i}-S^{\star}\rb\rn_{\infty}\leq C_{\mathrm{thresh}}\lsb \frac{\mu r}{n}\sigma_1^{\star} +(1+\gamma)C_N^{(1)}\sigma\sqrt{\log n}\rsb.
$$

\noindent\textbf{Part \RomanNumeralCaps{1}: Bound for $\ln E^{0,\infty} \rn_2$.} First consider $\ln E_1^{0,i} \rn_2$,
\begin{equation}\label{eq:S_init_op}
\begin{aligned} 
\ln E_1^{0,i} \rn_2 = &\ln p^{-1}\Po^{(-i)}\lb S^{0,i}-S^{\star}\rb\rn_2 \\
\leq & p^{-1}\cdot\lb 2\alpha p n\rb \cdot C_{\mathrm{thresh}}\lb \frac{\mu r}{n}\sigma_1^{\star}+2C_N^{(1)}\sigma\sqrt{\log n}\rb \\
\leq &\frac{1}{2C_0}\frac{\sigma_{r}^{\star}}{\sqrt{\kappa}}\gamma+\lb 4C_{\mathrm{thresh}}C_N^{(1)}\alpha n\sqrt{\log n}\rb\sigma,
\end{aligned}
\end{equation}
where the first inequality follows from applying Lemma~\ref{lem:S_op} to the $2\alpha p$-sparse matrix $\Po^{(-i)}\lb S^{0,i}-S^{\star}\rb$, and the second inequality holds if 
$
\alpha\leq\frac{1}{4C_0}\frac{1}{\kappa^{1.5} \mu r}\cdot\frac{\gamma}{C_{\mathrm{thresh}}}.
$ 

For $\ln E_2^{0,i} \rn_2$, one has
\begin{equation}\label{eq:op_init}
\begin{aligned}
\ln E_2^{0,i} \rn_2 = &\ln\Ho^{(-i)}\lb-L^{\star}\rb\rn_2
\leq \ln\Ho\lb L^{\star}\rb\rn_2\\
\leq & c_{12}\lb \frac{\log n}{p}\ln L^{\star}\rn_{\infty}+\sqrt{\frac{\log n}{p}}\cdot\max\left\{\ln L^{\star}\rn_{2,\infty},\ln \lb L^{\star}\rb^T\rn_{2,\infty}\right\}\rb\\
\leq & c_{12}\lb \frac{\log n}{p}\frac{\mu r}{n}\sigma_1^{\star}+\sqrt{\frac{\log n}{p}}\sqrt{\frac{\mu r}{n}}\sigma_1^{\star}\rb\\
\leq & 2c_{12}\sqrt{\frac{\mu r\log n}{np}}\sigma_1^{\star}\leq \frac{1}{2C_0}\frac{\sigma_{r}^{\star}}{\sqrt{\kappa}}\gamma,
\end{aligned}
\end{equation}
where second inequality holds with high probability due to Lemma \ref{lem:init}, and the last two inequality holds if 
$
p\geq\frac{16c_{12}^2C_0^2}{\gamma^2}\cdot\frac{\kappa^{3}\mu r\log n}{n}.
$

For $\ln E_3^{0,i} \rn_2$, by Lemma~\ref{lem:noise},
\begin{equation}\label{eq:noise_init}
\begin{aligned}
    \ln E_3^{0,i} \rn_2 \leq p^{-1}\ln \Po\lb N\rb\rn_2\leq \lb C_N^{(2)}\sqrt{\frac{n}{p}}\rb\sigma.
\end{aligned}
\end{equation}

Therefore, with $C_N$ defined in \eqref{eq:C_N},
\begin{equation*}
\begin{aligned}
\ln E^{0,i} \rn_2 \leq & \ln E_1^{0,i} \rn_2+\ln E_2^{0,i} \rn_2+\ln E_3^{0,i} \rn_2 \\
\leq & \frac{1}{C_0}\frac{\sigma_{r}^{\star}}{\sqrt{\kappa}}\gamma+C_{N}\lb 4C_{\mathrm{thresh}}\alpha n\sqrt{\log n}+\sqrt{\frac{n}{p}}\rb\sigma \\
\leq & \frac{1}{C_0}\frac{\sigma_{r}^{\star}}{\sqrt{\kappa}}\gamma+4C_{\mathrm{noise}}\sigma.
\end{aligned}
\end{equation*}

One can see that $\| E^{0,\infty}\|_2\leq \frac12\sigma_r^{\star}$ if $C_{\mathrm{noise}}\sigma\leq \frac{1}{2C_0}\sigma_r^{\star}$ and $C_0\geq 6$, and could be smaller as one tunes $C_0$.


\noindent\textbf{Part \RomanNumeralCaps{2}: Bound for $\ln \De^{1,\infty}\rn_{2,\infty}$.} 
Since $\ln E^{0,\infty}\rn_2\leq\frac12\sigma_r^{\star}$, applying Lemma~\ref{lem:incoherence} and one can get
\begin{align*}
\ln \De^{1,i}\rn_{2,\infty} \leq & 18\kappa\frac{\ln E^{0,i}\rn_2}{\sigma_r^{\star}}\sqrt{\frac{\mu r}{n}}+\max\lcb\ln E^{0,i}V^{1,i}\rn_{2,\infty},\ln \lb E^{0,i}\rb^T U^{1,i}\rn_{2,\infty}\rcb\cdot\ln \lb \Si^{1,i}\rb^{-1}\rn_2.
\end{align*}

We only derive the bound for $\ln E^{0,i}V^{1,i}\rn_{2,\infty}$ and the same bound for $\ln \lb E^{0,i}\rb^T U^{1,i}\rn_{2,\infty}$ can be obtained. Consider $1\leq m\leq n$ and $i\neq m$ since due to our definition of the leave-one-out sequences, $\ln e_m^TE^{0,i}V^{1,i}\rn_2 = 0
$ when $i=m$. Without loss of generality, we only consider the case  $0\leq i\leq n,$ and the proof can be done similarly when $i>n$.

For $\ln e_m^TE_1^{0,i}V^{1,i}\rn_2$, one has
\begin{align*}
    \ln e_m^TE_1^{0,i}V^{1,i}\rn_2 
    = & p^{-1}\ln e_m^T\Po^{(-i)}\lb S^{0,i}-S^{\star}\rb V^{1,i}\rn_2\\
    \leq & p^{-1}\cdot(2\alpha pn)\cdot C_{\mathrm{thresh}}\lb\frac{\mu r}{n}\sigma_{1}^{\star}+2C_{N}\sigma\sqrt{\log n}\rb\cdot \ln V^{1,i}\rn_{2,\infty}\\
    \leq & \lsb\frac{1}{2C_0}\frac{\sigma_{r}^{\star}}{\sqrt{\kappa}}\gamma+\lb 4C_{\mathrm{thresh}} C_{N}\alpha n\sqrt{\log n}\rb\sigma\rsb\lb\sqrt{\frac{\mu r}{n}}+\ln\De^{1,\infty}\rn_{2,\infty}\rb,
\end{align*}
where the last inequality follows from the bound in \eqref{eq:S_init_op}. 

For $\ln e_m^TE_2^{0,i}V^{1,i}\rn_2$, one has
\begin{align*}
    \ln e_m^TE_2^{0,i} V^{1,i}\rn_2 \leq & \ln e_m^TE_2^{0,i} \lb V^{1,i}-V^{1,m}G^{1,i,m} \rb\rn_2+\ln e_m^TE_2^{0,i} V^{1,m}G^{1,i,m}\rn_2 \\
    \leq & \ln E_2^{0,i}\rn_2\ln D^{1,i,m} \rn_{\mathrm{F}}+\ln e_m^TE_2^{0,i} V^{1,m} \rn_2\\
    \leq & \frac{\sigma_r^{\star}}{2C_0}\ln D^{1,i,m}\rn_{\mathrm{F}}+\ln e_m^T\Ho^{(-i)}\lb L^{\star}\rb V^{1,m} \rn_2,
\end{align*}
where in the last inequality the bound for the first term follows from \eqref{eq:op_init}, and the second term can be bounded using the property that $V^{1,m}$ is independent of the Bernoulli variables on the $m$-th row. For $k=1,\cdots,n$, define 
\[
v_k = \lb 1-\delta_{mk}/p\rb L_{mk}^{\star} V_{k,:}^{1,m}.
\]
One has
\begin{align*}
\ln v_k\rn_2 \leq &\frac1p\ln L^{\star} \rn_{\infty} \ln V^{1,m} \rn_{2,\infty},\\
\left| \sum_{k=1}^n\mathbb{E} \lsb\ln v_k\rn_2^2\rsb \right| \leq & \sum_{k=1}^n \frac1p \lb L_{mk}^{\star}\rb^2\ln V_{k,:}^{1,m} \rn_2^2 \leq \frac1p\ln L^{\star} \rn_{2,\infty}^2\ln V^{1,m} \rn_{2,\infty}^2.
\end{align*}
By Lemma \ref{lem:bernstein},
\begin{equation}\label{eq:vector_bern}
\begin{aligned}
& \ln e_m^T\Ho^{(-i)}\lb L^{\star}\rb V^{1,m} \rn_2 = \ln \sum_{k=1}^n v_k\rn_2\\
    \leq & C_1\cdot\lb\sqrt{\frac{\log n}{p}}\ln L^{\star} \rn_{2,\infty}\ln V^{1,m} \rn_{2,\infty}+\frac{\log n}{p}\ln L^{\star} \rn_{\infty} \ln V^{1,m} \rn_{2,\infty}\rb \\
    \leq & C_1\cdot\lb\sqrt{\frac{\mu r\log n}{np}}\sigma_1^{\star}+\frac{\mu r\log n}{np}\sigma_1^{\star}\rb\cdot\lb \sqrt{\frac{\mu r}{n}}+\ln \De^{1,\infty} \rn_{2,\infty}\rb  \\
    \leq & \frac{C_1}{2C_0}\frac{\sigma_{r}^{\star}}{\sqrt{\kappa}}\gamma\lb\sqrt{\frac{\mu r}{n}} + \ln \De^{1,\infty} \rn_{2,\infty}\rb,
\end{aligned}
\end{equation}
where the last inequality follows from the bound of $\sqrt{\frac{\mu r\log n}{np}}\sigma_1^{\star}$ in \eqref{eq:op_init} and the fact that the universal constant $c_{12}$ from Lemma \ref{lem:init} is greater than 1.

For $\ln e_m^TE_3^{0,i}V^{1,i}\rn_2$, one can similarly get the bound
\begin{align*}
    \ln e_m^TE_3^{0,i}V^{1,i}\rn_2 \leq & \ln E_3^{0,i}\rn_2 \ln D^{1,i,m} \rn_{\mathrm{F}} + \ln e_m^T E_3^{0,i}V^{1,m}\rn_2 \\
    \leq & \frac{\sigma_r^{\star}}{2C_0}\ln D^{1,i,m} \rn_{\mathrm{F}}+C_1\lb C_N\sqrt{\frac{n\log n}{p}}\rb\sigma\lb\sqrt{\frac{\mu r}{n}} + \ln \De^{1,\infty} \rn_{2,\infty}\rb, 
\end{align*}
where the bound for the first term in the last inequality follows from \eqref{eq:noise_init} by assuming 
$\lb C_N\sqrt{\frac{n}{p}}\rb\sigma\leq \frac{1}{2C_0}\sigma^{\star}_r$, and the bound for the second term is due to Lemma~\ref{lem:noise}.

Therefore when $C_1\geq 4$,
\begin{align*}
    \ln e_m^TE^{0,i}V^{1,i}\rn_2 \leq & \ln e_m^TE_1^{0,i}V^{1,i}\rn_2+\ln e_m^TE_2^{0,i}V^{1,i}\rn_2+\ln e_m^TE_3^{0,i}V^{1,i}\rn_2 \\
    \leq & \frac{1+C_1}{2C_0}\frac{\sigma_r^{\star}}{\sqrt{\kappa}}\gamma\lb\sqrt{\frac{\mu r}{n}}+\ln\De^{1,\infty} \rn_{2,\infty}\rb+C_1C_{\mathrm{noise}}\sigma\lb\sqrt{\frac{\mu r}{n}}+\ln\De^{1,\infty} \rn_{2,\infty}\rb+\frac{\sigma_r^{\star}}{C_0}\ln D^{1,\infty} \rn_{\mathrm{F}}.
\end{align*}
As a result,
\begin{equation}\label{eq:2_infty_init}
\begin{aligned}
    \ln E^{0,i}V^{1,i}\rn_{2,\infty}
    \leq & \lb\frac{1+C_1}{2C_0}\frac{\sigma_r^{\star}}{\sqrt{\kappa}}\gamma+C_1C_{\mathrm{noise}}\sigma\rb\lb\sqrt{\frac{\mu r}{n}}+\ln\De^{1,\infty} \rn_{2,\infty}\rb
    +\frac{\sigma_r^{\star}}{C_0}\ln D^{1,\infty} \rn_{\mathrm{F}}.
\end{aligned}
\end{equation}

It follows that
\begin{equation*}
\begin{aligned}
    \ln \De^{1,i} \rn_{2,\infty}
    \leq & 18\kappa\frac{\ln E^{0,\infty}\rn_2}{\sigma^{\star}_r}\sqrt{\frac{\mu r}{n}}+\frac{1+C_1}{C_0}\frac{\gamma}{\sqrt{\kappa}}\lb\sqrt{\frac{\mu r}{n}}+\ln\De^{1,\infty} \rn_{2,\infty}\rb\\
    &+2C_1C_{\mathrm{noise}}\lb\frac{\sigma}{\sigma^{\star}_r}\rb\lb\sqrt{\frac{\mu r}{n}}+\ln\De^{1,\infty} \rn_{2,\infty}\rb+\frac{2}{C_0}\ln D^{1,\infty} \rn_{\mathrm{F}} \\
    \leq & \lsb\frac{19+C_1}{C_0}\gamma+\lb 72+2C_1\rb\sqrt{\kappa} C_{\mathrm{noise}}\lb\frac{\sigma}{\sigma^{\star}_r}\rb\rsb\sqrt{\frac{\kappa\mu r}{n}}\\
    &+\lsb\frac{1+C_1}{C_0}+2C_1C_{\mathrm{noise}}\lb\frac{\sigma}{\sigma^{\star}_r}\rb\rsb\ln\De^{1,\infty} \rn_{2,\infty}+\frac{2}{C_0}\ln D^{1,\infty} \rn_{\mathrm{F}}. 
\end{aligned} 
\end{equation*}


\noindent\textbf{Part \RomanNumeralCaps{3}: Bound for $\ln D^{1,\infty} \rn_{\mathrm{F}}$.} For $i\in\{0,1,\cdots,2n\},~m\in\{1,\cdots,2n\}$,
\begin{align*}
\left\|D^{1, i, m}\right\|_{\mathrm{F}} \leq & \left\|F^{1, i}-F^{1, m} G^{1,0,m}G^{1,i,0}\right\|_{\mathrm{F}} \\
\leq & \left\|F^{1, i}-F^{1,0} G^{1,i,0}\right\|_{\mathrm{F}}+\left\|\left(F^{1,0} - F^{1, m}G^{1,0,m}\right) G^{1,i,0}\right\|_{\mathrm{F}}
= \ln D^{1,i,0} \rn_{\mathrm{F}}+\ln D^{1,0,m} \rn_{\mathrm{F}}.
\end{align*}
We only need to bound $\ln D^{1,0,m} \rn_{\mathrm{F}}$, since $\ln D^{1,m,0} \rn_{\mathrm{F}}=\ln D^{1,0,m} \rn_{\mathrm{F}}$. Let $A=\widehat{L^{\star}}+\widehat{E^{0,m}}$ and $\widetilde{A}=\widehat{L^{\star}}+\widehat{E^{0}}$ (see \eqref{eq:symmetry} for the definition of symmetrization). Suppose the eigenvalues of $A$ are in a descending order. According to Weyl's inequality,
$
|\lambda_r(A)-\sigma_r^{\star}|\leq\ln E^{0,m}\rn_2,~ 
|\lambda_{r+1}(A)|\leq\ln E^{0,m}\rn_2.
$
Define $\delta:=\lambda_r(A)-\lambda_{r+1}(A)$ and $W^{0,m}:=\widehat{E^{0}}-\widehat{E^{0,m}}$. Then by Lemma \ref{lem:perturb_F},
\begin{equation}\label{eq:WF_init}
\begin{aligned}
\ln D^{1,0,m}\rn_{\mathrm{F}} \leq \frac{\sqrt{2}\ln W^{0,m} F^{1,m}\rn_{\mathrm{F}}}{\delta-\ln W^{0,m}\rn_2}\leq &\frac{2}{\sigma_r^{\star}}\ln W^{0,m} F^{1,m}\rn_{\mathrm{F}},
\end{aligned}
\end{equation}
where the inequalities follow from the bound of $\ln E^{0,\infty}\rn_2$ assuming $C_{\mathrm{noise}}\sigma\leq \frac{1}{2C_0}\sigma_r^{\star}$ and $C_0\geq 42$.
It is evident that 
$$
\ln W^{0,m}F^{1,m}\rn_{\mathrm{F}}\leq\ln \lb E^0-E^{0,m}\rb V^{1,m}\rn_{\mathrm{F}}+\ln\lb E^{0}-E^{0,m}\rb^T U^{1,m}\rn_{\mathrm{F}}.
$$
In the following we only derive the bound for $\ln\lb E^{0}-E^{0,m}\rb V^{1,m}\rn_{\mathrm{F}}$, and the same bound can be obtained for $\ln\lb E^{0}-E^{0,m}\rb^T U^{1,m}\rn_{\mathrm{F}}$. Note that
\begin{align*}
\ln\lb E^{0}-E^{0,m}\rb V^{1,m}\rn_{\mathrm{F}}
\leq &\ln \lsb p^{-1}\Po^{(-m)}\lb S^{0,m}-S^{\star} \rb-p^{-1}\Po\lb S^{0}-S^{\star}\rb\rsb V^{1,m}\rn_{\mathrm{F}} \\
&+\ln\lb\Ho^{(-m)}-\Ho\rb\lb L^{\star}\rb V^{1,m}\rn_{\mathrm{F}}+p^{-1}\ln\lb\Po-\Po^{(-m)}\rb\lb N\rb V^{1,m}\rn_{\mathrm{F}}\\
\leq & p^{-1}\ln\lsb\Po^{(-m)}\lb S^{0,m}-S^{0}\rb+\lb\Po^{(-m)}-\Po\rb\lb S^{0}- S^{\star}\rb\rsb V^{1,m}\rn_{\mathrm{F}}\\
&+\ln\lb\Ho-\Ho^{(-m)}\rb\lb L^{\star}\rb V^{1,m}\rn_{\mathrm{F}}+p^{-1}\ln\lb\Po-\Po^{(-m)}\rb\lb N\rb V^{1,m}\rn_{\mathrm{F}}\\
= &\underbrace{p^{-1}\ln\lb\Po-\Po^{(-m)}\rb\lb S^{0}- S^{\star}\rb V^{1,m}\rn_{\mathrm{F}}}_{B_1}\\
&+\underbrace{\ln\lb\Ho-\Ho^{(-m)}\rb\lb L^{\star}\rb V^{1,m}\rn_{\mathrm{F}}}_{B_2}+\underbrace{p^{-1}\ln\lb\Po-\Po^{(-m)}\rb\lb N\rb V^{1,m}\rn_{\mathrm{F}}}_{B_3},
\end{align*}
where the equality holds since $\forall i\in\{0,\cdots,2n\}$, 
$
S^{0,i} = \T_{\xi^{0}}\lb\Po^{(-i)}\lb M\rb\rb,
$
therefore $\Po^{(-m)}\lb S^{0,m}-S^{0}\rb$ is a zero matrix.

\noindent\textbf{$\bullet$ Bounding $B_1$.} If $m\leq n$,
\begin{align*}
    B_1 = & p^{-1}\ln e_m^T\Po\lb S^0-S^{\star}\rb V^{1,m}\rn_2 \\
    \leq & p^{-1}\cdot(2\alpha pn)\cdot C_{\mathrm{thresh}}\lb\frac{\mu r}{n}\sigma_{1}^{\star}+2C_{N}\sigma\sqrt{\log n}\rb\cdot\lb\sqrt{\frac{\mu r}{n}}+\ln \De^{1,\infty} \rn_{2,\infty}\rb \\
    \leq &\lsb\frac{1}{2C_0}\frac{\sigma_{r}^{\star}}{\sqrt{\kappa}}\gamma+\lb 4C_{\mathrm{thresh}}C_{N}\alpha n\sqrt{\log n}\rb\sigma\rsb\lb\sqrt{\frac{\mu r}{n}}+\ln \De^{1,\infty} \rn_{2,\infty}\rb,
\end{align*}
where the last inequality follows from the bound in \eqref{eq:S_init_op}. If $m> n$,
\begin{align*}
    B_1 = & p^{-1}\ln \Po\lb S^0-S^{\star}\rb e_{(m-n)}e_{(m-n)}^T V^{1,m}\rn_{\mathrm{F}} \\
    \leq & p^{-1}\cdot(2\alpha pn)\cdot C_{\mathrm{thresh}}\lb\frac{\mu r}{n}\sigma_{1}^{\star}+2C_{N}\sigma\sqrt{\log n}\rb\cdot\lb\sqrt{\frac{\mu r}{n}}+\ln \De^{1,\infty} \rn_{2,\infty}\rb \\
    \leq &\lsb\frac{1}{2C_0}\frac{\sigma_{r}^{\star}}{\sqrt{\kappa}}\gamma+\lb 4C_{\mathrm{thresh}}C_{N}\alpha n\sqrt{\log n}\rb\sigma\rsb\lb\sqrt{\frac{\mu r}{n}}+\ln \De^{1,\infty} \rn_{2,\infty}\rb,
\end{align*}
where the first inequality holds since $\ln\Po\lb S^0-S^{\star}\rb e_{(m-n)}e_{(m-n)}^T V^{1,m}\rn_{\mathrm{F}}$ is no more than the sum of the $l_2$ norm of its rows.

\noindent\textbf{$\bullet$ Bounding $B_2$.} If $m\leq n$,
\begin{align*}
B_2 = \ln e_m^T\Ho\lb L^{\star}\rb V^{1,m}\rn_2\leq\frac{C_1}{2C_0}\frac{\sigma_{r}^{\star}}{\sqrt{\kappa}}\gamma\lb\sqrt{\frac{\mu r}{n}} + \ln \De^{1,\infty} \rn_{2,\infty}\rb,
\end{align*}
where the inequality comes from the bound we have derived in \eqref{eq:vector_bern}. If $m>n$,
\begin{align*}
B_2 = &\ln \Ho\lb L^{\star}\rb e_{(m-n)}e_{(m-n)}^T V^{1,m}\rn_{\mathrm{F}}\\
\leq&\ln \Ho\lb L^{\star}\rb\rn_2\cdot\lb\sqrt{\frac{\mu r}{n}} + \ln \De^{1,\infty} \rn_{2,\infty}\rb
\leq\frac{1}{2C_0}\frac{\sigma_{r}^{\star}}{\sqrt{\kappa}}\gamma\lb\sqrt{\frac{\mu r}{n}} + \ln \De^{1,\infty} \rn_{2,\infty}\rb,
\end{align*}
where the last inequality follows from \eqref{eq:op_init}. 

\noindent\textbf{$\bullet$ Bounding $B_3$.} If $m\leq n$,
\begin{align*}
    B_3 = p^{-1}\ln e_m^T\Po\lb N\rb V^{1,m}\rn_2
    \leq C_1\lb C_{N}\sqrt{\frac{n\log n}{p}} \rb\sigma \lb\sqrt{\frac{\mu r}{n}} + \ln \De^{1,\infty} \rn_{2,\infty}\rb
\end{align*}
due to Lemma~\ref{lem:noise}. If $m>n$,
\begin{equation}\label{eq:B3}
\begin{aligned}
    B_3 = & p^{-1}\ln \Po(N) e_{(m-n)}e_{(m-n)}^TV^{1,m}\rn_{\mathrm{F}} \\
    \leq & p^{-1}\ln \Po(N) \rn_2 \cdot \ln e_{(m-n)}^TV^{1,m}\rn_{2}
    \leq \lb C_N\sqrt{\frac{n}{p}}\rb\sigma\lb\sqrt{\frac{\mu r}{n}} + \ln \De^{1,\infty} \rn_{2,\infty}\rb,
\end{aligned}
\end{equation}
where the last inequality follows from Lemma~\ref{lem:noise}.

Combining the bounds of $B_1$ to $B_3$, one can get
\begin{equation}\label{eq:WF_init_bound}
\begin{aligned}
\ln W^{0,m} F^{1,m}\rn_{\mathrm{F}}
\leq & \lb\frac{1+C_1}{C_0}\frac{\sigma_{r}^{\star}}{\sqrt{\kappa}}\gamma+2C_1C_{\mathrm{noise}}\sigma\rb\lb\sqrt{\frac{\mu r}{n}}
+ \ln \De^{1,\infty} \rn_{2,\infty}\rb.
\end{aligned}
\end{equation}
As a result,
\begin{equation*}
\begin{aligned}
\ln D^{1,i,m} \rn_{\mathrm{F}} \leq & 2\cdot\max_m\ln D^{1,0,m} \rn_{\mathrm{F}}
\leq\lsb\frac{4+4C_1}{C_0}\frac{\gamma}{\sqrt{\kappa}}+8C_1C_{\mathrm{noise}}\lb\frac{\sigma}{\sigma^{\star}_r}\rb\rsb\lb\sqrt{\frac{\mu r}{n}} + \ln \De^{1,\infty} \rn_{2,\infty}\rb.
\end{aligned}
\end{equation*}

\subsection{Proof of Lemma~\ref{lem:L_infty_init}}\label{appen:L_infty_init}
Define $H_1^{i}:=\lb\mathcal{P}_{T^{1,i}}-\I\rb\lb L^{\star}\rb$, 
$H_2^{i}:=\mathcal{P}_{T^{1,i}}E^{0,i}$ and $H^i:=H_1^{i}+H_2^{i}$.
The focus is to verify the second assumption in Lemma~\ref{lem:L_infinity} for $H^{i}$. In the following we derive the bounds for 
$\ln \lsb H^{i}\lb H^{i}\rb^T \rsb^a U^{\star}\rn_{2,\infty},$
while the bounds for the other three terms can be obtained similarly.

When $a=0$,
$
\ln U^{\star} \rn_{2,\infty} \leq \sqrt{\frac{\mu r}{n}}.
$
When $a\geq 1$, for $1\leq m\leq n$,
$$
\ln e_m^T \lsb H^{i}\lb H^{i}\rb^T \rsb^a U^{\star}\rn_2\leq \ln e_m^T H^{i}\rn_2\cdot\ln H^i\rn_2^{2a-1}.
$$

\noindent\textbf{$\bullet$ Bounding $\ln e_m^T H^{i}\rn_2$.}  One has
\begin{align*}
    \ln e_m^TH_1^{i}\rn_2 = & \ln e_m^T\lb I-U^{1,i}\lb U^{1,i}\rb^T\rb\lb L^{\star}\rb\lb I-V^{1,i}\lb V^{1,i}\rb^T\rb \rn_2 \\
    = & \ln e_m^T\lb U^{\star}\lb U^{\star}\rb^T-U^{1,i}\lb U^{1,i}\rb^T\rb \lb L^{\star}-L^{1,i}\rb \lb I-V^{1,i}\lb V^{1,i}\rb^T\rb\rn_2 \\
    \leq & \ln e_m^T\lb U^{\star}\lb U^{\star}\rb^T-U^{1,i}\lb U^{1,i}\rb^T\rb\rn_2\cdot\ln L^{\star}-L^{1,i}\rn_2 \\
    \leq & \lb\ln e_m^TU^{\star}\rn_2+\ln e_m^TU^{1,i}\rn_2\rb\cdot2\|E^{0,i}\|_2 \\
    \leq & \lb 2\sqrt{\frac{\mu r}{n}}+\frac12\sqrt{\frac{\kappa\mu r}{n}}\rb \cdot2\lb \frac{1}{C_0}\frac{\sigma_r^{\star}}{\sqrt{\kappa}}\gamma+4C_{\mathrm{noise}}\sigma\rb\\
    \leq & \lb\frac{5}{C_0}\sigma_r^{\star}\gamma+20\sqrt{\kappa}C_{\mathrm{noise}}\sigma\rb\cdot\sqrt{\frac{\mu r}{n}},
\end{align*}
where the bound $\ln \De^{1,\infty} \rn_{2,\infty}\leq\frac12\sqrt{\frac{\kappa\mu r}{n}}$ is used in the third inequality. Moreover, 
\begin{align*}
    \ln e_m^TH_2^{i}\rn_2 \leq & \ln e_m^TU^{1,i}\lb U^{1,i}\rb^TE^{0,i}\lb I-V^{1,i}\lb V^{1,i}\rb^T\rb \rn_2 +\ln e_m^TE^{0,i}V^{1,i}\lb V^{1,i}\rb^T\rn_2\\
    \leq & \ln e_m^TU^{1,i}\rn_2\cdot\ln E^{0,i}\rn_2 + \ln e_m^TE^{0,i}V^{1,i}\rn_2 \\
    \leq & \frac32\sqrt{\frac{\kappa\mu r}{n}} \cdot\lb \frac{1}{C_0}\frac{\sigma_r^{\star}}{\sqrt{\kappa}}\gamma+4C_{\mathrm{noise}}\sigma\rb \\
    & + \lb\frac{1+C_1}{2C_0}\frac{\sigma_r^{\star}}{\sqrt{\kappa}}\gamma+C_1C_{\mathrm{noise}}\sigma\rb\cdot\frac32\sqrt{\frac{\kappa\mu r}{n}}+\frac{1}{C_0}\lb \frac{8C_1}{C_0}\sigma_r^{\star}\gamma+12\sqrt{\kappa}C_1C_{\mathrm{noise}}\sigma\rb\cdot\sqrt{\frac{\mu r}{n}}\\
    \leq & \lb\frac{4+C_1}{C_0}\sigma_r^{\star}\gamma+2\sqrt{\kappa}C_1C_{\mathrm{noise}}\sigma\rb\cdot\sqrt{\frac{\mu r}{n}},
\end{align*}
where in the third inequality the bound for the second term readily follows from \eqref{eq:2_infty_init} and the bound of $\ln D^{1,\infty} \rn_{\mathrm{F}}$, and the last inequality holds if $C_0\geq 8C_1$ and $C_1\geq 15$. Therefore,
\begin{align*}
    \ln e_m^TH^{i}\rn_2 \leq \lb\frac{2C_1}{C_0}\sigma_r^{\star}\gamma+4\sqrt{\kappa}C_1C_{\mathrm{noise}}\sigma\rb\cdot\sqrt{\frac{\mu r}{n}}.
\end{align*}

\noindent\textbf{$\bullet$ Bounding $\ln H^{i}\rn_2$.}   Note that 
\begin{align*}
    \ln H^{i}\rn_2 \leq & \ln H_1^{i}\rn_2+\ln H_2^{i}\rn_2 \\
    \leq & \frac{\ln L^{1,i}-L^{\star}\rn_2^2}{\sigma_r^{\star}}+\sqrt{\frac43}\ln E^{0,i}\rn_2 \\
    \leq & \ln E^{0,i}\rn_2\lb\frac{4\ln E^{0,i}\rn_2}{\sigma_r^{\star}} + \sqrt{\frac43} \rb
    \leq 2\ln E^{0,i}\rn_2 =\frac{2}{C_0}\frac{\sigma_r^{\star}}{\sqrt{\kappa}}\gamma+8C_{\mathrm{noise}}\sigma,
\end{align*}
where in the second inequality the bound for the first term and second term follows from Lemma~\ref{lem:T_diff} and Lemma~\ref{lem:T_op}, respectively. Choose $
v := \frac{2C_1}{C_0}\sigma_r^{\star}\gamma+4\sqrt{\kappa}C_1C_{\mathrm{noise}}\sigma
$ and $v\leq \frac14 \sigma_r^{\star}$ if $C_{\mathrm{noise}}\sigma\leq \frac{1}{2C_0}\frac{\sigma_r^{\star}}{\sqrt{\kappa}}$ and $C_0\geq 16C_1$. We then get
$$
\ln e_m^T \lsb H^{i}\lb H^{i}\rb^T \rsb^a U^{\star}\rn_2 \leq v^{2a}\cdot\sqrt{\frac{\mu r}{n}}.
$$

After validating the assumptions of   Lemma~\ref{lem:L_infinity}, one can now obtain 
\begin{align*}
    \ln L^{1,i}-L^{\star}\rn_{\infty}\leq &\frac{\mu r}{n}\lb 5\ln H^{i}\rn_2+24v\rb \\
    \leq & \frac{\mu r}{n}\lsb\lb \frac{10}{C_0}\frac{\sigma_r^{\star}}{\sqrt{\kappa}}\gamma+40C_{\mathrm{noise}}\sigma\rb+\lb\frac{48C_1}{C_0}\sigma_r^{\star}\gamma+96\sqrt{\kappa}C_1C_{\mathrm{noise}}\sigma\rb\rsb \\
    \leq & \lb\frac{\mu r}{n}\sigma_r^{\star}\rb\gamma+100C_1C_{\mathrm{noise}}\sigma\frac{\sqrt{\kappa}\mu r}{n},
\end{align*}
where the last inequality holds if $C_1\geq 10$ and $C_0\geq 49C_1$.

\section{Proof of Lemmas in Section \ref{sec:induction}}\label{sec:proofs_induction}

\subsection{Proof of Lemma~\ref{lem:op_induct}}\label{proof:lem:op_induct}
For $\ln E_1^{t,i} \rn_2$,

\begin{equation}\label{eq:S_bound}
\begin{aligned}
    \left\|E_1^{t,i}\right\|_2 = &\left\|p^{-1}\Pti\mathcal{P}_\Omega^{(-i)} \lb S^{t,i}-S^\star\rb \right\|_2\\
    \leq & p^{-1}\cdot \sqrt{\frac{4}{3}}\left\|\mathcal{P}_\Omega^{(-i)}\lb S^{t,i}-S^\star\rb \right\|_2\\
    \leq &\sqrt{\frac{4}{3}} \cdot p^{-1} \cdot (2\alpha p n) 
    \cdot C_{\mathrm{thresh}}\lsb\lb\frac{\mu r}{n}\sigma_{1}^{\star}\rb\gamma^t+2C_N\sigma\sqrt{\log n}\rsb\\
    \leq & \frac{1}{2C_0} \frac{\sigma_r^\star}{\kappa}\gamma^{t+1}+
    5C_{N}\lb C_{\mathrm{thresh}}\alpha n\sqrt{\log n}\rb\sigma,
\end{aligned}
\end{equation}
where the first inequality follows from Lemma~\ref{lem:T_op} and the last inequality holds if
$
\alpha \leq \frac{1}{5 C_{0}} \frac{1}{\kappa^{2} \mu r}\cdot \frac{\gamma}{C_{\text {thresh }}}.
$

For $\ln E_2^{t,i} \rn_2$, one has
\begin{align*}
     \left\| E_{2}^{t, i} \right\|_2
    = & \left\|\left(\mathcal{I}-\Pti\left( \I-\Ho^{(-i)}  \right)\right) \lb L^{t,i}-L^\star\rb\right\|_2 \\
    = & \underbrace{\left\|\left(\mathcal{I}-\Pti\right)\lb L^{t,i}-L^\star\rb\right\|_2}_{\psi_1} 
    + \underbrace{\left\| \Pti\mathcal{H}_\Omega^{(-i)}\lb L^{t,i}-L^\star\rb\right\|_2}_{\psi_2}.
\end{align*}
By Lemma~\ref{lem:T_diff},
\begin{align*}
    \psi_1 \leq \frac{\ln L^{t,i}-L^\star\rn_2^2}{\sigma_r^{\star}}
    \leq &2\ln E^{t-1,i}\rn_2\cdot\frac{2\ln E^{t-1,i}\rn_2}{\sigma_r^{\star}} \\
    \leq & \lb\frac{2}{C_0} 
    \frac{\sigma_r^\star}{\sqrt{\kappa}} \gamma^{t}+12C_{\mathrm{noise}}\sigma\rb\cdot\frac{4}{C_0}\frac{\gamma}{\sqrt{\kappa}} \\
    \leq &\frac{1}{4C_0} \frac{\sigma_r^\star}{\kappa} \gamma^{t+1}+\frac12 C_{\mathrm{noise}}\sigma,
\end{align*}
where the bound $\ln E^{t-1,\infty}\rn_2 \leq \frac{2}{C_0}\frac{\sigma_r^{\star}}{\sqrt{\kappa}}\gamma$ is used in the third inequality, and the last inequality holds provided that $C_0\geq 96$. On the other hand,
\begin{equation}\label{eq:psi_2}
\begin{aligned} 
    \psi_2 
    \leq & \sqrt{\frac{4}{3}}\cdot \left\|\mathcal{H}_\Omega^{(-i)}(L^{t,i}-L^\star)\right\|_2\\
    \leq & \sqrt{\frac{4}{3}}\cdot\ln\Ho\lb L^{t,i}-L^{\star}\rb\rn_2\\
    \leq & \sqrt{\frac{4}{3}}\cdot 2C_1\sqrt{\frac{n\log n}{p}}\lsb\lb\frac{\mu r}{n}\sigma_r^{\star}\rb\gamma^t+100C_1C_{\mathrm{noise}}\sigma\frac{\sqrt{\kappa}\mu r}{n} \rsb\\
    \leq &\frac{1}{4C_0}\frac{\sigma_{r}^{\star}}{\kappa}\gamma^{t+1}+\frac12 C_{\mathrm{noise}}\sigma,
\end{aligned}
\end{equation}
where the first inequality is due to Lemma~\ref{lem:T_op} again; the third inequality follows from Lemma~\ref{lem:Ho} and applying Lemma~\ref{lem:bernstein} to get, with high probability, 
$$
\ln\Ho\lb\bm{1}\bm{1}^T\rb\rn_2\leq C_1\cdot\lb \sqrt{\frac{n\log n}{p}}+\frac{\log n}{p}\rb\leq 2C_1\sqrt{\frac{n\log n}{p}}
$$
if $p\geq \frac{\log n}{n}$; and the last inequality in \eqref{eq:psi_2} holds provided that
$
p\geq\frac{86C_1^2C_0^2}{\gamma^2}\cdot\frac{\kappa^2\mu^2r^2\log n}{n}
$
and $C_0\geq 50C_1$.

For $\ln E_3^{t,i} \rn_2$,
\begin{align*}
    \ln E_3^{t,i} \rn_2 &= \ln p^{-1}\Pti\Po^{(-i)}\lb N\rb\rn_2
   \leq p^{-1}\cdot \sqrt{\frac{4}{3}}\ln \Po\lb N\rb\rn_2
   \leq \sqrt{\frac{4}{3}}\lb C_N\sqrt{\frac{n}{p}}\rb\sigma,
\end{align*}
where the last inequality follows from Lemma~\ref{lem:noise}. Therefore,
\begin{align*}
\ln E^{t,i} \rn_2 \leq \ln E_1^{t,i} \rn_2+\ln E_2^{t,i} \rn_2+ \ln E_3^{t,i} \rn_2\leq \frac{1}{C_0}\frac{\sigma_{r}^{\star}}{\kappa}\gamma^{t+1}+6C_{\mathrm{noise}}\sigma.
\end{align*}

\subsection{Proof of Lemma~\ref{lem:2_infty_induction}}\label{sec:2_infty_induction} 
\noindent\textbf{Part \RomanNumeralCaps{1}: Bound for $\ln E_1^{t,i}\rn_{2,\infty}$} For $1\leq m\leq n$, consider $\left\|e_{m}^{T} E_{1}^{t, i} \right\|_2$.
\begin{align*}
    \left\|e_{m}^{T} E_{1}^{t, i} \right\|_2
    = & p^{-1} \left\|e_{m}^{T}\Pti \mathcal{P}_\Omega^{(-i)}\lb S^{t,i}-S^\star\rb\right\|_2 \\
    \leq & p^{-1} \left\|e_{m}^{T} U^{t,i}\lb U^{t,i} \rb^T \mathcal{P}_\Omega^{(-i)}\lb S^{t,i}-S^\star\rb \lb I-V^{t,i}\lb V^{t,i} \rb^T\rb\right\|_2\\
    & + p^{-1} \left\|e_{m}^{T} \mathcal{P}_\Omega^{(-i)}(S^{t,i}-S^\star)V^{t,i}\lb V^{t,i} \rb^T\right\|_2\\
    \leq & \left\|U^{t,i}\right\|_{2,\infty}\cdot p^{-1} \left\|\mathcal{P}_\Omega^{(-i)}\lb S^{t,i}-S^\star\rb\right\|_2
    + p^{-1} \left\|e_{m}^{T}  \mathcal{P}_\Omega^{(-i)}\lb S^{t,i}-S^\star\rb V^{t,i}\right\|_2\\
    \leq & \lb 1+\frac{6C_1}{C_0}\rb\sqrt{\frac{\kappa\mu r}{n}}
    \cdot p^{-1}\cdot\lb 2\alpha p n\rb\cdot C_{\mathrm{thresh}}\lsb \lb\frac{\mu r}{n}\sigma_{1}^{\star}\rb\gamma^t+2C_N\sigma\sqrt{\log n}\rsb\\
    & + p^{-1} \cdot (2\alpha p n)\cdot C_{\mathrm{thresh}}\lsb \lb\frac{\mu r}{n}\sigma_{1}^{\star}\rb\gamma^t+2C_N\sigma\sqrt{\log n}\rsb \cdot \lb 1+\frac{6C_1}{C_0}\rb\sqrt{\frac{\kappa\mu r}{n}}\\
    \leq & 5\alpha n \cdot C_{\mathrm{thresh}} \left(\frac{\mu r}{n} \sigma_1^\star \right)\gamma^t\cdot\sqrt{\frac{\kappa\mu r}{n}}+\lb 9C_{\mathrm{thresh}}C_N\alpha n\sqrt{\log n}\rb\sigma\cdot\sqrt{\frac{\kappa\mu r}{n}}\\
    \leq & \frac{1}{C_0}\left(\sigma_r^\star 
    \sqrt{\frac{\mu r}{n}}\right)\gamma^{t+1}+9C_N\lb C_{\mathrm{thresh}}\alpha n\sqrt{\log n}\rb\sigma\sqrt{\frac{\kappa\mu r}{n}},
\end{align*}
where in the third inequality the bound for $\left\|U^{t,i}\right\|_{2,\infty}$ follows since $\ln \De^{t,\infty}\rn_{2,\infty}\leq \frac{6C_1}{C_0}\sqrt{\frac{\kappa\mu r}{n}}\gamma$, the fourth inequality holds if $C_0\geq 48C_1$, and the last inequality holds provided that 
$
\alpha \leq \frac{1}{5 C_{0}} \frac{1}{\kappa^{1.5} \mu r} \cdot \frac{\gamma}{C_{\text {thresh }}}.
$ 

\noindent\textbf{Part \RomanNumeralCaps{2}: Bound for $\ln E_2^{t,i}\rn_{2,\infty}$} First note that for $1\leq m\leq n$,
\begin{align*}
     \left\|e_{m}^{T} E_{2}^{t, i} \right\|_2
    = & {\left\|e_{m}^{T}\left(\mathcal{I}-\Pti\left( \I-\Ho^{(-i)}  \right)\right)(L^{t,i}-L^\star)\right\|_2} \\
    \leq & {\left\|e_{m}^{T}\left(\mathcal{I}-\Pti\right)\lb L^{t,i}-L^\star\rb \right\|_2 
    +  \left\|e_{m}^{T} \Pti\mathcal{H}_\Omega^{(-i)}\lb L^{t,i}-L^\star\rb\right\|_2}\\
    = & \left\|e_{m}^{T}\lb I-U^{t,i}\lb U^{t,i} \rb^T\rb\lb L^{t,i}-L^\star\rb\lb I-V^{t,i}\lb V^{t,i} \rb^T\rb \right\|_2 \\
    &+\left\|e_{m}^{T} \Pti\mathcal{H}_\Omega^{(-i)}\lb L^{t,i}-L^\star\rb\right\|_2\\
    \leq & \left\|e_{m}^T\lb U^\star \lb U^\star \rb^T-U^{t,i}\lb U^{t,i} \rb^T\rb \lb L^{t,i}-L^\star\rb \lb I-V^{t,i}\lb V^{t,i} \rb^T\rb\right\|_2\\
    &+\left\|e_m^T U^{t,i}\lb U^{t,i} \rb^T\mathcal{H}_\Omega^{(-i)}\lb L^{t,i}-L^\star\rb\lb I-V^{t,i}\lb V^{t,i} \rb^T\rb \right\|_2 \\
    & + \left\| e_m^T \mathcal{H}_\Omega^{(-i)}\lb L^{t,i}-L^\star\rb V^{t,i}\lb V^{t,i} \rb^T \right\|_2\\
    \leq& {\underbrace{\left\|U^{t,i}\lb U^{t,i} \rb^T-U^\star \lb U^\star \rb^T\right\|_{2,\infty}\cdot\left\|L^{t,i}-L^\star\right\|_2}_{\beta_1}}\\
    &+ {\underbrace{\left\|U^{t,i}\rn_{2,\infty}\cdot\ln\mathcal{H}_\Omega^{(-i)}\lb L^{t,i}-L^\star\rb\right\|_2}_{\beta_2}}+ \underbrace{\left\| e_m^T \mathcal{H}_\Omega^{(-i)}(L^{t,i}-L^\star)V^{t,i}\right\|_2}_{\beta_3}.    
\end{align*}

\noindent\textbf{$\bullet$ Bounding $\beta_1$.}  One has
\begin{align*}
    \left\|U^{t,i}\lb U^{t,i} \rb^T-U^\star \lb U^\star \rb^T\right\|_{2,\infty}
    = &\left\|\lb U^{t,i}-U^\star G^{t,i}\rb\lb U^{t,i} \rb^T+U^\star G^{t,i}\lb U^{t,i}-U^\star G^{t,i}\rb^T\right\|_{2,\infty}\\
    \leq &\left\|\Delta^{t,\infty}\right\|_{2,\infty} 
    + \sqrt{\frac{\mu r}{n}} \cdot \left\| U^{t,i}-U^\star G^{t,i} \right\|_2\\
    \leq &\frac{6C_1}{C_0}\sqrt{\frac{\kappa\mu r}{n}}\gamma+\frac{8\sqrt{2}}{C_0}\sqrt{\frac{\mu r}{n}}\gamma\leq \frac{7C_1}{C_0}\sqrt{\frac{\kappa\mu r}{n}}\gamma,
\end{align*}
where the second inequality follow from Lemma~\ref{lem:op}, since
$$
\left\| U^{t,i}-U^\star G^{t,i} \right\|_2 \leq \left\| F^{t,i}-F^\star G^{t,i} \right\|_2 \leq \frac{4\sqrt{2}}{\sigma_r^\star}\left\|E^{t-1,i}\right\|_2,
$$
$\ln E^{t-1,\infty}\rn_2 \leq \frac{2}{C_0}\frac{\sigma_r^{\star}}{\sqrt{\kappa}}\gamma$, and the last inequality holds if $C_1\geq 12$. Consequently,
\begin{align*}
    \beta_1
    \leq & \frac{7C_1}{C_0}\sqrt{\frac{\kappa\mu r }{n}}\gamma
    \cdot \lb\frac{2}{C_0}\frac{\sigma_r^{\star}}{\sqrt{\kappa}}\gamma^{t}+12C_{\mathrm{noise}}\sigma\rb \leq \frac{1}{C_0}\left(\sigma_r^\star\sqrt{\frac{\mu r }{n}}\right)\gamma^{t+1}+C_{\mathrm{noise}}\sigma\sqrt{\frac{\kappa\mu r }{n}},
\end{align*}
where the last inequality holds provided that $C_0\geq 84C_1$.

\noindent\textbf{$\bullet$ Bounding $\beta_2$.} One has
\begin{align*}
    \beta_2
    \leq & \lb 1+\frac{6C_1}{C_0}\rb\sqrt{\frac{\kappa\mu r}{n}}
   \cdot\lb\frac{1}{4C_0}\frac{\sigma_{r}^{\star}}{\kappa}\gamma^{t+1}+\frac12 C_{\mathrm{noise}}\sigma\rb \\
   \leq &\frac{1}{C_0}\left(\sigma_r^\star\sqrt{\frac{\mu r }{n}}\right)\gamma^{t+1}+C_{\mathrm{noise}}\sigma\sqrt{\frac{\kappa\mu r}{n}},
\end{align*}
where the bound $\left\|\mathcal{H}_\Omega^{(-i)}(L^{t,i}-L^\star)\right\|_2
    \leq \frac{1}{4C_0}\frac{\sigma_{r}^{\star}}{\kappa}\gamma^{t+1}+\frac12 C_{\mathrm{noise}}\sigma$ (see \eqref{eq:psi_2}) in deriving Lemma~\ref{lem:op_induct} is used in the first inequality.

\noindent\textbf{$\bullet$ Bounding $\beta_3$.} Let $Q\in\mathbb{R}^{r\times r}$ be a rotation matrix to be specified later.
\begin{align*}
e_{m}^{T}\mathcal{H}_\Omega^{(-i)}\lb L^{t,i}-L^\star\rb V^{t, i}
= & e_{m}^{T} \mathcal{H}_{\Omega}^{(-i)}\left(L^{t, m}-L^{\star}\right) V^{t, m} Q
+e_{m}^{T} \mathcal{H}_{\Omega}^{(-i)}\left(L^{t, i}-L^{t, m}\right) V^{t, m} Q \\
& +e_{m}^{T} \mathcal{H}_\Omega^{(-i)}\lb L^{t,i}-L^\star\rb\left(V^{t, i}-V^{t, m} Q\right).
\end{align*}
We have
\begin{align*}
\beta_3 \leq&\ 
\underbrace{\left\|e_{m}^{T} \mathcal{H}_{\Omega}^{(-i)}\left(L^{t, m}-L^{\star}\right) V^{t, m} \right\|_2}_{\beta_{3,1}}
+\underbrace{\left\|e_{m}^{T} \mathcal{H}_{\Omega}^{(-i)}\left(L^{t, i}-L^{t, m}\right) V^{t, m} \right\|_2}_{\beta_{3,2}} \\
& +\underbrace{\left\|e_{m}^{T} \mathcal{H}_\Omega^{(-i)}\lb L^{t,i}-L^\star\rb\left(V^{t, i}-V^{t, m} Q\right)\right\|_2}_{\beta_{3,3}} .
\end{align*}
In the following, we consider the case $0\leq i\leq n$ and the proof can be done similarly when $i>n$.

\noindent\textbf{$-$ Bounding $\beta_{3,1}$.} For $k=1,\cdots,n$, define
$$
v_k = \lb 1-\delta_{mk}/p\rb \lb L_{mk}^{t, m}-L_{mk}^{\star}\rb V_{k,:}^{t,m}.
$$
Since $L^{t,m}$ and $V^{t,m}$ are independent with respect to the Bernoulli variables on the $m$-th row, applying Lemma~\ref{lem:bernstein},  we can get
\begin{equation}\label{eq:beta_31}
\begin{aligned}
    & \beta_{3,1} = \ln \sum_{k=1}^n v_k\rn_2\\
    \leq & C_1\cdot\lb\sqrt{\frac{\log n}{p}}\ln L^{t, m}- L^{\star} \rn_{2,\infty}+\frac{\log n}{p}\ln L^{t, m}-L^{\star} \rn_{\infty}\rb\cdot\ln V^{t,m}\rn_{2,\infty} \\
    \leq & C_1\cdot\lb\sqrt{\frac{n\log n}{p}}+\frac{\log n}{p}\rb\ln L^{t, m}-L^{\star} \rn_{\infty}\cdot\lb 1+\frac{6C_1}{C_0} \rb\sqrt{\frac{\kappa\mu r}{n}} \\
    \leq & C_1\cdot 2\sqrt{\frac{n\log n}{p}}\cdot\lsb\lb\frac{\mu r}{n}\sigma_r^{\star}\rb\gamma^t+100C_1C_{\mathrm{noise}}\sigma\frac{\sqrt{\kappa}\mu r}{n}\rsb\cdot \lb 1+\frac{6C_1}{C_0} \rb\sqrt{\frac{\kappa\mu r}{n}}  \\
    \leq & \frac{C_1}{C_0}\lb\sigma_r^{\star}\sqrt{\frac{\mu r}{n}}\rb\gamma^{t+1}+C_1C_{\mathrm{noise}}\sigma\sqrt{\frac{\kappa\mu r}{n}},
\end{aligned}
\end{equation}
where the last two inequalities hold provided that 
$
p\geq\frac{9C_{0}^2}{\gamma^2}\cdot\frac{\kappa\mu^2r^2\log n}{n}
$
and $C_0\geq 100C_1$.

\noindent\textbf{$-$ Bounding $\beta_{3,2}$.}  Note that
\begin{equation}\label{eq:L_diff_decomp}
\begin{aligned}  
L^{t, m}-L^{t, i} 
= & U^{t, m} \Si^{t, m}\lb V^{t, m}\rb^T-U^{t, i} \Si^{t, i} \lb V^{t, i}\rb^T \\ 
= & D_U^{t, m, i} \Si^{t, m} \lb V^{t, m}\rb^T+U^{t,i} S^{t, m,i}\lb V^{t, m}\rb^T - U^{t, i} \Si^{t, i} \lb D_V^{t, i, m}\rb^T,
\end{aligned}
\end{equation}
where
$D_U^{t,m,i} := U^{t,m}-U^{t,i}G^{t,m,i}$, $
S^{t,m,i} := G^{t,m,i}\Sigma^{t,m}-\Sigma^{t,i}G^{t,m,i}$ and
$D_V^{t,i,m} := V^{t,i}-V^{t,m}G^{t,i,m}= V^{t,i}-V^{t,m}\lb G^{t,m,i}\rb^T$.
Therefore,
\begin{align*}  
& \beta_{3,2} = \left\|e_{m}^{T} \mathcal{H}_{\Omega}^{(-i)}\left(L^{t, i}-L^{t, m}\right) V^{t, m} \right\|_2\\
\leq &\underbrace{\ln e_m^T\Ho^{(-i)}\lb D_U^{t, m, i} \Si^{t, m} \lb V^{t, m}\rb^T\rb V^{t,m}\rn_2}_{\beta_{3,2}^a}+\underbrace{\ln e_m^T\Ho^{(-i)}\lb U^{t,i} S^{t,m,i}\lb V^{t,m}\rb^T\rb V^{t,m}\rn_2}_{\beta_{3,2}^b}\\
&+\underbrace{\ln e_m^T\Ho^{(-i)}\lb U^{t, i} \Si^{t, i} \lb D_V^{t, i, m}\rb^T\rb V^{t,m}\rn_2}_{\beta_{3,2}^c}.
\end{align*}

Applying Lemma~\ref{lem:bound1} with
$A=D_U^{t, m, i} \Si^{t, m}$, $B=V^{t,m}$ and $C=V^{t,m}$, we get
\begin{align*} 
    \beta_{3,2}^a \leq \ln D_U^{t, m, i} \Si^{t, m} \rn_{2,\infty}\cdot \ln R^{(m)}\rn_2,
\end{align*}
where
$
R^{(m)} := \sum_{k=1}^n \lb 1- \delta_{mk}/p\rb \lb V^{t,m}_{k,:}\rb^T V^{t,m}_{k,:}:= \sum_{k=1}^n R_{k}^{(m)}.
$
Since
\begin{align*}
    \ln R_{k}^{(m)} \rn_2\leq & \frac1p \ln V^{t,m} \rn_{2,\infty}^2, \\
    \ln \sum_{k=1}^n\mathbb{E} \lsb \lb R_k^{(m)}\rb^T R_k^{(m)}\rsb\rn_2 \leq & \frac1p\ln\sum_{k=1}^n \lsb\ln V_{k,:}^{t,m}\rn_2^2\cdot\lb V_{k,:}^{t,m} \rb^T V_{k,:}^{t,m} \rsb\rn_2
    \leq \frac1p\ln V^{t,m} \rn_{2,\infty}^2,
\end{align*}
we can apply Lemma~\ref{lem:bernstein} to bound $\ln R^{(m)}\rn_2$, which requires separate discussion for the cases $t=1$ and $t\geq 2$, since $\ln V^{1,m}\rn_2=O(\sqrt{\frac{\kappa \mu r}{n}}),~\ln D_U^{1,m,i}\rn_2=O(\sqrt{\frac{\mu r}{n}})$ when $t= 1$ and $\ln V^{t,m}\rn_2=O(\sqrt{\frac{\mu r}{n}}),~\ln D_U^{t,m,i}\rn_2=O(\sqrt{\frac{\kappa \mu r}{n}})$ when $t\geq 2$. 
First consider the case $t=1$.
\begin{equation}\label{eq:bernstein_t=1}
\begin{aligned}
    \ln  R^{(m)} \rn_2 \leq & C_1\cdot\lb\sqrt{\frac{\log n}{p}}\ln V^{1,m} \rn_{2,\infty}+\frac{\log n}{p}\ln V^{1,m} \rn_{2,\infty}^2\rb\\
    \leq & 2C_1\cdot\lsb\sqrt{\frac{\log n}{p}}\cdot\lb 1+\frac{6C_1}{C_0}\rb\sqrt{\frac{\kappa\mu r}{n}}\rsb\leq \frac{\gamma}{25\kappa},
\end{aligned}
\end{equation}
where the last two inequalities hold if $C_0\geq 30C_1$ and
$
p\geq\frac{60^2C_1^2}{\gamma^2}\cdot\frac{\kappa^3\mu r\log n}{n}
$. Hence,
\begin{align*} 
    \beta_{3,2}^a
    \leq &\lsb\frac{8C_1}{C_0}\lb\sqrt{\frac{\mu r}{n}}\rb\gamma+12C_1C_{\mathrm{noise}}\lb\frac{\sigma}{\sigma_r^{\star}}\rb\sqrt{\frac{\kappa\mu r}{n}}\rsb\cdot\lb1+\frac{2}{C_0}\rb\sigma_1^{\star}\cdot\frac{\gamma}{25\kappa} \\
    \leq &\frac{C_1}{2C_0}\lb \sigma_r^{\star}\sqrt{\frac{\mu r}{n}}\rb\gamma^{2}+\frac12 C_1C_{\mathrm{noise}}\sigma\sqrt{\frac{\kappa\mu r}{n}},
\end{align*}
where the last inequality holds if $C_0\geq 48$. Then consider the case $t\geq 2$.
\begin{equation*}
\begin{aligned}
    \ln  R^{(m)} \rn_2
    \leq & 2C_1\cdot\lsb\sqrt{\frac{\log n}{p}}\cdot\lb 1+\frac{6C_1}{C_0}\rb\sqrt{\frac{\mu r}{n}}\rsb\leq \frac{\gamma}{50\kappa^{1.5}},
\end{aligned}
\end{equation*}
where the tight bound for $\ln V^{t,m}\rn_{2,\infty}$ follows from \eqref{eq:l_2_infty} and \eqref{eq:noise_bound2}, and the inequalities hold if 
$
p\geq\frac{120^2C_1^2}{\gamma^2}\cdot\frac{\kappa^3\mu r\log n}{n}.
$ Hence,
\begin{align*} 
    \beta_{3,2}^a
    \leq &\lsb\frac{8C_1}{C_0}\sqrt{\frac{\kappa\mu r}{n}}\gamma^{t}+24C_1C_{\mathrm{noise}}\lb\frac{\sigma}{\sigma_r^{\star}}\rb\sqrt{\frac{\kappa^2\mu r}{n}}\rsb\cdot\lb1+\frac{2}{C_0}\rb\sigma_1^{\star}\cdot\frac{\gamma}{50\kappa^{1.5}} \\
    \leq &\frac{C_1}{2C_0}\lb \sigma_r^{\star}\sqrt{\frac{\mu r}{n}}\rb\gamma^{t+1}+\frac12C_1 C_{\mathrm{noise}}\sigma\sqrt{\frac{\kappa\mu r}{n}}.
\end{align*}

The term  $\beta_{3,2}^b$ can be bounded similarly as $\beta_{3,2}^a$ by replacing $\ln D_U^{t, m, i} \Si^{t, m} \rn_{2,\infty}$ with $\ln U^{t,i} S^{t,m,i} \rn_{2,\infty}$, and thus we only need to bound $\ln S^{t,m,i} \rn_2$. Let $A=\widehat{L^{\star}}+\widehat{E^{t-1,m}}$ and $\widetilde{A}=\widehat{L^{\star}}+\widehat{E^{t-1,i}}$. The application of Lemma~\ref{lem:perturb_S} yields
\begin{align*}
\ln S^{t,m,i} \rn_2=&\ln \Sigma^{t,m}G^{t,i,m}-G^{t,i,m}\Sigma^{t,i}\rn_2\\
\leq & 4\kappa\cdot\ln E^{t-1,i}-E^{t-1,m}\rn_2
\leq \frac{8}{C_0}\frac{\sigma_{1}^{\star}}{\sqrt{\kappa}}\gamma^{t}+48\kappa C_{\mathrm{noise}}\sigma.
\end{align*}
With the same requirement for $p$ in bounding $\beta_{3,2}^a$, for $t\geq 1$ we can get
\begin{align*}
    \beta_{3,2}^b\leq\frac{1}{C_0}\lb \sigma_r^{\star}\sqrt{\frac{\mu r}{n}}\rb\gamma^{t+1}+2C_{\mathrm{noise}}\sigma\sqrt{\frac{\kappa\mu r}{n}}.
\end{align*}

{
For $\beta_{3,2}^c$, applying Lemma \ref{lem:bound1} again yields
\begin{align*} 
    \beta_{3,2}^c \leq \ln U^{t,i} \Si^{t,i} \rn_{2,\infty}\cdot \ln \lb D_V^{t,i,m}\rb^T H^{(m)}V^{t,m}\rn_2.
\end{align*}
Since $V^{t, m}$ is independent with respect to the random variables on the $m$-th row, the spectral norm of
$
H^{(m)}V^{t,m} = \sum_{k=1}^n (1-\delta_{mk}/p)\cdot e_k V^{t,m}_{k,:}
$
can be bounded by Lemma~\ref{lem:bernstein}, and with high probability we can get
\begin{equation}\label{eq:HV_2norm}
\begin{aligned}
\ln H^{(m)}V^{t,m} \rn_2 \leq & C_1\cdot\lb\sqrt{\frac{\max\{1,\ln V^{t,m}\rn_{2,\infty}^2\}\cdot\log n}{p}}+\frac{\ln V^{t,m}\rn_{2,\infty}\cdot\log n}{p}\rb \\
\leq & C_1\cdot\lsb\sqrt{\frac{r\log n}{p}}+\lb 1+\frac{6C_1}{C_0}\rb \sqrt{\frac{\kappa\mu r}{n}}\cdot\frac{\log n}{p}\rsb \\
\leq & C_1\cdot 3\sqrt{\frac{r\log n}{p}}
\leq \frac{\gamma}{60}\sqrt{\frac{n}{\kappa^3\mu r}},
\end{aligned}
\end{equation}
where the last two inequalities hold if $C_0\geq 6C_1$ and
$
p\geq \frac{180^2C_1^2}{\gamma^2}\cdot\frac{\kappa^3\mu r^2\log n}{n}.
$
Provided that $C_0\geq 30C_1$ and $C_0\geq 48$, when $t=1$, 
\begin{align*}
\beta_{3,2}^c\leq &\lb 1+\frac{6C_1}{C_0}\rb\sqrt{\frac{\kappa\mu r}{n}}\lb 1+\frac{2}{C_0}\rb\sigma_1^{\star}\cdot\lsb\frac{8C_1}{C_0}\sqrt{\frac{\mu r}{n}}\gamma+12C_1C_{\mathrm{noise}}\lb\frac{\sigma}{\sigma_r^{\star}}\rb\sqrt{\frac{\kappa\mu r}{n}}\rsb\cdot\frac{\gamma}{60}\sqrt{\frac{n}{\kappa^3\mu r}} \\
\leq &\frac{C_1}{2C_0}\lb\sigma_r^{\star}\sqrt{\frac{\mu r}{n}}\rb\gamma^{2}+\frac12 C_1C_{\mathrm{noise}}\sigma\sqrt{\frac{\kappa\mu r}{n}};
\end{align*}
and when $t\geq 2$,
\begin{align*}
\beta_{3,2}^c\leq &\lb 1+\frac{6C_1}{C_0}\rb\sqrt{\frac{\mu r}{n}}\lb 1+\frac{2}{C_0}\rb\sigma_1^{\star}\cdot\lsb\frac{8C_1}{C_0}\sqrt{\frac{\kappa\mu r}{n}}\gamma^t+24C_1C_{\mathrm{noise}}\lb\frac{\sigma}{\sigma_r^{\star}}\rb\sqrt{\frac{\kappa^2\mu r}{n}}\rsb\cdot\frac{\gamma}{60}\sqrt{\frac{n}{\kappa^3\mu r}} \\
\leq &\frac{C_1}{2C_0}\lb \sigma_r^{\star}\sqrt{\frac{\mu r}{n}}\rb\gamma^{t+1}+\frac12 C_1C_{\mathrm{noise}}\sigma\sqrt{\frac{\kappa\mu r}{n}}.
\end{align*}}

Combining the bounds of $\beta_{3,2}^a$ to $\beta_{3,2}^c$, we have
\[
\beta_{3,2}
\leq \frac{1+C_1}{C_{0}}\left(\sigma_{r}^{\star} \sqrt{\frac{\mu r}{n}}\right)\gamma^{t+1}+\lb 2+C_1\rb C_{\mathrm{noise}}\sigma\sqrt{\frac{\kappa\mu r}{n}}.
\]

\noindent\textbf{$-$ Bounding $\beta_{3,3}$.} Choose $Q = G^{t,i,m}$ and we get
\begin{align*}
    \beta_{3,3}
    \leq & \ln \Ho^{(-i)}\lb L^{t,i}-L^{\star}\rb\rn_2\cdot\ln D_V^{t,i,m}\rn_{\mathrm{F}} \\
    \leq & \frac{1}{2C_0}\frac{\sigma_r^{\star}}{\sqrt{\kappa}}\gamma\cdot \lsb\frac{8C_1}{C_0}\sqrt{\frac{\kappa\mu r}{n}}\gamma^{t}+24C_1C_{\mathrm{noise}}\lb\frac{\sigma}{\sigma_r^{\star}}\rb\sqrt{\frac{\kappa^2\mu r}{n}}\rsb\\
    \leq & \frac{1}{C_0}\lb \sigma_r^{\star}\sqrt{\frac{\mu r}{n}}\rb\gamma^{t+1}+C_{\mathrm{noise}}\sigma\sqrt{\frac{\kappa\mu r}{n}},
\end{align*}
where in the second inequality the loose bound of $\ln \Ho^{(-i)}\lb L^{t,i}-L^{\star}\rb\rn_2$ follows from \eqref{eq:psi_2} and \eqref{eq:noise_bound2}.

Combining the bounds for $\beta_{3,1}$ to $\beta_{3,3}$, one has
\begin{align*}
    \beta_3
    \leq & \frac{2+2C_1}{C_0}\lb\sigma_r^{\star}\sqrt{\frac{\mu r}{n}}\rb\gamma^{t+1}+\lb 3+2C_1\rb C_{\mathrm{noise}}\sigma\sqrt{\frac{\kappa\mu r}{n}}.
\end{align*}
Hence for $1\leq m\leq n$, 
\begin{align*}
    \ln e_m^TE_2^{t,i} \rn_2
    \leq & \beta_1+\beta_2+\beta_3\leq\frac{4+2C_1}{C_0}\lb \sigma_r^{\star}\sqrt{\frac{\mu r}{n}}\rb\gamma^{t+1}+\lb 5+2C_1\rb C_{\mathrm{noise}}\sigma\sqrt{\frac{\kappa\mu r}{n}}.
\end{align*}

\noindent\textbf{Part \RomanNumeralCaps{3}: Bound for $\ln E_3^{t,i}\rn_{2,\infty}$} For $1\leq m\leq n$,
\begin{align*}
    \left\|e_{m}^{T} E_{3}^{t, i} \right\|_2
    = & p^{-1} \left\|e_{m}^{T}\Pti \mathcal{P}_\Omega^{(-i)}\lb N\rb\right\|_2 \\
    \leq & p^{-1} \left\|e_{m}^{T} U^{t,i}\lb U^{t,i} \rb^T \mathcal{P}_\Omega^{(-i)}\lb N \rb \lb I-V^{t,i}\lb V^{t,i} \rb^T\rb\right\|_2+p^{-1} \left\|e_{m}^{T} \mathcal{P}_\Omega^{(-i)}(N)V^{t,i}\lb V^{t,i} \rb^T\right\|_2\\
    \leq & \left\|U^{t,i}\right\|_{2,\infty}\cdot p^{-1} \left\|\mathcal{P}_\Omega^{(-i)}\lb N \rb\right\|_2+ p^{-1} \left\|e_{m}^{T}  \mathcal{P}_\Omega^{(-i)}\lb N \rb V^{t,i}\right\|_2\\
    \leq &\left\|U^{t,i}\right\|_{2,\infty}\cdot p^{-1} \left\|\mathcal{P}_\Omega^{(-i)}\lb N \rb\right\|_2+p^{-1} \left\|e_{m}^{T}  \mathcal{P}_\Omega^{(-i)}\lb N \rb D_V^{t,i,m}\right\|_2+p^{-1} \left\|e_{m}^{T}  \mathcal{P}_\Omega^{(-i)}\lb N \rb V^{t,m}\right\|_2\\
    \leq & p^{-1} \left\|\mathcal{P}_\Omega^{(-i)}\lb N \rb\right\|_2\cdot\lb \left\|U^{t,i}\right\|_{2,\infty}+\ln D_V^{t,i,m}\rn_{\mathrm{F}}\rb+p^{-1} \left\|e_{m}^{T}  \mathcal{P}_\Omega^{(-i)}\lb N \rb V^{t,m}\right\|_2 \\
    \leq & \lb C_N\sqrt{\frac{n}{p}}\rb\sigma\cdot\lsb\lb 1+\frac{6C_1}{C_0}\rb\sqrt{\frac{\kappa\mu r}{n}}+\frac{9C_1}{C_0}\sqrt{\frac{\kappa\mu r}{n}}\rsb + C_1\lb C_N\sqrt{\frac{n\log n}{p}}\rb\sigma\cdot\lb 1+\frac{6C_1}{C_0}\rb\sqrt{\frac{\kappa\mu r}{n}}\\
    \leq & 2C_1\lb C_N\sqrt{\frac{n\log n}{p}}\rb\sigma\sqrt{\frac{\kappa\mu r}{n}},
\end{align*}
where in the third inequality $D_V^{t,i,m}:=V^{t,i}-V^{t,m}G^{t,i,m}$, the fifth inequality holds with high probability due to Lemma~\ref{lem:noise}, and the last inequality holds when $C_1\geq 2$.

Combining the bounds of $\ln E_1^{t,i}\rn_{2,\infty}$, $\ln E_2^{t,i}\rn_{2,\infty}$ and $\ln E_3^{t,i}\rn_{2,\infty}$, we get
\begin{equation*}
\begin{aligned}
    \ln E^{t,i} \rn_{2,\infty}
    \leq & \frac{5+2C_1}{C_0}\lb \sigma_r^{\star}\sqrt{\frac{\mu r}{n}}\rb\gamma^{t+1}+\lb 5+4C_1\rb C_{\text{noise}}\sigma\sqrt{\frac{\kappa\mu r}{n}}.
\end{aligned}
\end{equation*}

\subsection{Proof of Lemma~\ref{lem:D_bounds}}\label{proof:lem:D_bounds}

\noindent\textbf{Part \RomanNumeralCaps{1}: Bound for $\ln \zeta\rn_{\mathrm{F}}$.}
\begin{align*}
    & \zeta = p^{-1}\mathcal{P}_{T^{t,m}}\mathcal{P}_\Omega^{(-m)}\lb S^{t,m}-S^\star\rb 
    -p^{-1}\mathcal{P}_{T^t}\mathcal{P}_\Omega\lb S^t-S^\star\rb\\
    = & \underbrace{p^{-1}\mathcal{P}_{T^{t,m}}\mathcal{P}_\Omega^{(-m)}\lb S^{t,m}-S^t\rb}_{\zeta_1}
      + \underbrace{p^{-1}\left(\mathcal{P}_{T^{t,m}}-\mathcal{P}_{T^t}\right)\mathcal{P}_\Omega^{(-m)}\lb S^{t}-S^\star\rb}_{\zeta_2}+\underbrace{p^{-1}\mathcal{P}_{T^t}\left(\mathcal{P}_\Omega^{(-m)} - \mathcal{P}_\Omega\right)\lb S^t-S^\star\rb}_{\zeta_3}.
\end{align*}

\noindent\textbf{$\bullet$ Bounding $\zeta_{1}$.}
\begin{align*}
     \ln \zeta_1\rn_{\mathrm{F}}
    \leq & p^{-1}\left\|\lb U^{t,m} \rb^T\mathcal{P}_\Omega^{(-m)}\lb S^{t,m}-S^t\rb\right\|_{\mathrm{F}}+ p^{-1}\left\|\mathcal{P}_\Omega^{(-m)}\lb S^{t,m}-S^t\rb V^{t,m}\right\|_{\mathrm{F}}\\
    = & \underbrace{p^{-1}\left\|\lsb\mathcal{P}_\Omega^{(-m)}\lb S^{t,m}-S^t\rb\rsb^T U^{t,m}\right\|_{\mathrm{F}}}_{\zeta_{1,1}}+\underbrace{p^{-1}\left\|\mathcal{P}_\Omega^{(-m)}\lb S^{t,m}-S^t\rb V^{t,m}\right\|_{\mathrm{F}}}_{\zeta_{1,2}}.
\end{align*}
We only derive the bound for $\zeta_{1,2}$ in the following, since $\zeta_{1,1}$ has the same bound.

Let $\Omega_{S^{*}}^{(-m)}$ be $\Omega_{S^{*}}$ without the indices from the $m$-th row if $1 \leq m \leq n$, and $\Omega_{S^{\star}}$ without the indices from the $(m-n)$-th column if $n+1 \leq m \leq 2 n$. For $(i, j) \in[n] \times[n]$, define $w_{i j}=1$ if $(i, j) \in \Omega_{S^{\star}}^{(-m)}$ and $w_{i j}=0$ otherwise. Then,
\begin{align*}
\left\|\mathcal{P}_{\Omega}^{(-m)}\left(S^{t, m}-S^{t}\right) V^{t, m}\right\|_{\mathrm{F}}^{2}
= & \sum_{i=1}^{n} \sum_{j=1}^{r}\left[\sum_{k=1}^{n} w_{i k}\left(S^{t, m}-S^{t}\right)_{i k} V_{k j}^{t, m}\right]^{2} \\
\leq &\sum_{i=1}^{n} \sum_{j=1}^{r}\left[\sum_{k=1}^{n} w_{i k}\left(S^{t, m}-S^{t}\right)_{i k}^{2}\right]\left[\sum_{k=1}^{n} w_{i k}\left(V_{k j}^{t, m}\right)^{2}\right].
\end{align*}
Note that
\begin{align*}
\sum_{j=1}^r\sum_{k=1}^{n} w_{i k}\left(V_{k j}^{t, m}\right)^{2} =\sum_{k=1}^{n} w_{i k} \ln V_{k,:}^{t, m}\rn_2^2 \leq 2 \alpha p n \cdot\left\|V^{t, m}\right\|_{2, \infty}^{2}.
\end{align*}
As a result,
\begin{align*}
\left\|\mathcal{P}_{\Omega}^{(-m)}\left(S^{t, m}-S^{t}\right) V^{t, m}\right\|_{\mathrm{F}}^{2}
\leq &  \sum_{i=1}^{n}\left[\sum_{k=1}^{n} w_{i k}\left(S^{t, m}-S^{t}\right)_{i k}^{2}\right] \cdot \sum_{j=1}^{r}\left[\sum_{k=1}^{n} w_{i k}\left(V_{k j}^{t, m}\right)^{2}\right] \\
\leq&  \sum_{i=1}^{n}\left[\sum_{k=1}^{n} w_{i k}\left(S^{t, m}-S^{t}\right)_{i k}^{2}\right] \cdot\left(2 \alpha p n \left\|V^{t, m}\right\|_{2, \infty}^{2}\right) \\
= & \left\|\mathcal{P}_{\Omega_{S^{\star}}}^{(-m)}\left(S^{t, m}-S^{t}\right)\right\|_{\mathrm{F}}^{2} \cdot\left(2 \alpha p n \left\|V^{t, m}\right\|_{2, \infty}^{2}\right) .
\end{align*}
The term $\left\|\mathcal{P}_{\Omega_{S^{\star}}}^{(-m)}\left(S^{t, m}-S^{t}\right)\right\|_{\mathrm{F}}$ can be bounded as follows.
\begin{align*}
\left\|\mathcal{P}_{\Omega_{S^{\star}}}^{(-m)}\left(S^{t, m}-S^{t}\right)\right\|_{\mathrm{F}}
= & \left\|\mathcal{P}_{\Omega_{S^{\star}}^{(-m)}}\left(\mathcal{T}_{\xi^{t}}\left(M-L^{t, m}\right)-\mathcal{T}_{\xi^{t}}\left(M-L^{t}\right)\right)\right\|_{\mathrm{F}} \\
\leq & K \cdot\left\|\mathcal{P}_{\Omega_{S^{\star}}}^{(-m)}\left(\left(M-L^{t, m}\right)-\left(M-L^{t}\right)\right)\right\|_{\mathrm{F}} \\
\leq & K\left\|\mathcal{P}_{\Omega_{S^{\star}}}\left(L^{t, m}-L^{t}\right)\right\|_{\mathrm{F}} \\
\leq & K\left\|\mathcal{P}_{\Omega_{S^{\star}}}\left(D_{U}^{t, m, 0} \Sigma^{t, m}\left(V^{t, m}\right)^{T}\right)\right\|_{\mathrm{F}}+K\left\|\mathcal{P}_{\Omega_{S^{\star}}}\left(U^{t} S^{t, m, 0}\left(V^{t, m}\right)^{T}\right)\right\|_{\mathrm{F}} \\
& +K\left\|\mathcal{P}_{\Omega_{S^{\star}}}\left(U^{t} \Sigma^{t}\left(D_{V}^{t, 0, m}\right)^{T}\right)\right\|_{\mathrm{F}} \\
\leq & K \sqrt{2\alpha p n} \cdot \left\|D_{U}^{t, m, 0} \Sigma^{t, m}\right\|_{\mathrm{F}}\left\|V^{t, m}\right\|_{2,\infty} + K\sqrt{2\alpha p n} \cdot\left\|U^{t} S^{t, m, 0}\right\|_{\mathrm{F}}\left\|V^{t, m}\right\|_{2,\infty}\\
& + K\sqrt{2\alpha p n} \cdot \left\|U^{t}\right\|_{2,\infty}\left\| D_{V}^{t, 0, m} \Sigma^{t}\right\|_{\mathrm{F}} \\
\leq & K \sqrt{2\alpha p n} \cdot\lsb 2\left(1+\frac{2}{C_{0}}\right) \sigma_{1}^{\star}\left\|D^{t, m, 0}\right\|_{\mathrm{F}}+\left\|S^{t, m, 0}\right\|_{\mathrm{F}}\rsb
\cdot \left(1+\frac{6C_1}{C_0}\right)\sqrt{\frac{\kappa\mu r}{n}},
\end{align*}
where the first inequality follows from property \labelcref{P2} of the thresholding function (applied to each entry in $\Omega_{S^{\star}}^{(-m)}$), the fourth inequality follows from Lemma~\ref{lem:P_Omega_AB}, and in the last inequality $\ln E^{t-1,\infty}\rn_2 \leq \frac{2}{C_0}\frac{\sigma_r^{\star}}{\sqrt{\kappa}}\gamma$ is used again to bound $\ln\Sigma^{t, m} \rn_2$ and $\ln \Sigma^{t}\rn_2$.

Combining the above pieces together, we have
\begin{align*}
\zeta_{1,2}
& \leq p^{-1}\left\|\mathcal{P}_{\Omega_{S^{\star}}}^{(-m)}\left(S^{t, m}-S^{t}\right)\right\|_{\mathrm{F}} \cdot  \sqrt{2 \alpha p n} \left\|V^{t, m}\right\|_{2, \infty}  \\
& \leq 3K \cdot \lsb\left(2+\frac{4}{C_{0}}\right) \sigma_{1}^{\star}\left\|D^{t, m, 0}\right\|_{\mathrm{F}}+\left\|S^{t, m, 0}\right\|_{\mathrm{F}}\rsb \cdot\alpha\kappa\mu r\\
& \leq 12K\cdot\lsb\frac{8C_1}{ C_{0}} \left(\sigma_{1}^{\star} \sqrt{\frac{\kappa\mu r}{n}}\right) \gamma^{t}+24C_1 C_{\mathrm{noise}}\sigma\sqrt{\frac{\kappa^4\mu r}{n}}\rsb\cdot\alpha\kappa\mu r\\
& \leq \frac{C_1}{8C_{0}} \left(\sigma_{r}^{\star}\sqrt{\frac{\kappa\mu r}{n}}\right) \gamma^{t+1}+\frac18C_1C_{\mathrm{noise}}\sigma\sqrt{\frac{\kappa^2\mu r}{n}}, 
\end{align*}
where the second inequality holds if $C_0\geq 30C_1$, \eqref{eq:S_F} is used in the third inequality, and the last inequality holds if
$
\alpha \leq \frac{1}{\kappa^2 \mu r} \cdot \frac{\gamma}{2304 K}.
$

\noindent\textbf{$\bullet$ Bounding $\zeta_{2}$.} It is easy to verify that for any $Z\in\mathbb{R}^{n\times n}$,
\begin{equation}\label{eq:T_diff_decomp}
\begin{aligned}
    \left(\mathcal{P}_{T^t}-\mathcal{P}_{T^{t,m}}\right)\lb Z\rb
    = &\lb U^t\lb U^t \rb^T -U^{t,m}\lb U^{t,m} \rb^T \rb Z\lb I-V^t\lb V^t \rb^T \rb\\
    & + \lb I-U^{t,m}\lb U^{t,m} \rb^T \rb Z\lb V^t\lb V^t \rb^T -V^{t,m}\lb V^{t,m} \rb^T \rb.
\end{aligned}
\end{equation}
Furthermore,
\begin{align*}
    \left\|U^t\lb U^t \rb^T -U^{t,m}\lb U^{t,m} \rb^T \right\|_\mathrm{F}
    =&\left\|\lb U^t-U^{t,m}G^{t,0,m}\rb\lb U^t \rb^T +U^{t,m}G^{t,0,m} \lb U^t-U^{t,m}G^{t,0,m}\rb^T\right\|_\mathrm{F}\\
    \leq& 2\left\| U^t-U^{t,m}G^{t,0,m} \right\|_\mathrm{F}
    \leq 2\left\| D^{t,0,m} \right\|_\mathrm{F},
\end{align*}
and $\left\|V^t\lb V^t \rb^T -V^{t,m}\lb V^{t,m} \rb^T \right\|_\mathrm{F}$ has the same bound. As a result,
\begin{align*}
    \ln \zeta_{2} \rn_{\mathrm{F}}
    \leq & \left\|U^t\lb U^t \rb^T -U^{t,m}\lb U^{t,m} \rb^T \right\|_{\mathrm{F}} \cdot p^{-1}\left\|\mathcal{P}_\Omega^{(-m)}\lb S^{t}-S^\star\rb \right\|_2\\
    & +p^{-1}\left\|\mathcal{P}_\Omega^{(-m)}\lb S^{t}-S^\star\rb \right\|_2 \cdot \left\|V^t\lb V^t \rb^T -V^{t,m}\lb V^{t,m} \rb^T \right\|_{\mathrm{F}}\\
    \leq & 4\left\| D^{t,0,m} \right\|_\mathrm{F}\cdot p^{-1}\left\|\mathcal{P}_\Omega^{(-m)}\lb S^{t}-S^\star\rb \right\|_2 \\
    \leq & \lsb\frac{32 C_1}{C_{0}}\sqrt{\frac{\kappa\mu r}{n}}\gamma^{t} + 96C_1 C_{\mathrm{noise}}\lb\frac{\sigma}{\sigma_r^{\star}}\rb\sqrt{\frac{\kappa^2\mu r}{n}}\rsb
    \cdot \frac{1}{C_0}\frac{\sigma_{r}^{\star}}{\sqrt{\kappa}}\gamma\\
    \leq & \frac{1}{C_{0}}\left(\sigma_r^\star\sqrt{\frac{\mu r}{n}}\right) \gamma^{t+1}+C_{\mathrm{noise}}\sigma\sqrt{\frac{\kappa\mu r}{n}},
\end{align*}
where in the third inequality the loose bound of the spectral norm follows from \eqref{eq:S_bound} and \eqref{eq:noise_bound2}, and the last inequality holds if $C_0\geq 96C_1$. 

\noindent\textbf{$\bullet$ Bounding $\zeta_{3}$.}
\begin{align*}
    \ln \zeta_3\rn_{\mathrm{F}}
    \leq & p^{-1}\ln \lb U^{t}\rb^T\left(\mathcal{P}_\Omega^{(-m)} - \mathcal{P}_\Omega\right)\lb S^t-S^\star\rb \rn_{\mathrm{F}}+p^{-1}\ln \left(\mathcal{P}_\Omega^{(-m)} - \mathcal{P}_\Omega\right)\lb S^t-S^\star\rb V^t \rn_{\mathrm{F}} \\
    = & \underbrace{p^{-1}\ln \lsb\left(\mathcal{P}_\Omega-\mathcal{P}_\Omega^{(-m)}\right)\lb S^t-S^\star\rb\rsb^T U^t \rn_{\mathrm{F}}}_{\zeta_{3,1}}+\underbrace{p^{-1}\ln \left(\mathcal{P}_\Omega-\mathcal{P}_\Omega^{(-m)}\right)\lb S^t-S^\star\rb V^t \rn_{\mathrm{F}}}_{\zeta_{3,2}}.  
\end{align*}
We only derive the bound for $\zeta_{3,2}$, since $\zeta_{3,1}$ has the same bound. If $m\leq n$, 
$$
\zeta_{3,2} = p^{-1}\ln e_m^T\Po\lb S^t-S^{\star}\rb V^{t}\rn_2,
$$ 
and if $m>n$,
$$\zeta_{3,2} = p^{-1}\ln \Po\lb S^t-S^{\star}\rb e_{(m-n)}e_{(m-n)}^T V^{t}\rn_{\mathrm{F}}.$$
In both cases,
\begin{align*}
    \zeta_{3,2}\leq &p^{-1}\cdot\lb 2\alpha p n\rb\cdot C_{\mathrm{thresh}}\lsb \lb\frac{\mu r}{n}\sigma_{1}^{\star}\rb\gamma^t+2C_N\sigma\sqrt{\log n}\rsb
    \cdot\lb 1+\frac{6C_1}{C_0}\rb\sqrt{\frac{\kappa\mu r}{n}} \\
    \leq &\frac{1}{C_0}\lb \sigma_{r}^{\star}\sqrt{\frac{\kappa\mu r}{n}}\rb\gamma^{t+1}+5C_N\lb C_{\mathrm{thresh}}\alpha n\sqrt{\log n}\rb\sigma\sqrt{\frac{\kappa\mu r}{n}},
\end{align*}
where the last inequality holds if $C_0\geq 24C_1$ and
$
\alpha \leq \frac{1}{3C_0}\frac{1}{\kappa\mu r}\cdot\frac{\gamma}{C_{\mathrm{thresh}}}.
$

\noindent\textbf{Part \RomanNumeralCaps{2}: Bound for $\ln \xi\rn_{\mathrm{F}}$.}
\begin{align*}
    &\xi = \left( \mathcal{I}-p^{-1}\mathcal{P}_{T^t}\mathcal{P}_\Omega\right)\lb L^t-L^\star\rb -\left( \mathcal{I}-\mathcal{P}_{T^{t,m}}\left(\I-\Ho^{(-m)}\right)\right)\lb L^{t,m}-L^\star\rb \\
    = &\left(\mathcal{I}-\mathcal{P}_{T^t}\right)\lb L^t-L^\star\rb 
     + \mathcal{P}_{T^t}\Ho\lb L^t-L^\star\rb \\
    & -\left(\mathcal{I}-\mathcal{P}_{T^{t,m}}\right)\lb L^{t,m}-L^\star\rb 
     - \mathcal{P}_{T^{t,m}}\Ho^{(-m)}\lb L^{t,m}-L^\star\rb \\
     = & \left( \mathcal{P}_{T^t}-\mathcal{P}_{T^{t,m}}\right)\lb L^\star\rb
     + \left( \mathcal{P}_{T^t}-\mathcal{P}_{T^{t,m}} \right)\Ho\lb L^t-L^\star\rb\\
     & + \mathcal{P}_{T^{t,m}}\Ho\lb L^t-L^\star\rb 
     - \mathcal{P}_{T^{t,m}}\Ho^{(-m)}\lb L^{t,m}-L^\star\rb \\
     = & \underbrace{\left( \mathcal{P}_{T^t}-\mathcal{P}_{T^{t,m}}\right)\lb L^\star\rb }_{\xi_{1}}
     + \underbrace{\left( \mathcal{P}_{T^t}-\mathcal{P}_{T^{t,m}} \right)\Ho\lb L^t-L^\star\rb }_{\xi_{2}}\\
     & + \underbrace{\mathcal{P}_{T^{t,m}}\Ho\lb L^t-L^{t,m}\rb}_{\xi_{3}} 
     + \underbrace{\mathcal{P}_{T^{t,m}}\lb\Ho-\Ho^{(-m)}\rb\lb L^{t,m}-L^\star\rb }_{\xi_{4}}.
\end{align*}

\noindent\textbf{$\bullet$ Bounding $\xi_{1}$.}  According to \eqref{eq:T_diff_decomp},
\begin{align*}
    \left\|\xi_{1}\right\|_{\mathrm{F}}
    \leq & \left\|U^t\lb U^t \rb^T -U^{t,m}\lb U^{t,m} \rb^T \right\|_\mathrm{F}
    \cdot \left\|L^\star-L^{t}\right\|_2\\
    &+\left\|L^\star-L^{t,m}\right\|_2\cdot
    \left\|V^t\lb V^t \rb^T -V^{t,m}\lb V^{t,m} \rb^T \right\|_\mathrm{F}\\
    \leq & 2  \left\|D^{t,0,m}\right\|_{\mathrm{F}}
    \cdot2\left\|E^{t-1}\right\|_2
     +  2\left\|E^{t-1,m}\right\|_2 
    \cdot 2\left\|D^{t,0,m}\right\|_{\mathrm{F}} \\
    \leq & 8\cdot\lsb\frac{8 C_1}{C_{0}}\sqrt{\frac{\kappa\mu r}{n}}\gamma^{t}+24C_1 C_{\mathrm{noise}}\lb\frac{\sigma}{\sigma_r^{\star}}\rb\sqrt{\frac{\kappa^2\mu r}{n}}\rsb 
    \cdot \frac{2}{C_{0}}\frac{\sigma_{r}^{\star}}{\sqrt{\kappa}}\gamma \\
    \leq & \frac{1}{C_0}\left(\sigma_r^\star\sqrt{\frac{\mu r}{n}}\right) \gamma^{t+1}+C_{\mathrm{noise}}\sigma\sqrt{\frac{\kappa\mu r}{n}},
\end{align*}
provided that $C_0\geq 384 C_1$.

\noindent\textbf{$\bullet$ Bounding $\xi_{2}$.} Following the decomposition \eqref{eq:T_diff_decomp} and one can get
\begin{align*}
    \left\|\xi_{2}\right\|_{\mathrm{F}}
    \leq & \left\|U^t\lb U^t \rb^T -U^{t,m}\lb U^{t,m} \rb^T \right\|_{\mathrm{F}} \cdot\left\|\mathcal{H}_\Omega \lb L^t-L^\star\rb \right\|_2\\
    & +\left\|\mathcal{H}_\Omega \lb L^t-L^\star\rb \right\|_2\cdot\left\|V^t\lb V^t \rb^T -V^{t,m}\lb V^{t,m} \rb^T \right\|_{\mathrm{F}} \\
    \leq & 4 \left\|D^{t,0,m}\right\|_{\mathrm{F}} 
    \cdot \left\|\mathcal{H}_\Omega \lb L^t-L^\star\rb \right\|_2\\
    \leq & \lsb\frac{32 C_1}{C_{0}}\sqrt{\frac{\kappa\mu r}{n}}\gamma^{t} + 96C_1 C_{\mathrm{noise}}\lb\frac{\sigma}{\sigma_r^{\star}}\rb\sqrt{\frac{\kappa^2\mu r}{n}}\rsb
    \cdot \frac{1}{2C_0}\frac{\sigma_{r}^{\star}}{\sqrt{\kappa}}\gamma\\
    \leq & \frac{1}{C_{0}}\left(\sigma_r^\star \sqrt{\frac{\mu r}{n}}\right) \gamma^{t+1}+C_{\mathrm{noise}}\sigma\sqrt{\frac{\kappa\mu r}{n}},
\end{align*}
where the loose bound of $\left\|\mathcal{H}_\Omega \lb L^t-L^\star\rb \right\|_2$ follows \eqref{eq:psi_2} and \eqref{eq:noise_bound2}.

\noindent\textbf{$\bullet$ Bounding $\xi_{3}$.} 
\begin{align*}
    \ln\xi_{3}\rn_{\mathrm{F}} = & \ln\mathcal{P}_{T^{t,m}}\Ho\lb L^t-L^{t,m}\rb\rn_{\mathrm{F}}\\
    \leq & \ln \lb U^{t,m}\rb^T\Ho\lb L^t-L^{t,m}\rb\rn_{\mathrm{F}}+\ln \Ho\lb L^t-L^{t,m}\rb V^{t,m}\rn_{\mathrm{F}} \\
    = & \underbrace{\ln \lsb\Ho\lb L^t-L^{t,m}\rb\rsb^TU^{t,m}\rn_{\mathrm{F}}}_{\xi_{3,1}}+\underbrace{\ln \Ho\lb L^t-L^{t,m}\rb V^{t,m}\rn_{\mathrm{F}}}_{\xi_{3,2}}.
\end{align*}
We only derive the bound for $\xi_{3,2}$ in the following, since $\xi_{3,1}$ has the same bound. Following the decomposition in \eqref{eq:L_diff_decomp}, we can get
\begin{align*}
    \xi_{3,2} \leq & \underbrace{\ln\Ho\lb D_U^{t, m, 0} \Si^{t, m} \lb V^{t, m}\rb^T\rb V^{t,m}\rn_{\mathrm{F}}}_{\xi_{3,2}^a}+\underbrace{\ln\Ho\lb U^{t} S^{t, m,0}\lb V^{t, m}\rb^T\rb V^{t,m}\rn_{\mathrm{F}}}_{\xi_{3,2}^b} \\
    & + \underbrace{\ln\Ho\lb U^{t}\Si^{t} \lb D_V^{t, 0, m}\rb^T \rb V^{t,m}\rn_{\mathrm{F}}}_{\xi_{3,2}^c}.
\end{align*}

For $\xi_{3,2}^a$, the application of Lemma~\ref{lem:bound1} yields that
\begin{align*}
\xi_{3,2}^a
\leq & \ln D_U^{t, m, 0} \Si^{t, m} \rn_{\mathrm{F}}\cdot\max_{1\leq j\leq n} \ln R^{(j)} \rn_2,
\end{align*}
where
$$
R^{(j)} := \sum_{k=1}^n \lb 1- \delta_{jk}/p\rb \lb V^{t,m}_{k,:}\rb^T V^{t,m}_{k,:}.
$$ 
For each $j\in\{1,\cdots,n\}$, denoting $
H^{(j)} := \diag\lb 1-\delta_{j1}/p,\cdots,1-\delta_{jn}/p\rb\in\mathbb{R}^{n\times n}
$,
\begin{align*}
& \ln R^{(j)} \rn_2
= \ln\lb V^{t, m}\rb^T H^{(j)}V^{t,m}\rn_2 \\
= & \ln\lb D_V^{t,m,j}+V^{t, j}G^{t,m,j}\rb^T H^{(j)}\lb D_V^{t,m,j}+V^{t,j}G^{t,m,j}\rb\rn_2 \\
\leq & \ln\lb D_V^{t,m,j}\rb^T H^{(j)} D_V^{t,m,j}\rn_2+2\ln\lb D_V^{t,m,j}\rb^T H^{(j)}V^{t,j}\rn_2+\ln\lb V^{t, j}\rb^T H^{(j)} V^{t,j}\rn_2.
\end{align*}

Applying Lemma~\ref{lem:bernstein} to bound the 2-norm of
$
H^{(j)} = \sum_{k=1}^n (1-\delta_{jk}/p)\cdot e_ke_k^T,
$
and with high probability we can get
$$
\ln H^{(j)} \rn_2 \leq C_1\cdot\lb\sqrt{\frac{\log n}{p}}+\frac{\log n}{p}\rb\leq 2C_1\frac{\log n}{p}.
$$
Therefore,
\begin{align*}
\ln\lb D_V^{t,m,j}\rb^T H^{(j)} D_V^{t,m,j}\rn_2 \leq & \ln H^{(j)}\rn_2 \ln D_V^{t,m,j}\rn_{\mathrm{F}}^2
\leq 2C_1\frac{\log n}{p}\cdot\lb\frac{9C_1}{C_0} \sqrt{\frac{\kappa\mu r}{n}}\gamma\rb^2
\leq \frac{\gamma}{800\kappa},
\end{align*}
where in the second inequality $\ln D^{t,\infty}\rn_{\mathrm{F}}\leq\frac{9C_1}{C_0}\sqrt{\frac{\kappa\mu r}{n}}\gamma$ is used, and the last inequality holds if $C_0\geq 90 C_1$ and
$
p\geq 16C_1\cdot\frac{\kappa^2\mu r\log n}{n}.
$

Since $V^{t, j}$ is independent with respect to the random variables on the $j$-th row, similarly as \eqref{eq:HV_2norm} we can get with high probability 
\begin{align*}
\ln H^{(j)}V^{t,j} \rn_2 
\leq \frac{1}{20}\sqrt{\frac{n}{\kappa^3\mu r}},
\end{align*}
provided that
$
p\geq 60^2C_1^2\cdot\frac{\kappa^3\mu r^2\log n}{n}.
$
As a result,
\begin{align*}
\ln\lb D_V^{t,m,j}\rb^T H^{(j)}V^{t,j}\rn_2 
\leq & \frac{9C_1}{C_0} \sqrt{\frac{\kappa\mu r}{n}}\gamma\cdot\frac{1}{20}\sqrt{\frac{n}{\kappa^3\mu r}}\leq \frac{\gamma}{800\kappa},
\end{align*}
as long as $C_0\geq 360 C_1$.

Due to the independence again,
\begin{align*}
\ln\lb V^{t, j}\rb^T H^{(j)} V^{t,j}\rn_2 = & \ln\sum_{k=1}^n \lb 1- \delta_{jk}/p\rb \lb V^{t,j}_{k,:}\rb^T  V^{t,j}_{k,:}\rn_2\leq \frac{\gamma}{800\kappa},
\end{align*}
where the last bound follows from the same argument for \eqref{eq:bernstein_t=1}, provided that
$
p \geq \frac{1800^2C_1^2}{\gamma^2}\cdot\frac{\kappa^3\mu r\log n}{n}
$ and $C_0\geq 48C_1$. Provided that $C_0\geq 48$,
\begin{align*}
    \xi_{3,2}^a\leq &\ln D_U^{t, m, 0} \Si^{t, m} \rn_{\mathrm{F}} \cdot \frac{\gamma}{200\kappa} \\
    \leq &\lsb\frac{8C_1}{C_0}\sqrt{\frac{\kappa\mu r}{n}}\gamma^t+24C_1 C_{\mathrm{noise}}\lb\frac{\sigma}{\sigma_r^{\star}}\rb\sqrt{\frac{\kappa^2\mu r}{n}}\rsb\cdot\lb 1+\frac{2}{C_0}\rb\sigma_1^{\star}\cdot \frac{\gamma}{200\kappa}\\
    \leq &\frac{C_1}{8C_0}\lb \sigma_r^{\star}\sqrt{\frac{\kappa\mu r}{n}}\rb\gamma^{t+1}+\frac18C_1 C_{\mathrm{noise}}\sigma\sqrt{\frac{\kappa^2\mu r}{n}}.
\end{align*}

For $\xi_{3,2}^b$, we need a bound for $\ln S^{t,m,0} \rn_{\mathrm{F}}$. We remind the reader that the bound for $\|D^{t,\infty}\|_{\mathrm{F}}$ is indeed provided through the bound for
$\ln W^{t-1,m} F^{t,m}\rn_{\mathrm{F}}$, see \eqref{eq:WF_init} for the case when $t=1$ and  \eqref{eq:WF_induction} for the case when $t>1$. Therefore, in the $t$-th iteration, we have already proved that
$$
\ln W^{t-1,m} F^{t,m}\rn_{\mathrm{F}}
\leq\frac{2C_1}{C_0}
\lb\sigma_r^{\star}\sqrt{\frac{\kappa\mu r}{n}}\rb\gamma^t+6C_1C_{\mathrm{noise}}\sigma\sqrt{\frac{\kappa^2\mu r}{n}}.
$$
For the base case, such a bound can be established by substituting the bound of $\ln \De^{1,\infty}\rn_{2,\infty}$ into \eqref{eq:WF_init_bound}. For the induction steps, such a bound follows \eqref{eq:WF_induct_bound} with the change of the numbering from $t$ to $t-1$, since $\ln W^{t-1,m}F^{t,m}\rn_{\mathrm{F}}\leq 2\ln E^{t-1}-E^{t-1,m}\rn_{\mathrm{F}}$. Applying Lemma~\ref{lem:perturb_S} with $A=\widehat{L^{\star}}+\widehat{E^{t-1,m}}$ and $\widetilde{A}=\widehat{L^{\star}}+\widehat{E^{t-1}}$ yields that
\begin{equation}\label{eq:S_F}
\begin{aligned}
\ln S^{t,m,0} \rn_{\mathrm{F}} = &\ln \Sigma^{t,m}G^{t,0,m}-G^{t,0,m}\Sigma^{t}\rn_{\mathrm{F}}\\
\leq & 4\kappa\cdot\ln W^{t-1,m} F^{t,m}\rn_{\mathrm{F}}
\leq \frac{8C_1}{C_0}\lb\sigma_1^{\star}\sqrt{\frac{\kappa\mu r}{n}}\rb\gamma^{t}+24C_1 C_{\mathrm{noise}}\sigma\sqrt{\frac{\kappa^4\mu r}{n}}.
\end{aligned}
\end{equation}
Due to such a bound of $\ln S^{t,m,0}\rn_{\mathrm{F}}$, $\xi_{3,2}^b$ has the same bound as $\xi_{3,2}^a$. 

For $\xi_{3,2}^c$,  
\begin{align*}
\xi_{3,2}^c = & \max_{\ln Z\rn_{\mathrm{F}}=1} \left\langle\Ho\lb U^{t}\Si^{t} \lb D_V^{t, 0, m}\rb^T \rb V^{t,m},~Z\right\rangle \\
= & \max_{\ln Z\rn_{\mathrm{F}}=1} \left\langle\Ho\lb U^{t}\Si^{t} \lb D_V^{t, 0, m}\rb^T \rb,~Z\lb V^{t,m}\rb^T\right\rangle \\
\leq & \ln\Ho\lb \bm{1}\bm{1}^T\rb\rn_2 \ln U^{t}\Si^{t} \rn_{2,\infty}\ln D_V^{t, 0, m} \rn_{\mathrm{F}}\ln V^{t,m} \rn_{2,\infty},
\end{align*}
where the inequality follows from  Lemma~\ref{lem:bound2}. When $t=1$,
\begin{align*}
\xi_{3,2}^c\leq & \ln\Ho\lb \bm{1}\bm{1}^T\rb\rn_2 \ln U^{1}\Si^{1} \rn_{2,\infty}\ln V^{1,m} \rn_{2,\infty}\ln D_V^{1, 0, m} \rn_{\mathrm{F}}\\
\leq & 2C_1\sqrt{\frac{n\log n}{p}}\cdot\lb 1+\frac{6C_1}{C_0}\rb\sqrt{\frac{\kappa\mu r}{n}}\lb 1+\frac{2}{C_0}\rb\sigma_1^{\star}\cdot\lb 1+\frac{6C_1}{C_0}\rb\sqrt{\frac{\kappa\mu r}{n}}\\
&\cdot\lsb\frac{8C_1}{C_0}\sqrt{\frac{\mu r}{n}}\gamma+12C_1 C_{\mathrm{noise}}\lb\frac{\sigma}{\sigma_r^{\star}}\rb\sqrt{\frac{\kappa\mu r}{n}}\rsb \\
\leq & \frac{1}{C_0}\lb\sigma_r^{\star}\sqrt{\frac{\kappa\mu r}{n}}\rb\gamma^2+C_{\mathrm{noise}}\sigma\sqrt{\frac{\kappa^2\mu r}{n}},
\end{align*}
provided that $C_0\geq 72C_1$, $C_0\geq 58$ and
$
p\geq\frac{30^2 C_1^4}{\gamma^2}\cdot\frac{\kappa^3\mu^2 r^2\log n}{n}.
$
When $t\geq 2$,
\begin{align*}
\xi_{3,2}^c\leq & \ln\Ho\lb \bm{1}\bm{1}^T\rb\rn_2 \ln U^{t}\Si^{t} \rn_{2,\infty}\ln V^{t,m} \rn_{2,\infty}\ln D_V^{t, 0, m} \rn_{\mathrm{F}}\\
\leq & 2C_1\sqrt{\frac{n\log n}{p}}\cdot\lb 1+\frac{6C_1}{C_0}\rb\sqrt{\frac{\mu r}{n}}\lb 1+\frac{2}{C_0}\rb\sigma_1^{\star}\cdot\lb 1+\frac{6C_1}{C_0}\rb\sqrt{\frac{\mu r}{n}}\\
&\cdot\lsb\frac{8C_1}{C_0}\sqrt{\frac{\kappa\mu r}{n}}\gamma^t+24C_1 C_{\mathrm{noise}}\lb\frac{\sigma}{\sigma_r^{\star}}\rb\sqrt{\frac{\kappa^2\mu r}{n}}\rsb \\
\leq & \frac{1}{C_0}\lb\sigma_r^{\star}\sqrt{\frac{\kappa\mu r}{n}}\rb\gamma^{t+1}+C_{\mathrm{noise}}\sigma\sqrt{\frac{\kappa^2 \mu r}{n}},
\end{align*}
where the tight bounds for $\ln U^{t}\rn_{2,\infty}$ and $\ln V^{t,m}\rn_{2,\infty}$ follow from \eqref{eq:l_2_infty} and \eqref{eq:noise_bound2}, and the last inequality holds provided that
$
p\geq\frac{60^2 C_1^4}{\gamma^2}\cdot\frac{\kappa^2\mu^2 r^2\log n}{n}.
$

Combining the bounds of $\xi_{3,2}^a$ to $\xi_{3,2}^c$, we have
$$
\xi_{3} \leq \frac{4+C_1}{2C_0}\lb\sigma_r^{\star}\sqrt{\frac{\kappa\mu r}{n}}\rb\gamma^{t+1}+\lb 2+\frac{1}{2}C_1\rb C_{\mathrm{noise}}\sigma\sqrt{\frac{\kappa^2\mu r}{n}}.
$$

\noindent\textbf{$\bullet$ Bounding $\xi_{4}$.} Consider $m\leq n$. When $m> n$, the proof can be done similarly.
\begin{align*}
\ln\xi_{4}\rn_{\mathrm{F}} = & \ln\mathcal{P}_{T^{t,m}}\lb\Ho-\Ho^{(-m)}\rb\lb L^{t,m}-L^\star\rb\rn_{\mathrm{F}} \\
\leq & \ln \lb U^{t,m}\rb^T \lb\Ho-\Ho^{(-m)}\rb\lb L^{t,m}-L^\star\rb \rn_{\mathrm{F}}+\ln \lb\Ho-\Ho^{(-m)}\rb\lb L^{t,m}-L^\star\rb  V^{t,m}\rn_{\mathrm{F}} \\
= & \ln \lb U^{t,m}\rb^T e_m e_m^T\Ho\lb L^{t,m}-L^\star\rb \rn_{\mathrm{F}}+\ln e_m^T\Ho\lb L^{t,m}-L^\star\rb  V^{t,m}\rn_{2} \\
\leq &\ln U^{t,m}\rn_{2,\infty}\cdot\ln \Ho\lb L^{t,m}-L^\star\rb  \rn_2+\ln e_m^T\Ho\lb L^{t,m}-L^\star\rb  V^{t,m}\rn_{2} \\
\leq & \lb 1+\frac{6C_1}{C_0}\rb\sqrt{\frac{\kappa\mu r}{n}}\cdot\lb\frac{1}{4C_0}\frac{\sigma_r^{\star}}{\kappa}\gamma^{t+1}+\frac12 C_{\mathrm{noise}}\sigma\rb+\frac{C_1}{10C_0}\lb\sigma_r^{\star}\sqrt{\frac{\mu r} {n}}\rb\gamma^{t+1}+\frac{1}{10}C_1C_{\mathrm{noise}}\sigma\sqrt{\frac{\kappa\mu r} {n}} \\
\leq & \frac{C_1}{8C_0}\lb \sigma_r^{\star}\sqrt{\frac{\mu r} {n}}\rb\gamma^{t+1}+\frac{1}{8}C_1C_{\mathrm{noise}}\sigma\sqrt{\frac{\kappa\mu r} {n}}, 
\end{align*}
where in the third inequality the bound for the second term follows from \eqref{eq:beta_31} if 
$
p\geq\frac{900C_{0}^2}{\gamma^2}\cdot\frac{\kappa\mu^2r^2\log n}{n}
$ and $C_0\geq 100C_1$, and the last inequality holds if $C_1\geq 40$.

\noindent\textbf{Part \RomanNumeralCaps{3}: Bound for $\ln \tau\rn_{\mathrm{F}}$.}
\begin{align*}
    & \tau = p^{-1}\mathcal{P}_{T^t}\mathcal{P}_\Omega\lb N \rb
    -p^{-1}\mathcal{P}_{T^{t,m}}\mathcal{P}_\Omega^{(-m)}\lb N \rb \\
    = & \underbrace{p^{-1}\mathcal{P}_{T^t}\left(\mathcal{P}_\Omega - \mathcal{P}_\Omega^{(-m)}\right)\lb N\rb}_{\tau_1}
      + \underbrace{p^{-1}\left(\mathcal{P}_{T^t}-\mathcal{P}_{T^{t,m}}\right)\mathcal{P}_\Omega^{(-m)}\lb N\rb}_{\tau_2}.
\end{align*}

\noindent\textbf{$\bullet$ Bounding $\tau_{1}$.}
\begin{align*}
    \ln \tau_1\rn_{\mathrm{F}} \leq & p^{-1}\ln \lb U^{t}\rb^T\left(\mathcal{P}_\Omega - \mathcal{P}_\Omega^{(-m)}\right)\lb N\rb \rn_{\mathrm{F}}+p^{-1}\ln \left(\mathcal{P}_\Omega - \mathcal{P}_\Omega^{(-m)}\right)\lb N\rb V^t \rn_{\mathrm{F}} \\
    = & \underbrace{p^{-1}\ln \lsb\left(\mathcal{P}_\Omega - \mathcal{P}_\Omega^{(-m)}\right)\lb N\rb \rsb^T U^t \rn_{\mathrm{F}}}_{\tau_{1,1}}+\underbrace{p^{-1}\ln \left(\mathcal{P}_\Omega - \mathcal{P}_\Omega^{(-m)}\right)\lb N\rb V^t \rn_{\mathrm{F}}}_{\tau_{1,2}}.  
\end{align*}
We only derive the bound for $\tau_{1,2}$, since $\tau_{1,1}$ has the same bound. If $m\leq n$,
\begin{align*}
    \tau_{1,2} = & p^{-1}\ln e_m^T\Po\lb N\rb V^{t}\rn_2 \\
    \leq & p^{-1}\ln e_m^T\Po\lb N\rb D_V^{t,0,m}\rn_2+p^{-1}\ln e_m^T\Po\lb N\rb V^{t,m}\rn_2 \\
    \leq & p^{-1}\ln \Po\lb N\rb \rn_2\cdot\ln D_V^{t,0,m}\rn_{\mathrm{F}}+p^{-1}\ln e_m^T\Po\lb N\rb V^{t,m}\rn_2 \\
    \leq &\lb C_N\sqrt{\frac{n}{p}}\rb\sigma\cdot\frac{9C_1}{C_0}\sqrt{\frac{\kappa\mu r}{n}}+C_1\lb C_N\sqrt{\frac{n\log n}{p}}\rb\sigma\cdot\lb 1+\frac{6C_1}{C_0}\rb\sqrt{\frac{\kappa\mu r}{n}}\\
    \leq & \lb 1+\frac{33}{32} C_1\rb\lb C_N\sqrt{\frac{n\log n}{p}}\rb\sigma\sqrt{\frac{\kappa\mu r}{n}},
\end{align*}
where the third inequality follows from Lemma~\ref{lem:noise} and $\ln D^{t,\infty}\rn_{\mathrm{F}}\leq\frac{9C_1}{C_0}\sqrt{\frac{\kappa\mu r}{n}}\gamma$, and the last inequality holds provided that $C_{0}\geq 192C_1$. If $m>n$, following a similar argument as in \eqref{eq:B3} and one can get
\begin{align*}
    \tau_{1,2} = p^{-1}\ln \Po\lb N\rb e_{(m-n)}e_{(m-n)}^T V^{t}\rn_{\mathrm{F}} \leq 2\lb C_N\sqrt{\frac{n}{p}}\rb\sigma\sqrt{\frac{\kappa\mu r}{n}}.
\end{align*}

\noindent\textbf{$\bullet$ Bounding $\tau_{2}$.} Following the decomposition \eqref{eq:T_diff_decomp} again,
\begin{align*}
    \ln \tau_{2} \rn_{\mathrm{F}}
    \leq & 4\left\| D^{t,0,m} \right\|_\mathrm{F}\cdot p^{-1}\left\|\mathcal{P}_\Omega^{(-m)}\lb N\rb \right\|_2 \\
    \leq & 4\cdot\frac{9C_1}{C_0}\sqrt{\frac{\kappa\mu r}{n}}
    \cdot \lb C_N\sqrt{\frac{n}{p}}\rb\sigma \leq \lb C_N\sqrt{\frac{n}{p}}\rb\sigma\sqrt{\frac{\kappa\mu r}{n}}.
\end{align*}

Combining the bounds of of $\zeta_{1}$ to $\zeta_{3}$ for the outlier part, $\xi_{1}$ to $\xi_{4}$ for the low-rank part and $\tau_{1}$ to $\tau_{2}$ for the noise part, we get
\begin{align*}
    \left\|E^{t}-E^{m}\right\|_{\mathrm{F}} \leq & \ln \zeta\rn_{\mathrm{F}} + \ln \xi\rn_{\mathrm{F}} + \ln \tau\rn_{\mathrm{F}} \\
    \leq &\frac{56+7C_1}{8C_0}\lb \sigma_{r}^{\star}\sqrt{\frac{\kappa\mu r}{n}}\rb\gamma^{t+1} + \lb 8+\frac{47}{16}C_1\rb C_{\text{noise}}\sigma\sqrt{\frac{\kappa^2\mu r}{n}}
    \\
    \leq & \frac{C_1}{C_0}\lb\sigma_{r}^{\star}\sqrt{\frac{\kappa\mu r}{n}}\rb\gamma^{t+1}+3C_1C_{\text{noise}}\sigma\sqrt{\frac{\kappa^2\mu r}{n}},
\end{align*}
where the last inequality holds provided that $C_1\geq 128$.

\end{appendix}

\end{document}